\documentclass[]{externalized}
\usepackage[T1]{fontenc}
\usepackage[utf8]{inputenc}
\usepackage{microtype}
\usepackage{inconsolata}
\usepackage{graphicx}
\usepackage{hyperref}
\usepackage{url}
\usepackage{booktabs}
\usepackage{amsfonts}
\usepackage{amsmath}
\usepackage{amssymb}
\usepackage{amsthm}
\usepackage{mathtools}
\usepackage{enumitem}
\usepackage{multirow}
\usepackage{subcaption}
\usepackage{xcolor}
\usepackage{listings}
\usepackage{natbib}
\usepackage{multicol}
\usepackage{pifont}
\usepackage{xspace}
\usepackage{wrapfig}
\usepackage{float}
\usepackage{tabularx}
\usepackage{ragged2e}
\usepackage{longtable}
\usepackage{colortbl}

\DeclareTextFontCommand{\tabletextbf}{\fontseries{bx}\selectfont}
\AtBeginEnvironment{tabular}{\let\textbf\tabletextbf}
\AtBeginEnvironment{tabular*}{\let\textbf\tabletextbf}
\AtBeginEnvironment{tabularx}{\let\textbf\tabletextbf}
\AtBeginEnvironment{longtable}{\let\textbf\tabletextbf}

\providecommand{\noopsort}[1]{}
\AtBeginDocument{%
  \providecommand\BibTeX{{%
    \normalfont B\kern-0.5em{\scshape i\kern-0.25em b}\kern-0.8em\TeX}}}

\makeatletter
\DeclareRobustCommand\onedot{\futurelet\@let@token\@onedot}
\def\@onedot{\ifx\@let@token.\else.\null\fi}

\makeatother

\definecolor{codegreen}{rgb}{0,0.6,0}
\definecolor{codegray}{rgb}{0.5,0.5,0.5}
\definecolor{codepurple}{rgb}{0.58,0,0.82}
\definecolor{backcolour}{rgb}{0.95,0.95,0.92}

\lstdefinestyle{mystyle}{
  backgroundcolor=\color{backcolour},
  commentstyle=\color{codegreen},
  keywordstyle=\color{magenta},
  numberstyle=\tiny\color{codegray},
  stringstyle=\color{codepurple},
  basicstyle=\ttfamily\footnotesize,
  breakatwhitespace=false,
  breaklines=true,
  captionpos=b,
  keepspaces=true,
  numbers=left,
  numbersep=5pt,
  showspaces=false,
  showstringspaces=false,
  showtabs=false,
  tabsize=2
}

\usepackage{amsmath,amsfonts,bm}

\def\eqref#1{equation~\ref{#1}}
\def\1{\bm{1}}

\DeclareMathAlphabet{\mathsfit}{\encodingdefault}{\sfdefault}{m}{sl}
\SetMathAlphabet{\mathsfit}{bold}{\encodingdefault}{\sfdefault}{bx}{n}

\newcommand{\brandlabel}[2]{%
  \raisebox{-0.15em}{\includegraphics[height=1em,width=1em,keepaspectratio]{#1}}%
  \hspace{0.35em}#2}
\newlength{\resourceiconheight}
\newcommand{\resourcelabel}[3]{%
  \begingroup
  \settoheight{\resourceiconheight}{M}%
  \includegraphics[height=\resourceiconheight,trim=#1,clip]{#2}%
  \hspace{0.35em}#3%
  \endgroup}
\newcommand{\modelname}[2]{\brandlabel{figures/model-icons/model-#1.png}{#2}}
\newcommand{\harnessname}[2]{\brandlabel{figures/harness-icons/#1.png}{#2}}
\newcommand{\checkerbenchtitlelogo}{%
  \raisebox{-0.62em}[0pt][0pt]{\includegraphics[height=2em,trim=85bp 306bp 85bp 270bp,clip]{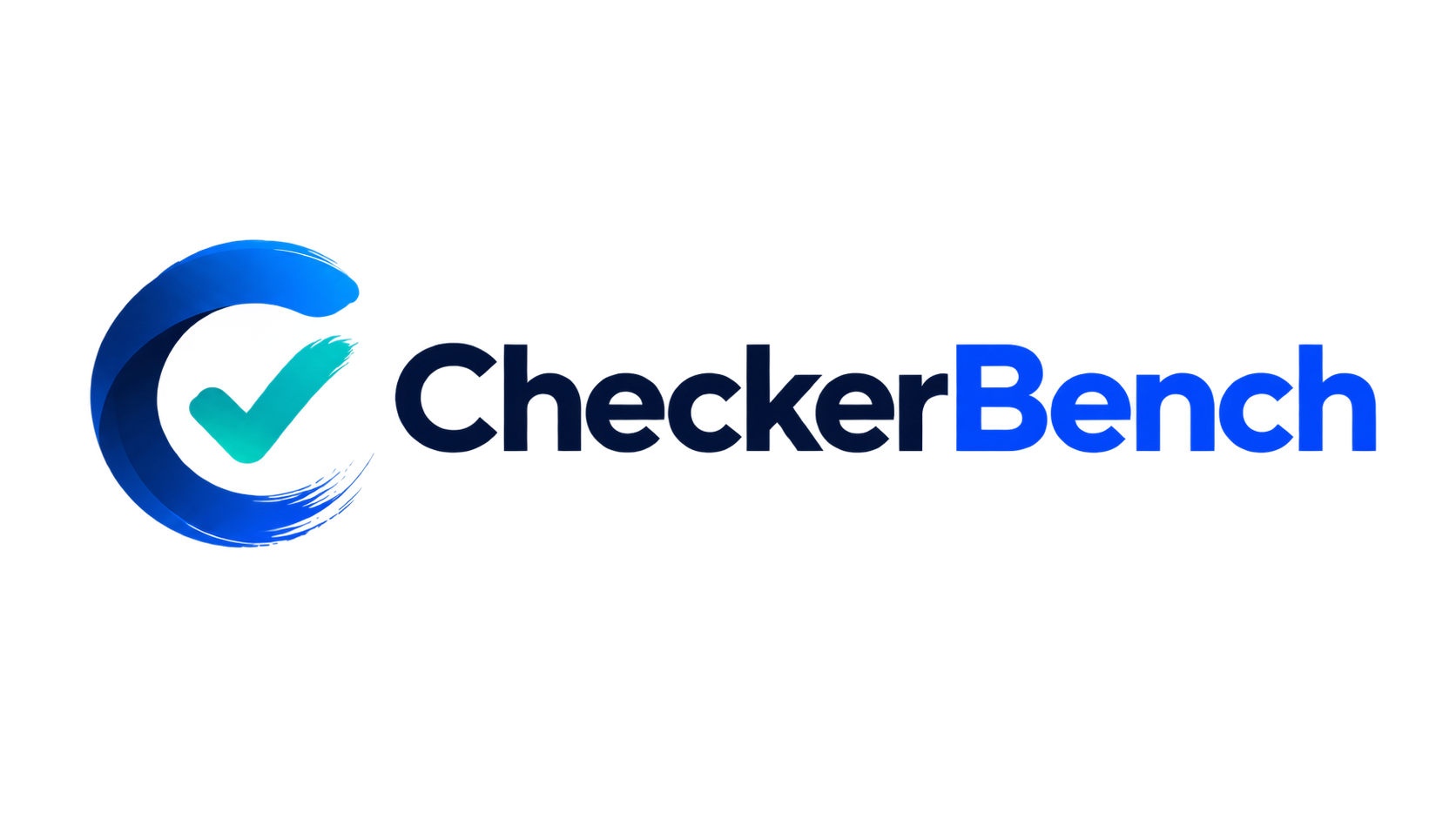}}}
\newcommand{\checkerbenchname}[1]{%
  \raisebox{-0.15em}{\includegraphics[height=1em,width=1em,keepaspectratio,trim=85bp 306bp 1222bp 270bp,clip]{figures/logo/checkerbench-logo-transparent.png}}%
  \hspace{0.35em}#1}

\input{sections/float-placement}
\usepackage{needspace}
\usepackage{fontawesome5}

\usepackage{tikz}
\usetikzlibrary{arrows.meta,calc,positioning}

\AddToHook{shipout/foreground}{%
  \ifnum\value{page}=2\relax
    \begin{tikzpicture}[remember picture,overlay]
      \node[anchor=north east,inner sep=0pt,outer sep=0pt]
        at ([xshift=\dimexpr1in+\hoffset+\oddsidemargin+\textwidth\relax,
             yshift=-1.05cm]current page.north west) {%
          \begin{minipage}{0.46\textwidth}
            \normalfont\rmfamily\itshape\fontsize{9}{10.5}\selectfont
            \setlength{\parindent}{0pt}%
            \setlength{\parskip}{0pt}%
            \raggedright
            \mbox{``Let agents turn past flaws into reusable checks,}\\
            \mbox{so yesterday's failures can guard tomorrow's code.''}\par
            \vspace{5pt}%
            \hrule height 0.4pt
          \end{minipage}%
        };
    \end{tikzpicture}%
  \fi
}

\title{\texorpdfstring{\checkerbenchtitlelogo}{CheckerBench}: Can Long-Horizon Agents Synthesize Static-Analysis Checkers?}
\author[*]{Hang He}
\author[*]{Li Wang}
\author{Hao Chen}
\author{Yuchen Shao}
\author{Yuling Shi}
\author{Lisheng Wang}
\author{Peiyang Liu}
\author{\\[-0.3ex]Goose Lin}
\author{Zaiyuan Wang}
\author{Haiying Sun}
\author{Ting Su}
\author[\mkern5mu\text{\faIcon[regular]{envelope}}]{Chengcheng Wan}

\contribution[*]{Equal Contribution}
\contribution[\text{\faIcon[regular]{envelope}}]{Corresponding Author}
\affiliation{East China Normal University, Humanlaya Data, Shanghai Jiao Tong University,\\[-0.2ex] Peking University, Shanghai Innovation Institute}

\checkdata[\protect\resourcelabel{0bp 0bp 0bp 0bp}{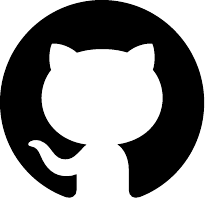}{GitHub}]{\protect\url{https://github.com/ahang0712/CheckerBench-Dataset}}
\checkdata[\protect\resourcelabel{15.25bp 18.5bp 13.75bp 24.25bp}{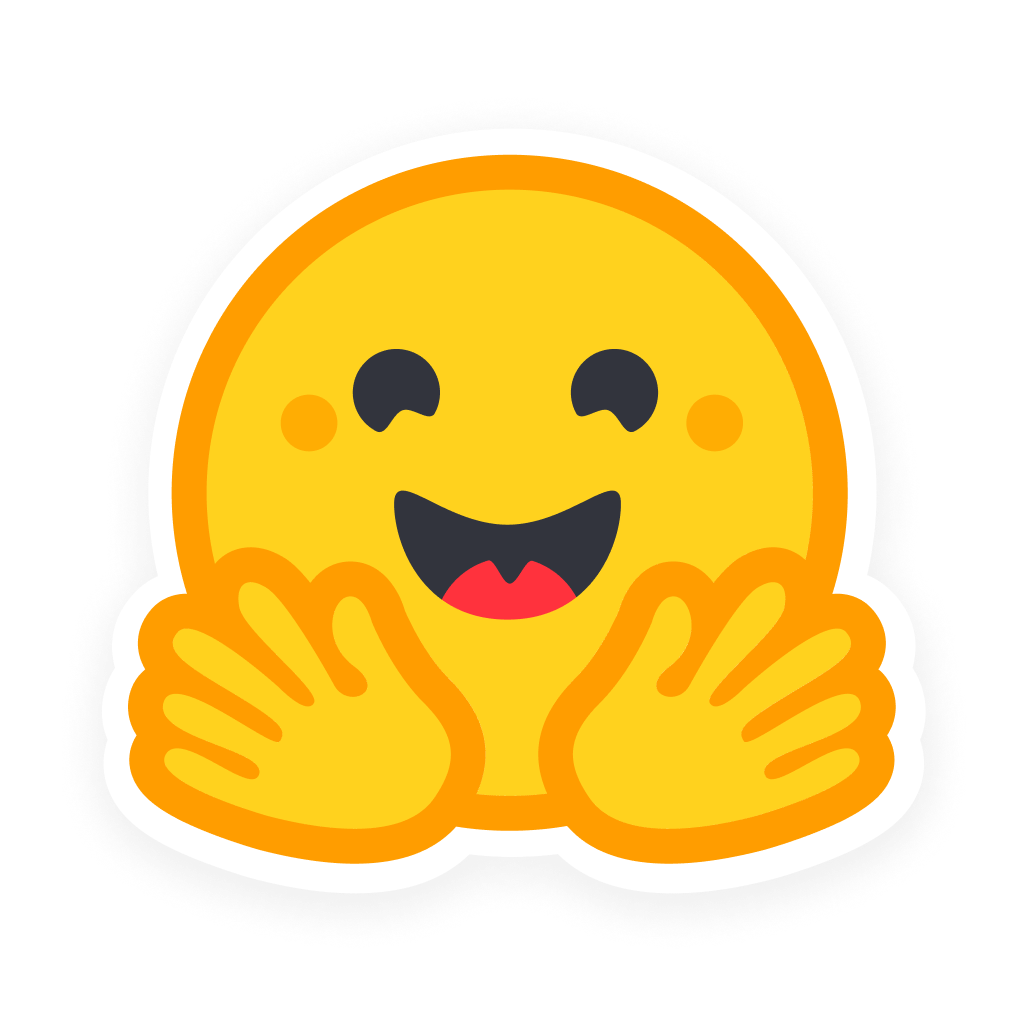}{Hugging Face}]{\protect\url{https://huggingface.co/datasets/Benchanything/checkerbench}}
\abstract{%
Static-analysis checker synthesis requires agents to interpret a defect
specification, inspect a repository, implement analyzer-specific logic, and
refine the checker through repeated compilation and analysis feedback. Existing
coding-agent benchmarks focus on tasks such as patch generation or vulnerability
detection and rarely assess whether an agent can develop a working checker in a
repository from start to finish. We introduce \textsc{CheckerBench},
an executable benchmark of 300 tasks derived from 297 CVEs across 167
repositories, 85 CWEs, and five language ecosystems. Each task
includes vulnerable and fixed revisions, a pinned analysis environment, and a
checker scaffold. We further introduce \textsc{CheckerLab}, a common evaluation
framework that independently rebuilds submitted checkers and measures
vulnerable--fixed diagnostic contrast, patch localization, false positives,
and tool use. Across 21 model--harness configurations and three independent repeats per configuration,
mean Pass@1 is 32.30\%, while the best reaches 45.33\%. These results show
that reliable, reusable checker development remains challenging for current
coding agents.
}

\begin{document}
\raggedbottom

\maketitle
% Rebuild artwork: python3 scripts/draw_table3_pass_at_1.py --paper.
\begingroup
\setlength{\intextsep}{0pt}
\begin{figure}[H]
\centering
\setlength{\abovecaptionskip}{4pt}
% Match the abstract frame, which extends 1.5pt past each text margin.
\makebox[\linewidth][c]{%
% Trim blank side and bottom padding while keeping the rotated labels intact.
\includegraphics[width=\dimexpr\linewidth+3pt\relax,trim=6.5bp 9bp 2.75bp 0,clip]{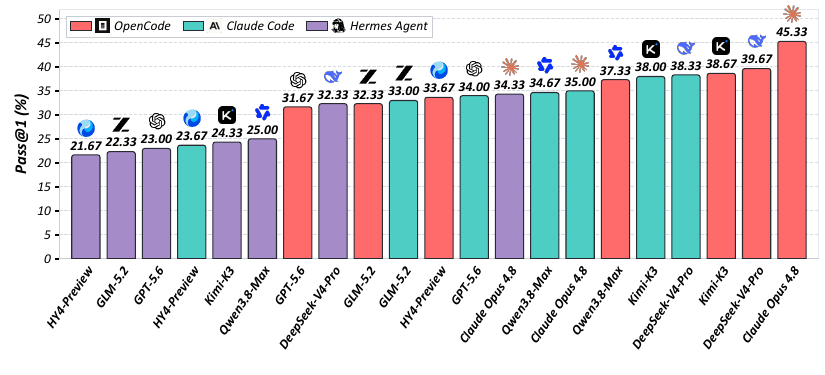}%
}\par
\caption{Pass@1 for 21 model--harness pairs on 300 \textsc{CheckerBench} tasks (Table~\ref{tab:overall-performance}).}
\label{fig:overall-pass-at-1}
\end{figure}
\endgroup

% Compact paragraph gaps in the main text; restore the template before references.
\setlength{\parskip}{4pt plus 1pt minus 1pt}
\section{Introduction}
\sloppy
\label{sec:introduction}

% Native editable source (Figure 2): output/powerpoint/standalone/Figure_2_editable.pptx
% Preserve the current direct PPT edits when exporting this artwork.
% Queued after Figure 1; top floats are suppressed on the title page.
\begin{figure}[tp]
\centering
\includegraphics[width=\linewidth,trim=0 8bp 0 16bp,clip]{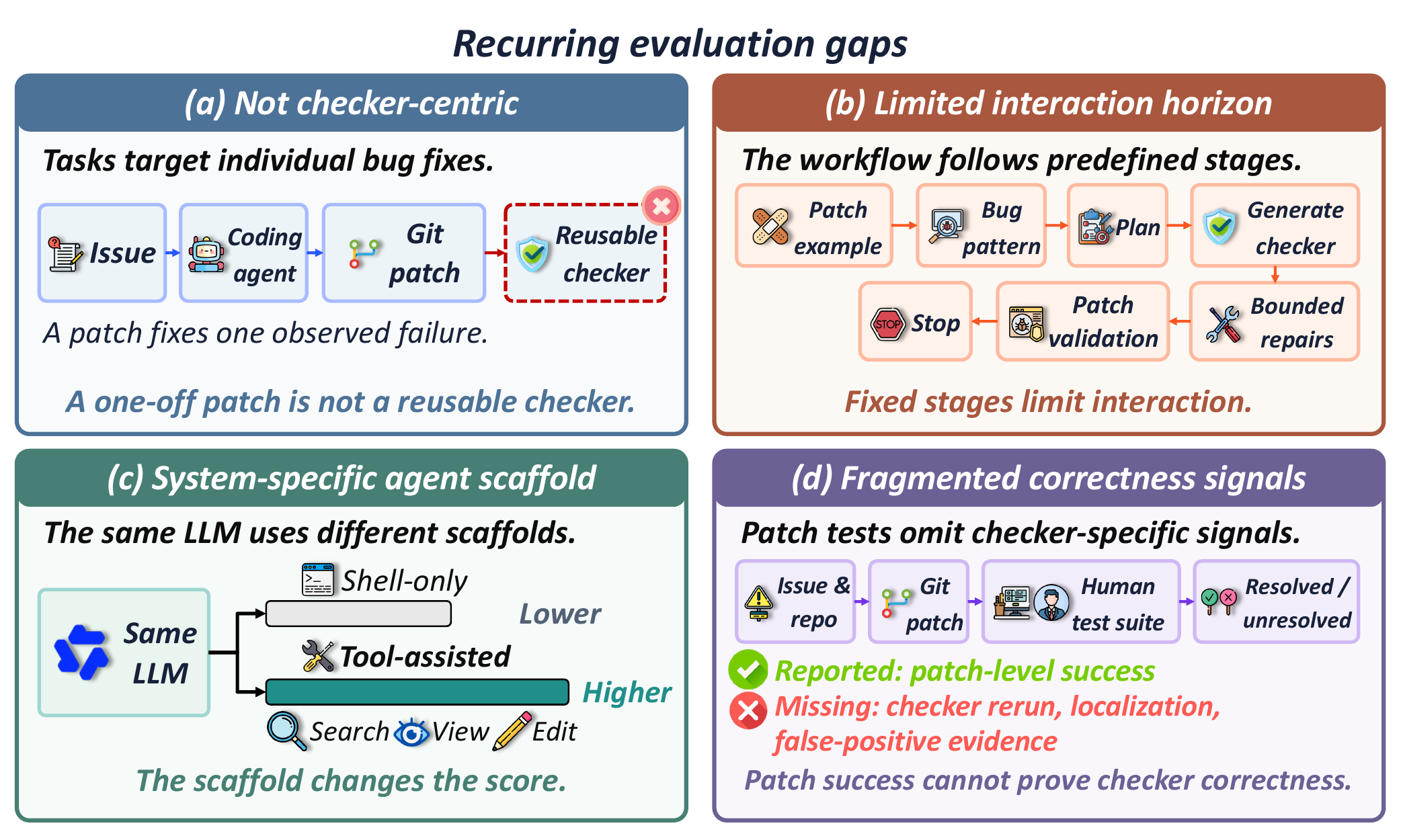}
\caption[Four recurring evaluation gaps]{\justifying Four recurring evaluation gaps:
(a) Not checker-centric~\citep{jimenez2024swebench};
(b) Limited interaction horizon~\citep{yang2025knighter};
(c) System-specific agent scaffold~\citep{yang2024sweagent}; and
(d) Fragmented correctness signals~\citep{pan2025swegym}.}
\label{fig:benchmark-limitations}
\par
\end{figure}

Static analysis provides a scalable approach for detecting complex program defects without executing the target programs
~\citep{li2025iris,he2025atomicity,chen2025safe,yang2025knighter,wang2026qlcoder}. However, existing static analyzers typically rely on manually designed \emph{checkers}, which require substantial expert effort to encode defect patterns and analysis strategies into tool-specific rules~\citep{garg2022rhosynth,yang2025knighter,wang2026qlcoder}. Recent advances in large language models (LLMs) have opened new opportunities for automated code understanding and defect detection. Nevertheless, directly applying LLMs to code snippets remains challenging due to high inference costs and limited access to repository-wide context, which can lead to incomplete reasoning and hallucinated findings~\citep{hao2023ev,mohajer2023skipanalyzer}.

To overcome these limitations, recent coding agents extend LLMs with the ability to interact with software repositories and external tools~\citep{chen2025toolforge,chen2026toolforge}, enabling iterative code modification, execution, and feedback-driven refinement. These capabilities provide a promising foundation for automating static-analysis checker development, which traditionally requires substantial manual engineering effort. Given a defect specification, an agent can navigate analyzer infrastructures, implement checker logic, integrate the checker into the build system, and iteratively refine its behavior based on compiler and analyzer feedback~\citep{jimenez2024swebench,wang2024openhands,ouyang2025repograph,xia2024agentless,wang2026repomemory}. However, synthesizing effective static-analysis checkers remains a challenging long-horizon task: agents must understand complex analyzer frameworks, make coordinated modifications across multiple files, and ensure that the resulting checker compiles, detects target defects, and maintains low false-positive rates~\citep{li2025iris,wu2024lemur,chen2025safe}. We refer to
this task as \textbf{\textit{long-horizon agentic static-analysis checker synthesis}}.

However, as illustrated in Figure~\ref{fig:benchmark-limitations}, existing
benchmarks are insufficient for evaluating long-horizon agentic
static-analysis checker synthesis.
\emph{First, current coding and vulnerability benchmarks are not checker-centric}:
repository-level benchmarks typically evaluate code completion, patch
generation, or feature implementation, while vulnerability benchmarks focus
on detection, classification, or repair rather than reusable checker
synthesis~\citep{liu2024repobench,jimenez2024swebench,zhou2026featurebench,
wang2024reposvul,wen2024vuleval,yildiz2025jitvul}.
\emph{Second, existing benchmarks provide limited task interaction}. They
either evaluate isolated tasks or decompose workflows into predefined stages,
preventing agents from performing the iterative planning, implementation,
and debugging required for checker development
~\citep{garg2022rhosynth,yang2025knighter,li2025mocq}.
\emph{Third, benchmarks differ substantially in environments, tool interfaces,
and evaluation protocols}, making results difficult to compare across
systems~\citep{yang2024sweagent,wang2024openhands,xia2024agentless}.
\emph{Finally, current evaluation metrics mainly measure patch-level
correctness or test-suite outcomes}, but do not capture checker-specific
properties such as vulnerability coverage, false-positive control, and
analyzer compatibility
~\citep{jimenez2024swebench,lee2025secbench,wang2026cybergym}.
These limitations motivate a unified benchmark that evaluates agents'
ability to autonomously synthesize reliable and reusable static-analysis
checkers.

To address these gaps, we introduce \textsc{CheckerBench}, the first benchmark
for evaluating long-horizon static-analysis checker synthesis.
\textsc{CheckerBench} contains 300 real-world tasks collected from 297 CVEs,
167 repositories, 85 CWEs, and five programming-language ecosystems, covering
diverse analyzer environments and defect patterns. Each task provides a
reproducible repository, an analyzer-specific environment, and reusable skills,
enabling agents to perform iterative development and validation rather than
isolated code generation.
We further develop \textsc{CheckerLab}, a unified evaluation framework that
assesses coding-agent systems under a common task specification and scoring
protocol. \textsc{CheckerLab} performs independent repeated generation and
frozen-candidate verification: each synthesized checker is rebuilt, executed
on vulnerable and fixed revisions, and evaluated based on diagnostic reduction,
patch localization, residual reports, and computational cost. Across 21
model--harness configurations, the average Pass@1 score is only 32.30\%, with
the best-performing configuration reaching 45.33\%
(Figure~\ref{fig:overall-pass-at-1}). These results demonstrate that
synthesizing reliable and reusable static-analysis checkers remains a
challenging problem for current coding agents.
In summary, our contributions are threefold.
\begin{itemize}[leftmargin=*, itemsep=0pt, topsep=2pt, parsep=0pt]
    \item \textbf{To the best of our knowledge, we are the first to build a
    comprehensive executable benchmark for evaluating repository-level,
    long-horizon static-analysis checker synthesis by coding agents.}
    \textsc{CheckerBench} contains 300 real-world tasks spanning multiple
    programming-language ecosystems and analyzer environments, enabling
    systematic evaluation of agentic checker development.

    \item \textbf{We develop a scalable benchmark construction pipeline that
    automatically reconstructs analyzer environments, packages synthesis
    tasks, and validates checker behavior.} This pipeline enables
    reproducible task creation beyond manually curated checker examples.

    \item \textbf{We present \textsc{CheckerLab}, a unified evaluation framework
    for coding-agent systems with a shared task specification and scoring
    protocol.} \textsc{CheckerLab} supports independent repeated generation
    and frozen-candidate verification, providing fine-grained measurements
    of checker correctness, patch localization, false-positive control, and
    computational cost.
\end{itemize}

% Enable this section if readers need a standalone introduction to
% static-analysis checkers and the synthesis workflow.
% \section{Background}
% \input{sections/2-background}

\section{Related Work}
\sloppy
\label{sec:related-work}
\newcommand{\featureyes}{\textcolor{green!50!black}{\ding{51}}}
\DeclareRobustCommand{\featurepartial}{%
  \begin{tikzpicture}[baseline=(partialmark.base)]
    \node[anchor=base west,inner sep=0pt,outer sep=0pt,text=blue]
      (partialmark) at (0,0) {\ding{51}};
    \draw[blue,line width=.045em,line cap=butt]
      let \p1=(partialmark.south west), \p2=(partialmark.north east) in
      ({\x1+.38*(\x2-\x1)},{\y1+.84*(\y2-\y1)}) --
      ({\x1+.79*(\x2-\x1)},{\y1+.25*(\y2-\y1)});
  \end{tikzpicture}}
\newcommand{\featureno}{\textcolor{red!70!black}{\ding{55}}}

\begin{table*}[tp]
\centering
\scriptsize
\caption{Systems and benchmarks for agentic
static-analysis checker synthesis. \featureyes{} denotes full support,
\featurepartial{} partial support, and \featureno{} no support.
Appendix~\ref{app:benchmark-feature-audit} defines the operational criteria.}
\label{tab:benchmark-comparison}
\setlength{\tabcolsep}{3.0pt}
\resizebox{\textwidth}{!}{
\begin{tabular}{@{}lccccccc@{}}
\toprule[1.5pt]
\textbf{System / Benchmark}
& \textbf{Repository}
& \textbf{Executable}
& \textbf{Multi-Harness}
& \textbf{Long-Horizon}
& \textbf{Checker}
& \textbf{Semantic}
& \textbf{False-Positive} \\
& \textbf{Context}
& \textbf{Environment}
& \textbf{Support}
& \textbf{Interaction}
& \textbf{Synthesis}
& \textbf{Verification}
& \textbf{Evaluation} \\
\midrule
RhoSynth~\cite{garg2022rhosynth}
& \featureno & \featureno & \featureno & \featureno
& \featureyes & \featureyes & \featureyes \\
KNighter~\cite{yang2025knighter}
& \featurepartial & \featureyes & \featureno & \featureno
& \featureyes & \featureyes & \featureyes \\
SWE-bench~\cite{jimenez2024swebench}
& \featureyes & \featureyes & \featureno & \featurepartial
& \featureno & \featurepartial & \featureno \\
SWE-Gym~\cite{pan2025swegym}
& \featureyes & \featureyes & \featureno & \featureyes
& \featureno & \featurepartial & \featureno \\
R2E-Gym~\cite{jain2025r2egym}
& \featureyes & \featureyes & \featureno & \featureyes
& \featureno & \featurepartial & \featureno \\
ReposVul~\cite{wang2024reposvul}
& \featurepartial & \featureno & \featureno & \featureno
& \featureno & \featurepartial & \featureno \\
VulEval~\cite{wen2024vuleval}
& \featurepartial & \featureno & \featureno & \featureno
& \featureno & \featurepartial & \featureno \\
JitVul~\cite{yildiz2025jitvul}
& \featurepartial & \featureno & \featureno & \featurepartial
& \featureno & \featureyes & \featureno \\
\midrule
\textsc{CheckerBench}
& \featureyes & \featureyes & \featureyes & \featureyes
& \featureyes & \featureyes & \featureyes \\
\bottomrule[1.5pt]
\end{tabular}
}
\end{table*}

\subsection{Static Analysis and Checker Synthesis}

Static analyzers rely on expert-authored checkers that encode defect patterns
and analyzer-specific logic~\citep{garg2022rhosynth,yang2025knighter}.
RhoSynth learns graph-based code-quality rules from examples, while Gordon
explores program-specific analyses from behavioral examples
~\citep{garg2022rhosynth,gordon2018programspecific}. LLM-based systems execute
analysis pseudocode or filter bug reports~\citep{hao2023ev,mohajer2023skipanalyzer}.
IRIS infers taint specifications for repository-wide analysis
~\citep{li2025iris}; MoCQ and CQLLM generate vulnerability patterns or CodeQL
queries~\citep{li2025mocq,wang2026cqllm}. QLCoder synthesizes CodeQL queries
through tool use and iterative feedback~\citep{wang2026qlcoder}. KNighter
synthesizes and refines Clang Static Analyzer checkers from Linux kernel
patches~\citep{yang2025knighter}. These systems target specific analyzers or
domains, motivating a common evaluation of general-purpose agents developing
and validating checkers across analyzer environments.

\subsection{Coding Agents for Software Engineering}

Coding agents combine LLMs with repository navigation, code editing, tool
execution, and feedback. SWE-agent and OpenHands support interactive
development, while RepoGraph adds repository structure
~\citep{yang2024sweagent,wang2024openhands,ouyang2025repograph}. RepoAudit
applies an agent to repository-level bug detection~\citep{guo2025repoaudit}.
These systems produce patches or bug findings. Checker synthesis requires
analyzer integration and validation through compiler and scan feedback on
vulnerable and fixed revisions.

\subsection{Benchmarks for Repository-Level Code Agents}

Repository-level benchmarks cover issue resolution (SWE-bench and SWE-bench
Pro), feature development (FeatureBench), and terminal tasks (Terminal-Bench)
~\citep{jimenez2024swebench,deng2025swebenchpro,zhou2026featurebench,
merrill2026terminalbench}. SWE-Gym and R2E-Gym provide executable environments
and verifiers for issue-resolution tasks~\citep{pan2025swegym,jain2025r2egym}.
ReposVul, VulEval, and JitVul target vulnerability detection; SEC-bench
evaluates proof-of-concept generation and patching
~\citep{wang2024reposvul,wen2024vuleval,yildiz2025jitvul,lee2025secbench}.
These works score task completion, patches, or vulnerability detection.
\textsc{CheckerBench}
evaluates reusable checker construction with held-out verification of builds,
vulnerable--fixed diagnostics, patch localization, and false positives
(Table~\ref{tab:benchmark-comparison}).

\section{The \textsc{CheckerBench} Benchmark}
\label{sec:CheckerBench}

% Queue the construction overview early enough to use this section's first page top.
% Native editable source (Figure 3): output/powerpoint/standalone/Figure_3_editable.pptx
% Review artwork: python3 scripts/draw_figure3_code_example.py
% Previous six-card figure preserved as figures/generated/figure2.pdf.
% Keep the pipeline and the six construction stages within the same two pages.
\begin{figure}[tp]
\centering
\includegraphics[width=\linewidth]{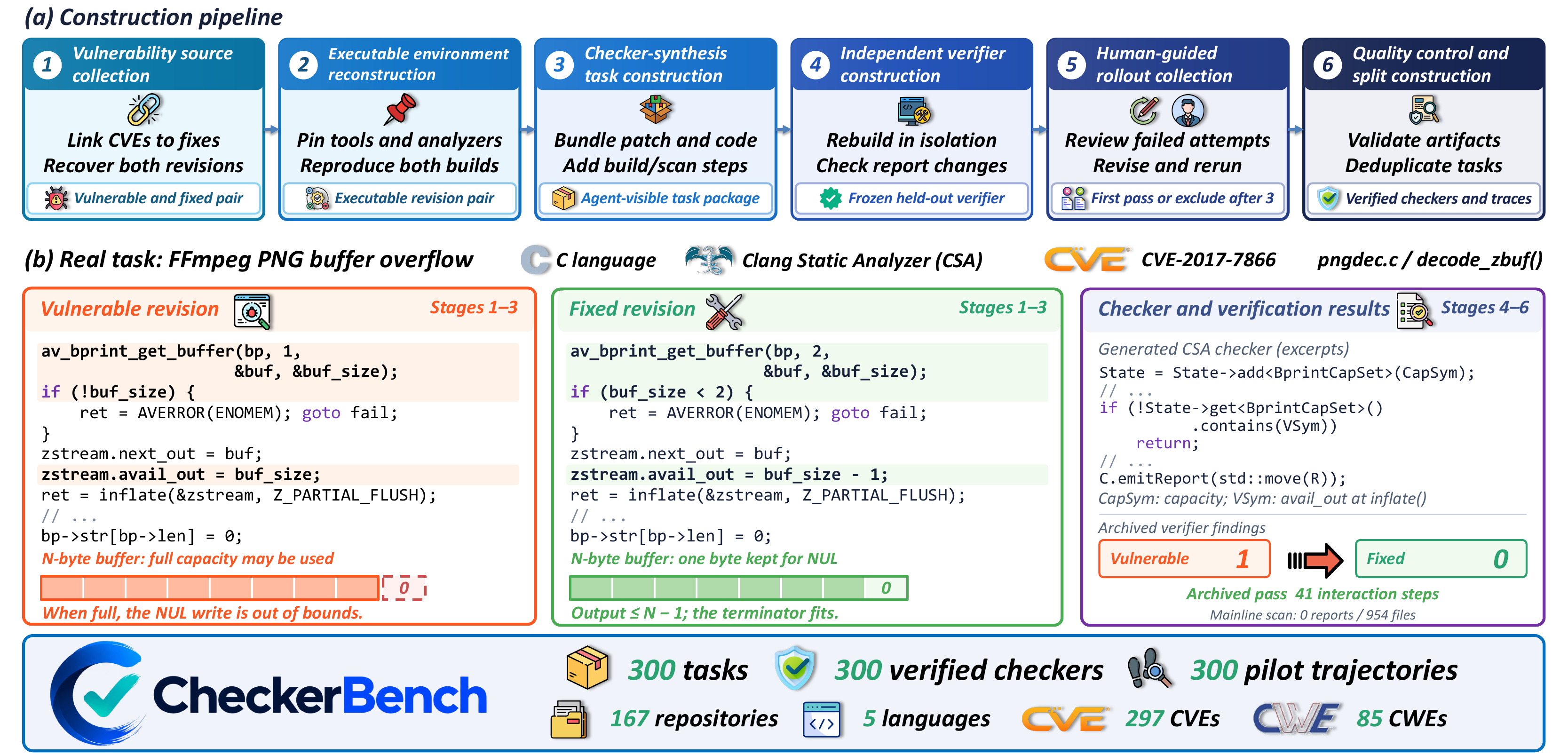}
\caption[\textsc{CheckerBench} construction pipeline and FFmpeg example]{%
\textsc{CheckerBench} construction.}
\label{fig:benchmark-construction}
\end{figure}

\paragraph{Problem formulation.}
\label{sec:task-definition}

Each agent-facing task
\(x_i=(p_i,\ell_i,\alpha_i,\mathcal{E}_i,\mathcal{Q}_i)\)
combines a vulnerability-fix instance
\(p_i=(r_i,c_i^{-},c_i^{+},\Delta_i)\) (repository, vulnerable/fixed
revisions, and patch) with its language, analyzer, reproducible version-pinned
environment, and synthesis instruction. All runs share the initial skill
library \(\mathcal{K}\); agents may edit checker artifacts but must preserve
both revisions.

Under an active generation time budget \(B\),
policy \(\pi_\theta\) in harness \(h\) yields trajectory
\(\tau_i^{(h,\theta)}=(o_0,a_0,\ldots,a_{N_i-1},o_{N_i})\)
and candidate checker \(q_i^{(h,\theta)}\). Here \(N_i\) counts interactions,
\(a_k\) is an action, and \(o_{k+1}\) its resulting observation~\citep{chai2025rlfactory}.
A released record is
\(d_i^{(h,\theta)}=(\mathcal{E}_i,\mathcal{Q}_i,\mathcal{V}_i,
\tau_i^{(h,\theta)},q_i^{(h,\theta)})\), with executable verifier
\(\mathcal{V}_i\) hidden during synthesis. The environment, instruction,
and verifier are fixed across model--harness configurations;
trajectories and checkers vary by rollout.

% Reserve the next page top for statistics instead of stacking both figures.
\suppressfloats[t]
% Editable Figure 4 source: output/powerpoint/editable/Figure_3_editable.pptx
% Export that PPT with scripts/render_ppt.applescript to figures/generated/figure3.pdf.
% Complete counts are reported in Appendix D; all three panels are pies.
\begin{figure}[!t]
    \centering
    \setlength{\abovecaptionskip}{2pt}
    % Trim the empty strip below the legends and keep the figure compact.
    \includegraphics[width=\linewidth,trim=0 48bp 0 12bp,clip]{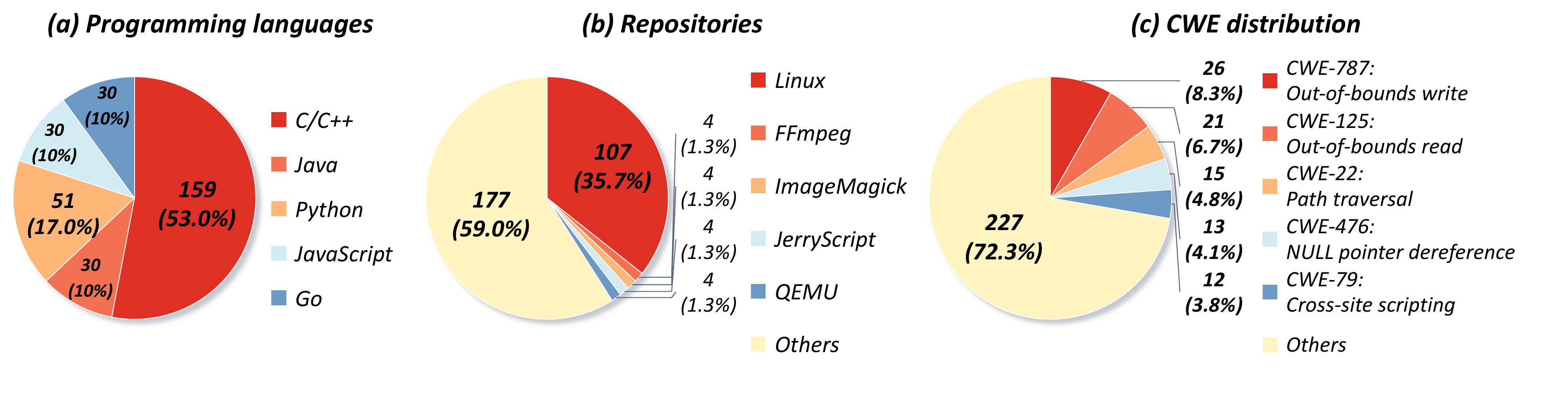}
    \caption{\textsc{CheckerBench} distributions by (a) language,
    (b) repository, and (c) CWE.}
    \label{fig:benchmark-statistics}
\end{figure}

\subsection{Benchmark Construction}
\label{sec:benchmark-construction}

Figure~\ref{fig:benchmark-construction} summarizes the six-stage pipeline:
recover executable vulnerability-fix pairs, package synthesis tasks, freeze
independent verifiers, and curate verified checker--trajectory pairs. All
candidates start from the same harness-component baseline and shared skill
library.

\paragraph{Stage~I: Vulnerability Source Collection.}
\label{sec:repository-collection}

We link public CVE records from the National Vulnerability
Database\footnote{\url{https://nvd.nist.gov/vuln}} to fixing commits in
open-source C/C++, Java, Python, JavaScript, and Go repositories. Each
candidate must have a retrievable vulnerable--fixed revision pair and a patch
that exposes a reusable static-analysis property. We exclude unresolved
provenance and incompatible redistribution terms, and retain CVE/CWE mappings
for auditing and stratified sampling.

\paragraph{Stage~II: Executable Environment Reconstruction.}
\label{sec:environment-reconstruction}

We reconstruct both revisions in isolated workspaces with pinned toolchains
and analyzers: Clang Static Analyzer (CSA) for C/C++ and CodeQL for the other
four languages. Revision-compatible build snapshots and compilation databases
recover the analyzable scope of CSA tasks; CodeQL tasks use pinned analysis
environments. Each pair includes affected functions and reproducible checker
compilation and scan commands. We reject pairs that cannot be reconstructed
or do not support stable differential analysis.

\paragraph{Stage~III: Checker-Synthesis Task Construction.}
\label{sec:task-packaging}

Each task exposes the patch, pre- and post-fix functions, build metadata,
compile/scan commands, and a minimal checker scaffold. Random workspace
identifiers and omitted CVE/CWE labels reduce provenance shortcuts. The shared
library \(\mathcal K\) supplies harness-specific workflow skills and read-only
analyzer documentation, example checker implementations, and semantic utilities;
it excludes task-specific checker solutions. The agent must infer the defect
mechanism, implement a reusable checker, and refine it through compilation
and analysis feedback, producing checker artifacts and an interaction trace.

\paragraph{Stage~IV: Independent Verifier Construction.}
\label{sec:verifier-construction}

We freeze a held-out verifier before rollout collection. It rebuilds each
submitted checker from source in a clean workspace and checks
vulnerable--fixed diagnostic contrast and patch relevance under
backend-specific rules. CSA
requires a disappearing report in a patch-modified function. Passing
additionally requires process and semantic quality, mainline false-positive
control, and mechanism and anti-hardcoding checks
(Appendix~\ref{app:verifier-rubric}).

\paragraph{Stage~V: Human-Guided Rollout Collection and Sampling.}
\label{sec:horizon-calibration}

Calibration uses fixed model, harness, tool, and budget settings. After a
failed attempt, a human diagnoses the verification evidence and revises
prompt guidance for a fresh rollout; the task, verifier, and thresholds stay
fixed. Each task permits one initial attempt and at most two rerolls, stopping
at the first pass and excluding tasks with no pass after three attempts. We
retain all attempts and interventions for audit. Checker--trace pairs enter
sampling only if the checker passes the frozen verifier and the trace contains
at least 40 interaction steps (agent decisions). Eligible pairs are sampled
jointly over repository, CWE, and horizon. The 40-step threshold
is a curation rule; these pilot interventions are separate from autonomous
main evaluation.

\paragraph{Stage~VI: Quality Control, Deduplication, and Splits.}
\label{sec:quality-control}
\label{sec:dataset-splits}

Automated checks establish artifact integrity, reproducible builds and scans,
verifier isolation, and package completeness; manual review confirms defect
semantics and patch-localized diagnostics. We remove duplicate revision pairs
and keep tasks connected by repository, CVE, shared revisions, or similar
patches in one split. Task-specific reference checkers, calibration traces,
and verifier details remain hidden from evaluation agents. Appendix~\ref{app:quality-control}
details these checks and split safeguards.

% Queue the framework overview early enough to accompany the transition below.
% Native editable source (Figure 5): output/powerpoint/standalone/Figure_5_editable.pptx
% Rebuild artwork: python3 scripts/draw_figure5_trajectory.py
% Allow this figure to share the page where the framework is introduced.
\begin{figure}[tp]
\centering
\includegraphics[width=\linewidth]{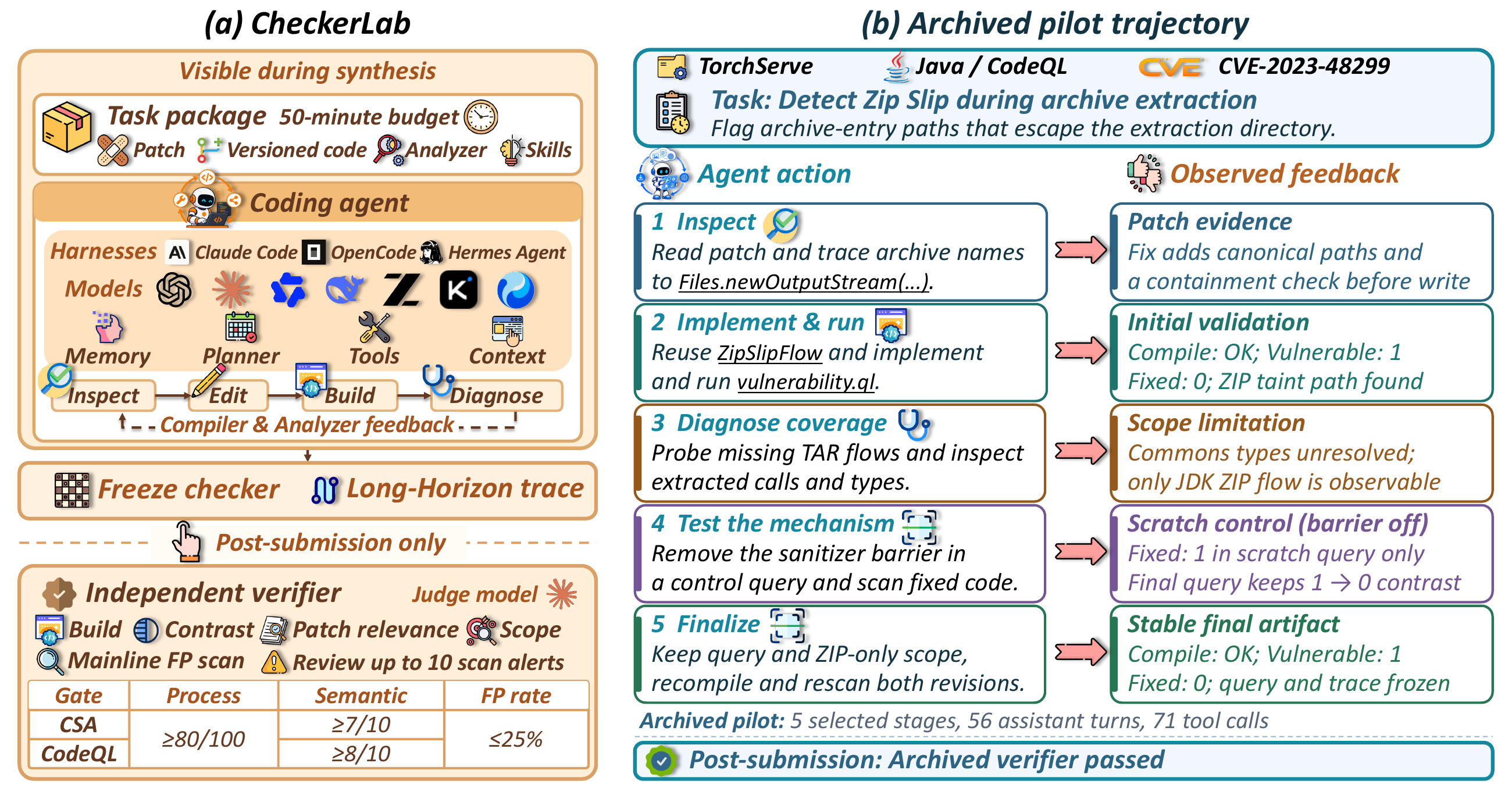}
\setlength{\abovecaptionskip}{4pt}
\caption[CheckerLab evaluation framework and pilot trajectory]{%
\textsc{CheckerLab} workflow and five selected stages from an archived TorchServe pilot.}
\label{fig:evaluation-architecture}
\end{figure}

\subsection{Benchmark Statistics}
\label{sec:benchmark-statistics}

\paragraph{Projects and tasks.}
% Counts cross-checked against figures/project_counts.csv, task_inventory.csv,
% and APPENDIX_D.md (repository-level normalization).
\textsc{CheckerBench} contains 300 tasks from 167 open-source repositories,
including Linux~\citep{linux2026github} (107 tasks),
FFmpeg~\citep{ffmpeg2026github} (4), and
ImageMagick~\citep{imagemagick2026github} (4).
Tasks cover C/C++ (159), Java (30), Python (51), JavaScript (30), and Go (30).
Static analysis environments are CSA~(159) and CodeQL~(141).

\paragraph{Trajectories and interaction horizons.}
We release 300 accepted calibration trajectories, one per task. Calibration
includes human intervention to diagnose failures and revise prompts.
All trajectories pass executable verification and the construction-time
semantic-score threshold. They are separate from independent main-experiment
repeats. Interaction steps reflect workload under the calibration
configuration~\citep{He2026VistaHopBL}.
Appendices~\ref{app:calibration-distributions} and~\ref{app:pilot-trajectories}
report trajectory statistics, including the assistant-turn and tool-call
distributions of the admitted pilot trajectories
(Figure~\ref{fig:pilot-trajectory-distributions}).

\paragraph{Defect and task coverage.}
\textsc{CheckerBench} covers 85 CWEs. The most frequent are CWE-787 (out-of-bounds write),
CWE-125 (out-of-bounds read), and CWE-22 (path traversal), as shown in
Figure~\ref{fig:benchmark-statistics}(c).
Each task also records \emph{repository scale}, \emph{integration scope},
\emph{verification evidence}, and \emph{calibration horizon}.
Appendix~\ref{app:dataset-composition} provides complete benchmark statistics.

% Queue the main results before the framework text to distribute page-top floats.
% Keep the overall results at a page top with the experiment discussion below.
% Inline changes follow the supplied reference: orange increases, teal decreases.
\begin{table*}[!t]
\centering
\scriptsize
% Keep the dense table neutral: white structural rows and faint gray striping.
\definecolor{resultheader}{HTML}{FFFFFF}
\definecolor{resultstripe}{HTML}{F7F7F7}
\definecolor{resultgroup}{HTML}{FFFFFF}
\definecolor{resultharness}{HTML}{EEEEEE}
\definecolor{resultup}{HTML}{D95638}
\definecolor{resultdown}{HTML}{008D9C}
\newcommand{\metricheader}[1]{%
  \begin{tabular}[c]{@{}c@{}}#1\end{tabular}}
% Reserve the same space for every value and subscript so decimals align.
\newcommand{\metricvalue}[2]{%
  \makebox[2.5em][r]{#1}\hspace{0.15pt}\makebox[17pt][l]{#2}}
\newcommand{\metricchange}[3]{%
  \raisebox{-0.35ex}{\fontsize{5}{5}\selectfont
    \textcolor{#1}{\ensuremath{#2}#3}}}
\newcommand{\deltaup}[1]{\metricchange{resultup}{\uparrow}{#1}}
\newcommand{\deltadown}[1]{\metricchange{resultdown}{\downarrow}{#1}}
\newcommand{\deltaequal}{\metricchange{gray}{=}{0.00}}
\caption[Model performance on CheckerBench across agent harnesses]{%
Performance of seven models across three agent harnesses on
\textsc{CheckerBench}.
\textbf{Bold} and \underline{underlined} values mark the best and second-best
models per metric within each harness. Model-row subscripts show numerical
changes relative to GPT-5.6 in the same harness:
\textcolor{resultup}{$\uparrow$ increases} and
\textcolor{resultdown}{$\downarrow$ decreases} (percentage points for percentage
metrics). Header arrows indicate the preferred direction.
Model-row metrics are means over three independent repeats.
Overall Average summarizes the 21 model--harness configurations.}
\label{tab:overall-performance}
\setlength{\tabcolsep}{1.4pt}
\renewcommand{\arraystretch}{1.2}
\begin{tabularx}{\textwidth}{l*{4}{>{\centering\arraybackslash}X}c*{4}{>{\centering\arraybackslash}X}}
% HY4/Hermes: one original-query scan timeout is counted as failure by user instruction.
% Build retains original evidence; Diff counts no success; missing stage scores stay null.
% The manually modified diagnostic query and its scores are excluded.
% Changes use the displayed values; average rows are summaries, not ranked models.
% Author-verified FP corrections: figures/performance/table3_fp_corrections.csv.
% Author-confirmed Pass@1 counts: figures/performance/table3_pass_corrections.csv.
\toprule[1.5pt]
\rowcolor{resultheader}
\metricheader{\textbf{Model / System}}
& \metricheader{\textbf{Pass@1}\\\textbf{(\%)} $\uparrow$}
& \metricheader{\textbf{Build}\\\textbf{(\%)} $\uparrow$}
& \metricheader{\textbf{Diff. SR}\\\textbf{(\%)} $\uparrow$}
& \metricheader{$\mathbf{FP}_{\mathrm{pass}}$\\\textbf{(\%)} $\downarrow$}
& \metricheader{\textbf{\#WN}}
& \metricheader{\textbf{Process}\\$\uparrow$}
& \metricheader{\textbf{Semantic}\\$\uparrow$}
& \metricheader{\textbf{Avg.}\\\textbf{Turns} $\downarrow$}
& \metricheader{\textbf{Avg. Tool}\\\textbf{Calls} $\downarrow$} \\
\midrule[0.6pt]
\rowcolor{resultharness}
\multicolumn{10}{c}{\harnessname{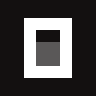}{\textbf{\textit{OpenCode}}}} \\
\modelname{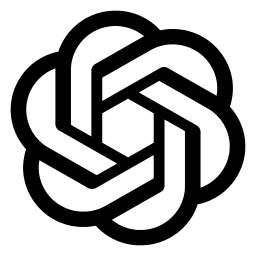}{GPT-5.6} & \metricvalue{31.67}{} & \metricvalue{\textbf{98.33}}{} & \metricvalue{\underline{85.33}}{} & \metricvalue{\textbf{0.00}}{} & 7 & \metricvalue{80.22}{} & \metricvalue{7.68}{} & \metricvalue{\textbf{29.19}}{} & \metricvalue{60.13}{} \\
\rowcolor{resultstripe}
\modelname{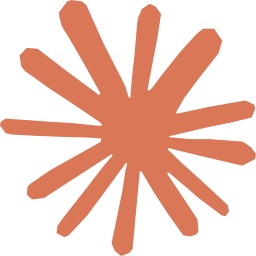}{Claude Opus 4.8} & \metricvalue{\textbf{45.33}}{\deltaup{13.66}} & \metricvalue{81.67}{\deltadown{16.66}} & \metricvalue{81.00}{\deltadown{4.33}} & \metricvalue{13.79}{\deltaup{13.79}} & 174 & \metricvalue{86.42}{\deltaup{6.20}} & \metricvalue{\underline{8.32}}{\deltaup{0.64}} & \metricvalue{44.78}{\deltaup{15.59}} & \metricvalue{64.35}{\deltaup{4.22}} \\
\modelname{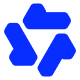}{Qwen3.8-Max} & \metricvalue{37.33}{\deltaup{5.66}} & \metricvalue{69.33}{\deltadown{29.00}} & \metricvalue{69.00}{\deltadown{16.33}} & \metricvalue{5.77}{\deltaup{5.77}} & 52 & \metricvalue{\textbf{90.15}}{\deltaup{9.93}} & \metricvalue{\textbf{8.46}}{\deltaup{0.78}} & \metricvalue{35.96}{\deltaup{6.77}} & \metricvalue{\underline{47.54}}{\deltadown{12.59}} \\
\rowcolor{resultstripe}
\modelname{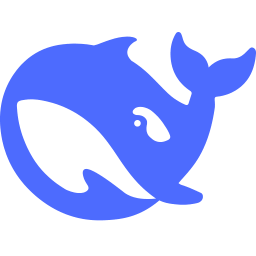}{DeepSeek-V4-Pro} & \metricvalue{\underline{39.67}}{\deltaup{8.00}} & \metricvalue{\underline{92.67}}{\deltadown{5.66}} & \metricvalue{\textbf{90.67}}{\deltaup{5.34}} & \metricvalue{2.56}{\deltaup{2.56}} & 78 & \metricvalue{84.89}{\deltaup{4.67}} & \metricvalue{8.12}{\deltaup{0.44}} & \metricvalue{51.09}{\deltaup{21.90}} & \metricvalue{65.28}{\deltaup{5.15}} \\
\modelname{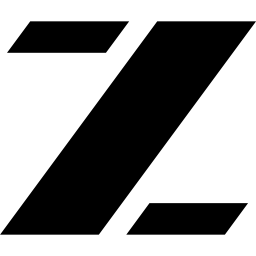}{GLM-5.2} & \metricvalue{32.33}{\deltaup{0.66}} & \metricvalue{77.33}{\deltadown{21.00}} & \metricvalue{75.33}{\deltadown{10.00}} & \metricvalue{10.34}{\deltaup{10.34}} & 58 & \metricvalue{86.73}{\deltaup{6.51}} & \metricvalue{8.11}{\deltaup{0.43}} & \metricvalue{41.21}{\deltaup{12.02}} & \metricvalue{61.02}{\deltaup{0.89}} \\
\rowcolor{resultstripe}
\modelname{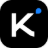}{Kimi-K3} & \metricvalue{38.67}{\deltaup{7.00}} & \metricvalue{83.33}{\deltadown{15.00}} & \metricvalue{81.67}{\deltadown{3.66}} & \metricvalue{\underline{1.59}}{\deltaup{1.59}} & 63 & \metricvalue{\underline{88.48}}{\deltaup{8.26}} & \metricvalue{8.30}{\deltaup{0.62}} & \metricvalue{\underline{32.42}}{\deltaup{3.23}} & \metricvalue{\textbf{43.53}}{\deltadown{16.60}} \\
\modelname{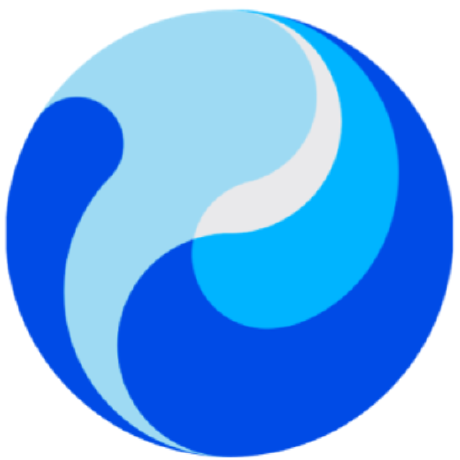}{HY4-Preview} & \metricvalue{33.67}{\deltaup{2.00}} & \metricvalue{72.67}{\deltadown{25.66}} & \metricvalue{71.33}{\deltadown{14.00}} & \metricvalue{12.38}{\deltaup{12.38}} & 105 & \metricvalue{86.90}{\deltaup{6.68}} & \metricvalue{\underline{8.32}}{\deltaup{0.64}} & \metricvalue{41.28}{\deltaup{12.09}} & \metricvalue{50.15}{\deltadown{9.98}} \\
\rowcolor{resultgroup}
\textbf{Average} & \metricvalue{\textbf{36.95}}{} & \metricvalue{\textbf{82.19}}{} & \metricvalue{\textbf{79.19}}{} & \metricvalue{\textbf{6.63}}{} & \textbf{76.71} & \metricvalue{\textbf{86.26}}{} & \metricvalue{\textbf{8.19}}{} & \metricvalue{\textbf{39.42}}{} & \metricvalue{\textbf{56.00}}{} \\
\midrule[0.4pt]
\rowcolor{resultharness}
\multicolumn{10}{c}{\harnessname{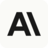}{\textbf{\textit{Claude Code}}}} \\
\modelname{openai}{GPT-5.6} & \metricvalue{34.00}{} & \metricvalue{\textbf{93.67}}{} & \metricvalue{\underline{83.00}}{} & \metricvalue{2.44}{} & 41 & \metricvalue{81.41}{} & \metricvalue{7.69}{} & \metricvalue{35.71}{} & \metricvalue{60.58}{} \\
\rowcolor{resultstripe}
\modelname{claude}{Claude Opus 4.8} & \metricvalue{35.00}{\deltaup{1.00}} & \metricvalue{63.67}{\deltadown{30.00}} & \metricvalue{62.33}{\deltadown{20.67}} & \metricvalue{2.33}{\deltadown{0.11}} & 43 & \metricvalue{86.88}{\deltaup{5.47}} & \metricvalue{8.21}{\deltaup{0.52}} & \metricvalue{\underline{32.60}}{\deltadown{3.11}} & \metricvalue{45.63}{\deltadown{14.95}} \\
\modelname{qwen}{Qwen3.8-Max} & \metricvalue{34.67}{\deltaup{0.67}} & \metricvalue{67.33}{\deltadown{26.34}} & \metricvalue{67.00}{\deltadown{16.00}} & \metricvalue{4.55}{\deltaup{2.11}} & 44 & \metricvalue{89.58}{\deltaup{8.17}} & \metricvalue{\underline{8.38}}{\deltaup{0.69}} & \metricvalue{33.91}{\deltadown{1.80}} & \metricvalue{\underline{41.57}}{\deltadown{19.01}} \\
\rowcolor{resultstripe}
\modelname{deepseek}{DeepSeek-V4-Pro} & \metricvalue{\textbf{38.33}}{\deltaup{4.33}} & \metricvalue{\underline{91.33}}{\deltadown{2.34}} & \metricvalue{\textbf{89.67}}{\deltaup{6.67}} & \metricvalue{\underline{1.72}}{\deltadown{0.72}} & 58 & \metricvalue{85.25}{\deltaup{3.84}} & \metricvalue{8.27}{\deltaup{0.58}} & \metricvalue{47.17}{\deltaup{11.46}} & \metricvalue{62.50}{\deltaup{1.92}} \\
\modelname{glm}{GLM-5.2} & \metricvalue{33.00}{\deltadown{1.00}} & \metricvalue{76.33}{\deltadown{17.34}} & \metricvalue{74.67}{\deltadown{8.33}} & \metricvalue{\textbf{0.00}}{\deltadown{2.44}} & 18 & \metricvalue{\underline{89.67}}{\deltaup{8.26}} & \metricvalue{\textbf{8.39}}{\deltaup{0.70}} & \metricvalue{38.75}{\deltaup{3.04}} & \metricvalue{49.93}{\deltadown{10.65}} \\
\rowcolor{resultstripe}
\modelname{kimi}{Kimi-K3} & \metricvalue{\underline{38.00}}{\deltaup{4.00}} & \metricvalue{83.67}{\deltadown{10.00}} & \metricvalue{82.33}{\deltadown{0.67}} & \metricvalue{2.74}{\deltaup{0.30}} & 73 & \metricvalue{87.79}{\deltaup{6.38}} & \metricvalue{8.34}{\deltaup{0.65}} & \metricvalue{\textbf{31.81}}{\deltadown{3.90}} & \metricvalue{\textbf{38.61}}{\deltadown{21.97}} \\
\modelname{hunyuan}{HY4-Preview} & \metricvalue{23.67}{\deltadown{10.33}} & \metricvalue{49.33}{\deltadown{44.34}} & \metricvalue{47.33}{\deltadown{35.67}} & \metricvalue{\textbf{0.00}}{\deltadown{2.44}} & 13 & \metricvalue{\textbf{90.83}}{\deltaup{9.42}} & \metricvalue{8.35}{\deltaup{0.66}} & \metricvalue{50.66}{\deltaup{14.95}} & \metricvalue{62.23}{\deltaup{1.65}} \\
\rowcolor{resultgroup}
\textbf{Average} & \metricvalue{\textbf{33.81}}{} & \metricvalue{\textbf{75.05}}{} & \metricvalue{\textbf{72.33}}{} & \metricvalue{\textbf{1.97}}{} & \textbf{41.43} & \metricvalue{\textbf{87.34}}{} & \metricvalue{\textbf{8.23}}{} & \metricvalue{\textbf{38.66}}{} & \metricvalue{\textbf{51.58}}{} \\
\midrule[0.4pt]
\rowcolor{resultharness}
\multicolumn{10}{c}{\harnessname{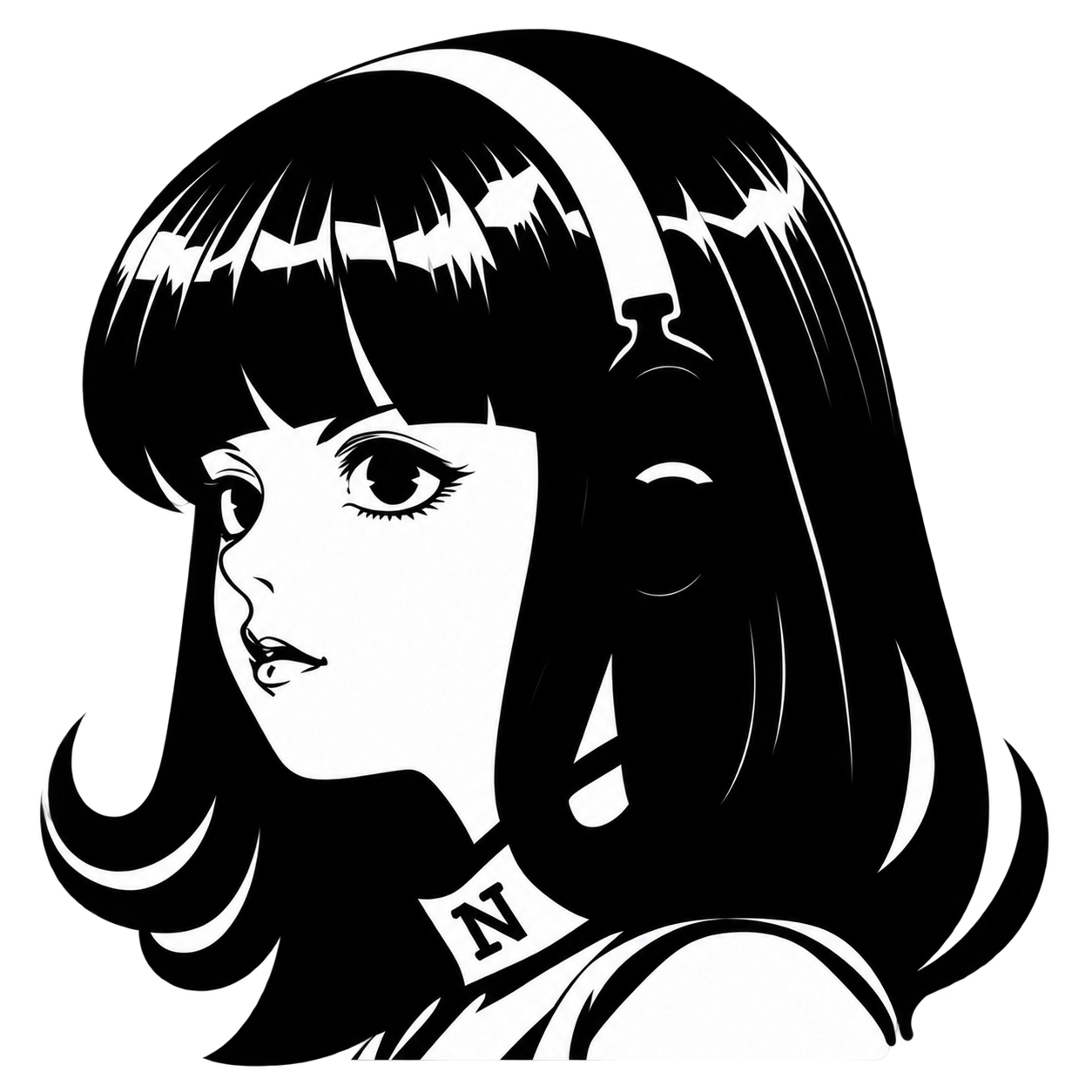}{\textbf{\textit{Hermes Agent}}}} \\
\modelname{openai}{GPT-5.6} & \metricvalue{23.00}{} & \metricvalue{\textbf{100.00}}{} & \metricvalue{64.00}{} & \metricvalue{\textbf{0.00}}{} & 6 & \metricvalue{79.85}{} & \metricvalue{7.50}{} & \metricvalue{\textbf{25.94}}{} & \metricvalue{\underline{46.91}}{} \\
\rowcolor{resultstripe}
\modelname{claude}{Claude Opus 4.8} & \metricvalue{\textbf{34.33}}{\deltaup{11.33}} & \metricvalue{80.33}{\deltadown{19.67}} & \metricvalue{78.67}{\deltaup{14.67}} & \metricvalue{5.45}{\deltaup{5.45}} & 55 & \metricvalue{85.58}{\deltaup{5.73}} & \metricvalue{8.32}{\deltaup{0.82}} & \metricvalue{44.84}{\deltaup{18.90}} & \metricvalue{61.89}{\deltaup{14.98}} \\
\modelname{qwen}{Qwen3.8-Max} & \metricvalue{25.00}{\deltaup{2.00}} & \metricvalue{44.00}{\deltadown{56.00}} & \metricvalue{42.33}{\deltadown{21.67}} & \metricvalue{11.11}{\deltaup{11.11}} & 27 & \metricvalue{\textbf{90.92}}{\deltaup{11.07}} & \metricvalue{\textbf{8.53}}{\deltaup{1.03}} & \metricvalue{40.55}{\deltaup{14.61}} & \metricvalue{51.95}{\deltaup{5.04}} \\
\rowcolor{resultstripe}
\modelname{deepseek}{DeepSeek-V4-Pro} & \metricvalue{\underline{32.33}}{\deltaup{9.33}} & \metricvalue{\underline{94.00}}{\deltadown{6.00}} & \metricvalue{\textbf{90.67}}{\deltaup{26.67}} & \metricvalue{\textbf{0.00}}{\deltaequal} & 12 & \metricvalue{85.29}{\deltaup{5.44}} & \metricvalue{8.03}{\deltaup{0.53}} & \metricvalue{39.82}{\deltaup{13.88}} & \metricvalue{62.37}{\deltaup{15.46}} \\
\modelname{glm}{GLM-5.2} & \metricvalue{22.33}{\deltadown{0.67}} & \metricvalue{79.67}{\deltadown{20.33}} & \metricvalue{78.00}{\deltaup{14.00}} & \metricvalue{\textbf{0.00}}{\deltaequal} & 3 & \metricvalue{82.87}{\deltaup{3.02}} & \metricvalue{7.81}{\deltaup{0.31}} & \metricvalue{42.85}{\deltaup{16.91}} & \metricvalue{63.22}{\deltaup{16.31}} \\
\rowcolor{resultstripe}
\modelname{kimi}{Kimi-K3} & \metricvalue{24.33}{\deltaup{1.33}} & \metricvalue{87.00}{\deltadown{13.00}} & \metricvalue{\underline{79.67}}{\deltaup{15.67}} & \metricvalue{\underline{4.76}}{\deltaup{4.76}} & 63 & \metricvalue{82.62}{\deltaup{2.77}} & \metricvalue{8.08}{\deltaup{0.58}} & \metricvalue{\underline{32.73}}{\deltaup{6.79}} & \metricvalue{\textbf{42.12}}{\deltadown{4.79}} \\
\modelname{hunyuan}{HY4-Preview} & \metricvalue{21.67}{\deltadown{1.33}} & \metricvalue{57.67}{\deltadown{42.33}} & \metricvalue{52.67}{\deltadown{11.33}} & \metricvalue{\textbf{0.00}}{\deltaequal} & 3 & \metricvalue{\underline{88.42}}{\deltaup{8.57}} & \metricvalue{\underline{8.44}}{\deltaup{0.94}} & \metricvalue{49.69}{\deltaup{23.75}} & \metricvalue{62.06}{\deltaup{15.15}} \\
\rowcolor{resultgroup}
\textbf{Average} & \metricvalue{\textbf{26.14}}{} & \metricvalue{\textbf{77.52}}{} & \metricvalue{\textbf{69.43}}{} & \metricvalue{\textbf{3.05}}{} & \textbf{24.14} & \metricvalue{\textbf{85.08}}{} & \metricvalue{\textbf{8.10}}{} & \metricvalue{\textbf{39.49}}{} & \metricvalue{\textbf{55.79}}{} \\
\midrule[0.6pt]
\rowcolor{resultheader}
\textbf{Overall Average} & \metricvalue{\textbf{32.30}}{} & \metricvalue{\textbf{78.25}}{} & \metricvalue{\textbf{73.65}}{} & \metricvalue{\textbf{3.88}}{} & \textbf{47.43} & \metricvalue{\textbf{86.23}}{} & \metricvalue{\textbf{8.17}}{} & \metricvalue{\textbf{39.19}}{} & \metricvalue{\textbf{54.46}}{} \\
% \midrule[0.6pt]
% Standalone multi-agent baseline; excluded from the 21-configuration average.
% Author-confirmed baseline: DeepSeek-V4-Pro on 300 tasks.
% TODO: insert verified metrics and document the baseline's budget and repeat count.
% \harnessname{knighter}{KNighter-Adapted} & \metricvalue{\pendingresult}{} & \metricvalue{\pendingresult}{} & \metricvalue{\pendingresult}{} & \metricvalue{\pendingresult}{} & \metricvalue{\pendingresult}{} & \metricvalue{\pendingresult}{} & \metricvalue{\pendingresult}{} & \metricvalue{\pendingresult}{} \\
\bottomrule[1.5pt]
\end{tabularx}
\end{table*}

\section{\textsc{CheckerLab}: Evaluation Framework}
\label{sec:evaluation-framework}

% Writing references: LocalSearchBench Sec. 4 (overview, roles, and metrics)
% and SWE-bench Sec. 2.2 (success criterion; technical details in the appendix).
\textsc{CheckerLab} separates interactive checker synthesis from independent
evaluation of the frozen checker and its trajectory~\citep{he2026localsearchbench}
(Figure~\ref{fig:evaluation-architecture}).

\paragraph{Evaluation workflow.}
\label{sec:agent-interface}
\label{sec:execution-harness}
\label{sec:verification}
Each run receives an isolated workspace with the task instruction, fixing
patch, code context, analyzer tools, and enabled skills; the reference checker
and held-out verifier remain hidden. Agents use native harness capabilities
and compilation and analysis feedback within a common 50-minute generation
budget. At termination, we freeze the submitted CSA checker or CodeQL query
source and record the trajectory and tool invocations.
The verifier then rebuilds the frozen checker from source in a clean environment
and checks vulnerable--fixed diagnostic contrast and patch relevance under
backend-specific rules (Section~\ref{sec:verifier-construction}). A broader
mainline scan assesses false positives without checker edits.
Claude Opus 5~\citep{anthropic2026opus5} judges process quality, vulnerability
semantics, and sampled mainline alerts in separate stages. A run passes only
after normal generation completion and satisfaction of all gates in
Equation~\ref{eq:verdict-gate}; Appendix~\ref{app:verifier-rubric} gives the full rubrics.

\paragraph{Metrics.}
\label{sec:metrics}
Our primary metric, Pass@1, is the percentage of tasks whose checker passes
the full evaluation after one independent synthesis run. For each of the
21 harness--model pairs, we evaluate the same 300 tasks in three independent
repeats and average the three rates equally. Generation failures, refusals,
and candidate-code errors count as failures.
\textit{Build} and \textit{Diff.\ SR} are the percentages of all tasks passing
independent compilation and differential verification, respectively.
\textit{Process} and \textit{Semantic} are mean valid scores on 100-point and
10-point scales. Within each repeat, the false-positive percentage
$\mathrm{FP}_{\mathrm{pass}}=100\sum\mathrm{FP}/\sum(\mathrm{TP}+\mathrm{FP})$
pools fully adjudicated recorded mainline alerts from passed checkers.
The acceptance gate uses each checker's sampled mainline alerts, with a
default FP cutoff of 25\% (Appendix~\ref{app:verifier-rubric}).
We compute each metric per repeat and average equally
(Appendix~\ref{app:additional-results}).
In Table~\ref{tab:overall-performance}, \#WN is the arithmetic mean of the
counts of fully adjudicated recorded mainline warnings from passed checkers
across the three repeats. Each repeat's count is
$\sum(\mathrm{TP}+\mathrm{FP})$.
The Average and Overall
Average rows report mean counts across seven models and all 21 configurations,
respectively. Warning counts are descriptive, with no ranking or change subscripts.
Appendix~\ref{app:fp-sampling} details alert selection, recorded warning
volume, and sensitivity to precision weighting and all-recorded-alert gates.
For interaction efficiency, we average turns, tool calls, and explicit skill calls over
successful trajectories within each repeat, including subagent activity.
We then average these means equally across repeats and report success counts
(Appendix~\ref{app:evaluation-protocol}).

% Keep the section heading with its setup text when the results table floats.
\Needspace{6\baselineskip}
\section{Experiments}
\label{sec:experiments}
\providecommand{\pendingresult}{\textcolor{gray!70!black}{--}}

% We first introduce the experimental setup (Sec.~\ref{sec:experimental-setup})
% and compare seven models across three agent harnesses
% (Sec.~\ref{sec:overall-performance}). We then examine potential prior exposure
% (Sec.~\ref{sec:data-contamination}) and evaluate the component ablations
% (Sec.~\ref{sec:ablation-study}). Finally, we analyze time-budget sensitivity
% (Sec.~\ref{sec:budget-sensitivity}) and checker-synthesis failure modes
% (Sec.~\ref{sec:failure-analysis}). Additional experimental details are provided
% in Appendices~\ref{app:evaluation-protocol} and~\ref{app:additional-results}.

\subsection{Experimental Setup}
\label{sec:experimental-setup}

% Writing references: EditBench Sec. 5, VERINA Sec. 5, and FeatureBench Sec. 4.1.1.
\paragraph{Models and agent harnesses.}
We evaluate 7 models using 3 agent harnesses, including
OpenCode~\citep{anomaly2026opencode},
Claude Code~\citep{anthropic2026claudecode}, and
\mbox{Hermes Agent~\citep{nousresearch2026hermesagent}}.
The models are GPT-5.6-Sol~\citep{openai2026gpt56}, Claude Opus 4.8~\citep{anthropic2026opus48},
Qwen3.8-Max~\citep{qwen2026qwen38release},
DeepSeek-V4-Pro-0813~\citep{deepseek2026v4}, GLM-5.2~\citep{zai2026glm52},
Kimi-K3~\citep{kimi2026k3}, and HY4-Preview~\citep{tencent2026hy4preview}.
% We additionally include KNighter-Adapted, a standalone baseline based on
% KNighter's multi-agent framework~\citep{yang2025knighter}, using DeepSeek-V4-Pro
% on the same 300 tasks.

\paragraph{Task execution.}
Each \textsc{CheckerLab} run uses an isolated E2B sandbox and workspace
containing the task instruction, fixing patch, relevant code context, and analyzer tools.
Agents can inspect code, implement checkers, and use compilation and analysis
feedback within a 50-minute active generation budget.
The submitted source and execution records are frozen at termination for
independent verification. Environment and execution details are provided
in Appendix~\ref{app:evaluation-protocol}.

\paragraph{Evaluation and judge.}
Each frozen checker is evaluated using backend-specific executable checks
and scoring rubrics. We use Claude Opus 5~\citep{anthropic2026opus5} as the LLM judge for false-positive
triage, process quality, and vulnerability semantics. Human validation gives
a pooled Cohen's $\kappa=0.85$ for judge--human agreement.
Appendix~\ref{app:human-validation} reports the sampling, labeling,
adjudication, and stage-specific agreement.
The scoring rubrics are detailed in Appendix~\ref{app:verifier-rubric}.
For the 21 model--harness configurations, we report mean Pass@1 over three
independent repeats on the same 300 tasks,
with additional diagnostic metrics defined in
Section~\ref{sec:metrics}. Efficiency metrics are averaged over successful
trajectories only.

% Queue the breakdown before its discussion so it can use the current page top.
% Add the dashed y=x reference: python3 scripts/add_turns_reference_line.py
\begin{figure}[tp]
\centering
\setlength{\abovecaptionskip}{2pt}
\includegraphics[width=\textwidth,trim=0 0 0 9bp,clip]{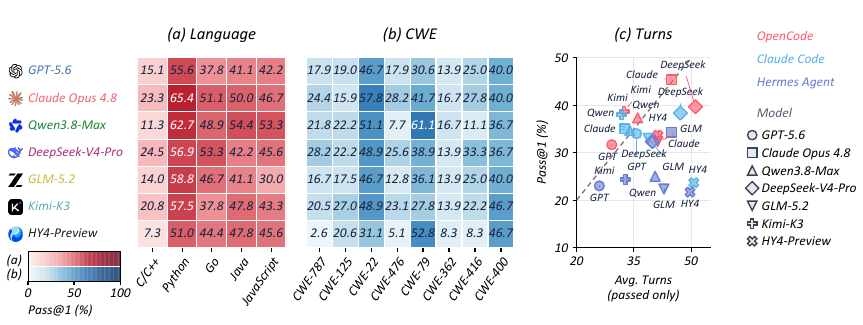}
\caption{Pass@1 by (a) language and (b) CWE, averaged equally across three harnesses.
(c) Pass@1 versus average turns on successful runs; colors denote harnesses and shapes denote models.}
\label{fig:model-performance-analysis}
\end{figure}

\subsection{Overall Performance}
\label{sec:overall-performance}
\label{sec:model-performance-analysis}

\paragraph{End-to-end success.}
Across 21 model--harness configurations, mean Pass@1 is 32.30\%; even the
best combination, Claude Opus 4.8 with OpenCode, reaches only 45.33\%
(Figure~\ref{fig:overall-pass-at-1}; Table~\ref{tab:overall-performance}).
Thus, more than half of tasks remain unsuccessful even for the strongest
configuration. Overall Build and Diff.\ SR are much higher, at 78.25\% and
73.65\%, respectively. The gap between Diff.\ SR and Pass@1 is
41.35 percentage points, much larger than the 4.60-point gap between
Build and Diff.\ SR.
For example, GPT-5.6 with Hermes Agent achieves 100.00\% Build but only
64.00\% Diff.\ SR and 23.00\% Pass@1.
The larger gap between differential verification and full acceptance
highlights the challenge of meeting the additional requirements for
semantic fidelity, process quality, false-positive control, and normal completion.

\paragraph{Model--harness dependence.}
OpenCode has the highest mean Pass@1 across models (36.95\%), followed by Claude Code
(33.81\%) and Hermes Agent (26.14\%), with a 10.81-percentage-point difference
between the highest and lowest averages. The best harness varies by model:
GPT-5.6 and GLM-5.2 attain their highest Pass@1 with Claude Code.
The leading model also changes across harnesses. Claude Opus 4.8 ranks first
with OpenCode and Hermes Agent, whereas DeepSeek-V4-Pro leads with Claude
Code (38.33\%, compared with 35.00\% for Claude Opus 4.8).
Kimi-K3 further illustrates the sensitivity to harness choice, declining
from 38.67\% with OpenCode to 24.33\% with Hermes Agent, a 14.34-point gap.
These shifts show that harness choice changes both success rates and
model rankings, making the complete model--harness configuration the
appropriate unit for comparison and selection.

\paragraph{Diagnostic quality and false positives.}
Qwen3.8-Max with Hermes Agent has the highest Process (90.92/100) and
Semantic (8.53/10) scores across all configurations, yet reaches only
25.00\% Pass@1 and 44.00\% Build.
OpenCode's higher mean Pass@1 is also accompanied by a higher
$\mathrm{FP}_{\mathrm{pass}}$ than Claude Code (6.63\% vs.\ 1.97\%).
Within OpenCode, Claude Opus 4.8 has 13.79\% $\mathrm{FP}_{\mathrm{pass}}$
with \#WN$=174$, compared with 0.00\% and \#WN$=7$ for GPT-5.6.
Both metrics summarize recorded, fully adjudicated alerts from passed checkers.
No evaluated configuration leads simultaneously in synthesis success,
process and semantic quality, and observed diagnostic precision.
Checker reliability should therefore be assessed across these dimensions,
with warning volume providing context for false-positive comparisons.

\paragraph{Language and defect-family variation.}
Figure~\ref{fig:model-performance-analysis}(a,b) averages each model's
results equally across the three harnesses. Every model has its lowest
language-specific Pass@1 in C/C++, where even the leading model,
DeepSeek-V4-Pro, reaches only 24.53\%. This stratum contains 159 of the 300
tasks and therefore strongly influences aggregate success.
The leaders change in other languages: Claude Opus 4.8 leads Python
(65.36\%), DeepSeek-V4-Pro leads Go (53.33\%), and Qwen3.8-Max leads Java
(54.44\%) and JavaScript (53.33\%). Qwen3.8-Max also outperforms
DeepSeek-V4-Pro on CWE-79, with the ordering reversed on CWE-787.
C/C++ uses CSA while the other languages use CodeQL, and CWE groups differ
in repository and language composition.
These results identify C/C++ tasks under CSA as a shared challenge and
reveal complementary model strengths across the evaluated task groups,
supporting performance breakdowns by language and defect family alongside
aggregate Pass@1.

\paragraph{Success and interaction efficiency.}
Kimi-K3 uses the fewest average tool calls on successful runs in all three
harnesses. With Claude Code, it achieves nearly the same Pass@1 as
DeepSeek-V4-Pro (38.00\% vs.\ 38.33\%), while using fewer turns
(31.81 vs.\ 47.17) and tool calls (38.61 vs.\ 62.50), a 38.2\% reduction
in tool calls
(Figure~\ref{fig:model-performance-analysis}(c)).
Kimi-K3 thus combines a competitive success rate with efficient tool use
on successful runs, highlighting the value of considering both Pass@1
and interaction efficiency when selecting a model--harness configuration.

\begin{figure}[tp]
\centering
\setlength{\abovecaptionskip}{2pt}
\includegraphics[width=\textwidth]{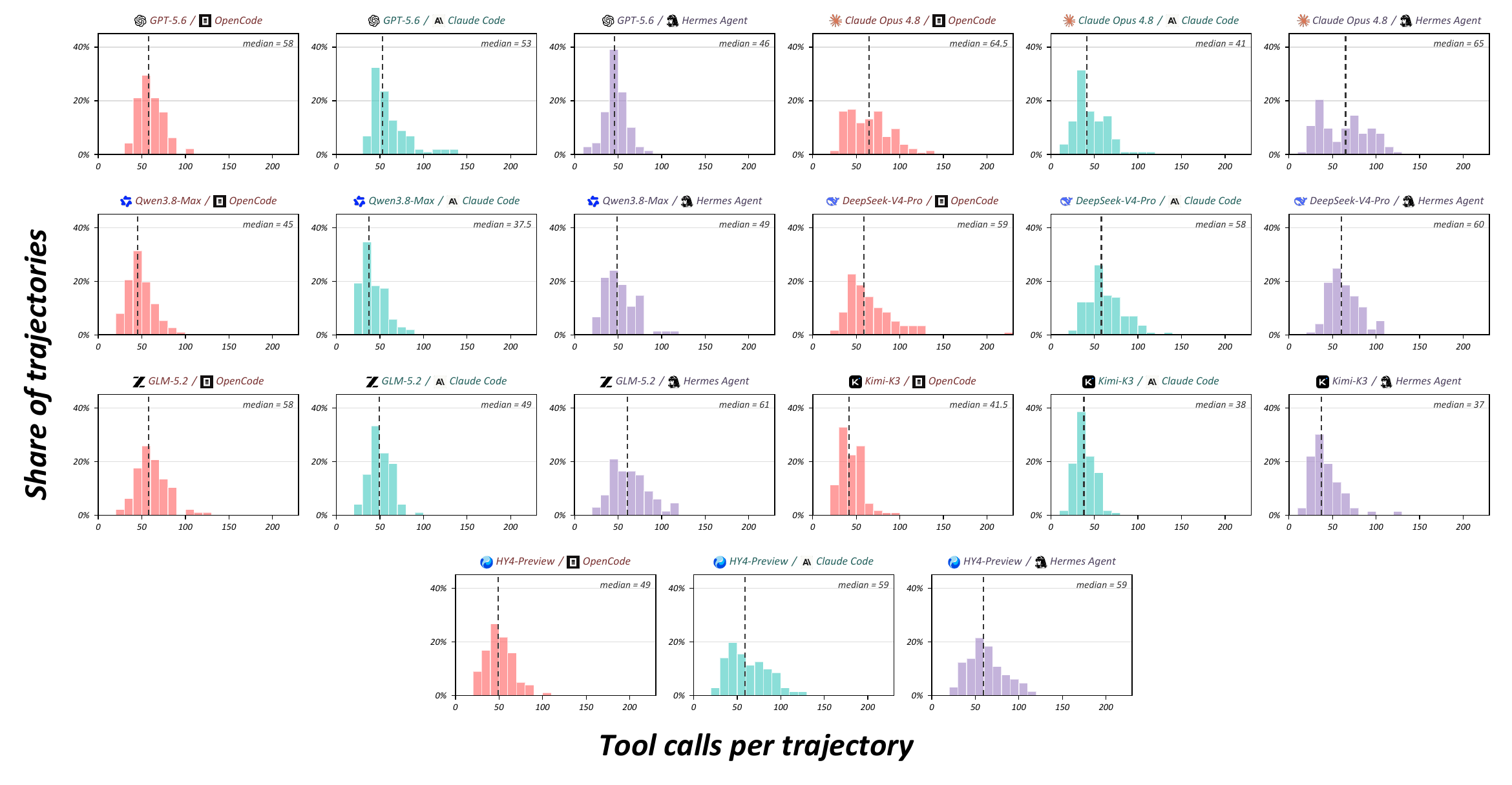}
\caption{Tool-call distributions over successful trajectories for the 21
model--harness combinations. Histogram bins have width~10; dashed lines
indicate medians.}
\label{fig:successful-tool-call-distributions}
\end{figure}

\subsection{Comparison with Baselines}
\label{sec:knighter-comparison}

\begingroup
\setlength{\intextsep}{4pt}
\setlength{\columnsep}{15pt}
\begin{wraptable}{r}{0.4\textwidth}
\centering
\small
\caption{Checker validity on KNighter's 61 Linux-kernel tasks.}
\label{tab:knighter-comparison}
\setlength{\tabcolsep}{3pt}
\renewcommand{\arraystretch}{1}
\begin{tabularx}{\linewidth}{@{}>{\raggedright\arraybackslash}Xcc@{}}
\toprule[1.5pt]
\textbf{Method} & \textbf{Valid} & \textbf{Rate} \\
\midrule[0.6pt]
\checkerbenchname{Ours} & \textbf{49/61} & \textbf{80.3\%} \\
\rowcolor{gray!5}
\harnessname{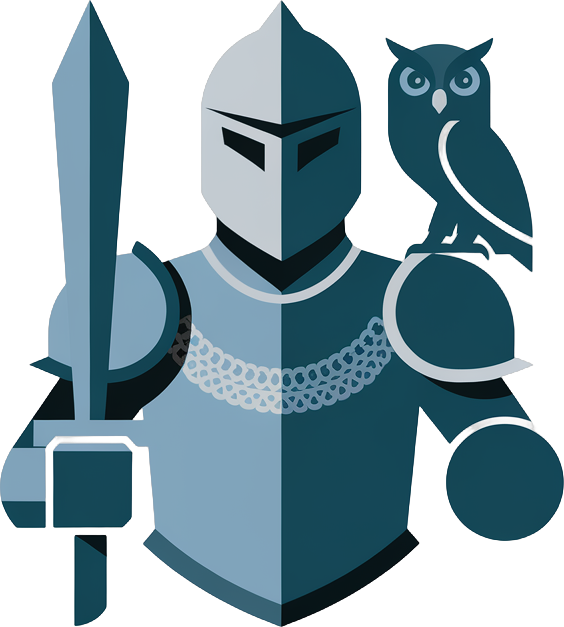}{KNighter} & \textbf{41/61} & \textbf{67.2\%} \\
\checkerbenchname{Ours} \textit{w/o skills} & \textbf{18/61} & \textbf{29.5\%} \\
\bottomrule[1.5pt]
\end{tabularx}
\end{wraptable}
We compare our skill-guided checker-synthesis method, implemented with
OpenCode, against KNighter on the 61 Linux-kernel tasks from KNighter's
original benchmark. All methods use DeepSeek-V4-Pro.
Our method achieves an 80.3\% valid-checker rate, compared with 67.2\% for
KNighter (Table~\ref{tab:knighter-comparison}).
Removing the skill library lowers our method's rate to 29.5\%.
KNighter organizes synthesis through stage-specific prompts and a checker
template, while our method integrates the shared reusable guidance as skills.
Its lead on KNighter's original tasks suggests that this skill-guided workflow
transfers well to an independently defined task set. Details in
Appendix~\ref{app:knighter-comparison}.
\par
\endgroup

\subsection{Ablation Study}
\label{sec:ablation-study}

We compare DeepSeek-V4-Pro on the same 300 tasks, using its original
OpenCode run as Full. As summarized by Table~\ref{tab:component-ablation}, Pass@1 is 39.67\%
for Full, 8.33\% without skills and strategy references, 13.33\% with
execution feedback hidden, and 0.33\% for one-shot generation without tools
or a harness. These gaps persist after excluding Process and normal-termination
requirements: artifact pass is 48.67\%, 18.67\%, 20.33\%, and 1.33\%,
respectively. One-shot performance also reflects invalid or incomplete
submissions in 237/300 tasks. Differences between historical and new run
conditions prevent attributing the gaps solely to the ablated components.
See details in Appendix~\ref{app:ablation-details}.

% Queue this figure before the bottom ablation table so it can use the next page top.
\begin{figure}[tp]
\centering
\setlength{\abovecaptionskip}{4pt}
\includegraphics[width=\textwidth,trim=0 0 0 9bp,clip]{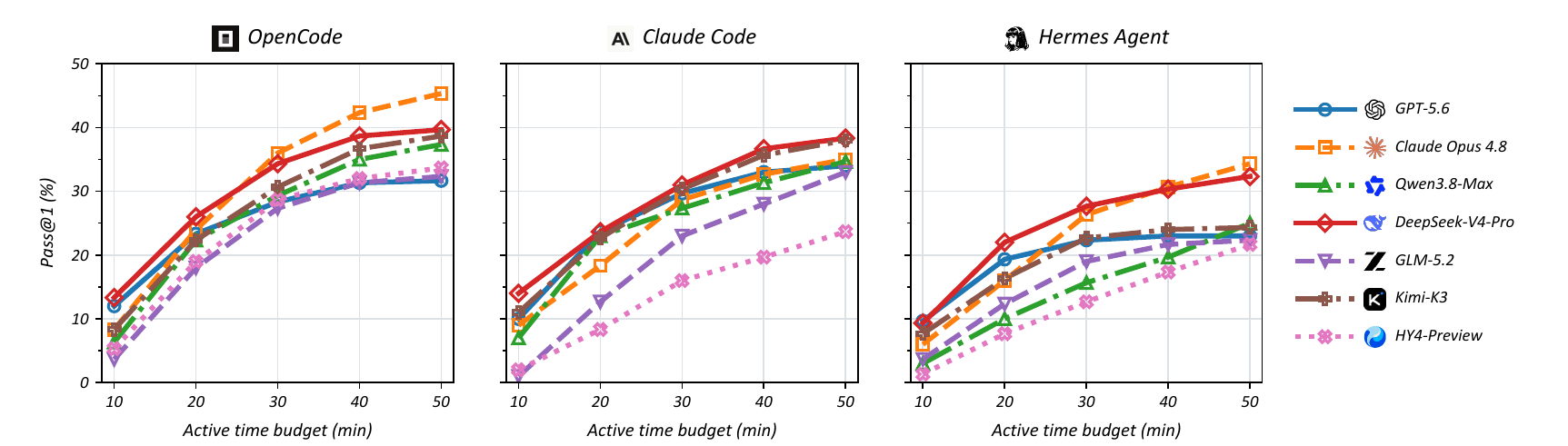}
\caption{Retrospective Pass@1 versus active generation time for three agent
harnesses.}
\label{fig:time-budget-sensitivity}
\end{figure}

% Match Table 3: aligned values, numerical change subscripts, and ranked styling.
% Place below the discussion so this table cannot precede the inline Table 3.
\begin{table}[!b]
\centering
\scriptsize
% Match Table 2's neutral treatment: white header and faint gray striping.
\definecolor{ablationheadercolor}{HTML}{FFFFFF}
\definecolor{ablationstripe}{HTML}{F7F7F7}
\definecolor{ablationup}{HTML}{D95638}
\definecolor{ablationdown}{HTML}{008D9C}
\newcommand{\ablationheader}[1]{%
  \begin{tabular}[c]{@{}c@{}}#1\end{tabular}}
\newcommand{\ablationconfig}[1]{%
  \begin{tabular}[c]{@{}l@{}}#1\end{tabular}}
\newcommand{\ablationvalue}[2]{%
  \makebox[2.5em][r]{#1}\hspace{0.15pt}\makebox[17pt][l]{#2}}
\newcommand{\ablationchange}[3]{%
  \raisebox{-0.35ex}{\fontsize{5}{5}\selectfont
    \textcolor{#1}{\ensuremath{#2}#3}}}
\newcommand{\ablationdeltaup}[1]{\ablationchange{ablationup}{\uparrow}{#1}}
\newcommand{\ablationdeltadown}[1]{\ablationchange{ablationdown}{\downarrow}{#1}}
\caption[Ablation performance on CheckerBench]{%
Performance comparison of DeepSeek-V4-Pro ablations on 300 \textsc{CheckerBench} tasks.
\textbf{Bold} and \underline{underlined} values mark the best and second-best
configurations per metric. Subscripts show numerical changes relative to Full:
\textcolor{ablationup}{$\uparrow$ increases} and
\textcolor{ablationdown}{$\downarrow$ decreases} (percentage points for percentage
metrics). Header arrows indicate the preferred direction. N/A denotes no alerts.}
\label{tab:component-ablation}
\setlength{\tabcolsep}{1.0pt}
\renewcommand{\arraystretch}{1.2}
\begin{tabularx}{\textwidth}{lcc*{8}{>{\centering\arraybackslash}X}}
% Changes use displayed values, as in Table 3. N/A has no numerical change.
\toprule[1.5pt]
\rowcolor{ablationheadercolor}
\ablationheader{\textbf{Configuration}}
& \ablationheader{\textbf{Skill}\\\textbf{lib.}}
& \ablationheader{\textbf{Exec.}\\\textbf{feedback}}
& \ablationheader{\textbf{Pass@1}\\\textbf{(\%)} $\uparrow$}
& \ablationheader{\textbf{Build}\\\textbf{(\%)} $\uparrow$}
& \ablationheader{\textbf{Diff. SR}\\\textbf{(\%)} $\uparrow$}
& \ablationheader{$\mathbf{FP}_{\mathrm{pass}}$\\\textbf{(\%)} $\downarrow$}
& \ablationheader{\textbf{Process}\\$\uparrow$}
& \ablationheader{\textbf{Semantic}\\$\uparrow$}
& \ablationheader{\textbf{Avg.}\\\textbf{Turns} $\downarrow$}
& \ablationheader{\textbf{Avg. Tool}\\\textbf{Calls} $\downarrow$} \\
\midrule[0.6pt]
\ablationconfig{Full} & \featureyes & \featureyes
& \ablationvalue{\textbf{39.67}}{}
& \ablationvalue{\textbf{92.67}}{}
& \ablationvalue{\textbf{90.67}}{}
& \ablationvalue{\textbf{2.56}}{}
& \ablationvalue{\textbf{84.89}}{}
& \ablationvalue{\textbf{8.12}}{}
% Full interaction means use all 300 trajectories (including failures), as in
% output/ablation-final-20260920/results/summary.csv in the source project.
& \ablationvalue{57.63}{}
& \ablationvalue{72.73}{} \\
\rowcolor{ablationstripe}
\ablationconfig{\textit{w/o} Skills} & \featureno & \featureyes
& \ablationvalue{8.33}{\ablationdeltadown{31.34}}
& \ablationvalue{\underline{79.33}}{\ablationdeltadown{13.34}}
& \ablationvalue{\underline{52.67}}{\ablationdeltadown{38.00}}
& \ablationvalue{\underline{6.25}}{\ablationdeltaup{3.69}}
& \ablationvalue{65.31}{\ablationdeltadown{19.58}}
& \ablationvalue{7.39}{\ablationdeltadown{0.73}}
& \ablationvalue{59.56}{\ablationdeltaup{1.93}}
& \ablationvalue{80.53}{\ablationdeltaup{7.80}} \\
\ablationconfig{\textit{w/o} Execution\\Feedback} & \featureyes & \featureno
& \ablationvalue{\underline{13.33}}{\ablationdeltadown{26.34}}
& \ablationvalue{67.33}{\ablationdeltadown{25.34}}
& \ablationvalue{41.67}{\ablationdeltadown{49.00}}
& \ablationvalue{25.00}{\ablationdeltaup{22.44}}
& \ablationvalue{\underline{79.93}}{\ablationdeltadown{4.96}}
& \ablationvalue{\underline{8.00}}{\ablationdeltadown{0.12}}
& \ablationvalue{\underline{53.58}}{\ablationdeltadown{4.05}}
& \ablationvalue{\underline{72.16}}{\ablationdeltadown{0.57}} \\
\rowcolor{ablationstripe}
\ablationconfig{\textit{w/o} Harness} & \featureno & \featureno
& \ablationvalue{0.33}{\ablationdeltadown{39.34}}
& \ablationvalue{3.33}{\ablationdeltadown{89.34}}
& \ablationvalue{3.33}{\ablationdeltadown{87.34}}
& \ablationvalue{\textnormal{N/A}}{}
& \ablationvalue{73.64}{\ablationdeltadown{11.25}}
& \ablationvalue{7.45}{\ablationdeltadown{0.67}}
& \ablationvalue{\textbf{1.00}}{\ablationdeltadown{56.63}}
& \ablationvalue{\textbf{0.00}}{\ablationdeltadown{72.73}} \\
\bottomrule[1.5pt]
\end{tabularx}
\end{table}

\subsection{Time-Budget Sensitivity}
\label{sec:budget-sensitivity}

We assess time-budget sensitivity by retrospectively applying 10, 20, 30, 40,
and 50-minute active generation cutoffs to the existing trajectories across
all 21 model--harness configurations. Only final artifacts completed by each
cutoff contribute, with all 300 tasks retained per configuration.
Figure~\ref{fig:time-budget-sensitivity} shows that aggregate Pass@1 rises from
7.24\% at 10 minutes to 26.05\% at 30 minutes and 32.30\% at 50 minutes,
with diminishing gains: the 40-minute cutoff captures 93.02\% of the successes
observed by 50 minutes, and the final 10 minutes add 2.25 percentage points.
These late gains are largest for Qwen3.8-Max and HY4-Preview.
The curves characterize completion timing under the original 50-minute
protocol; performance under separately imposed shorter deadlines remains
untested. Full metrics and timing rules appear in
Appendix~\ref{app:time-budget-sensitivity}.

\subsection{FP-Threshold Sensitivity}
\label{sec:fp-threshold-sensitivity}

\begin{figure}[tp]
\centering
\setlength{\abovecaptionskip}{4pt}
\includegraphics[width=\textwidth,trim=0 24bp 0 0,clip]{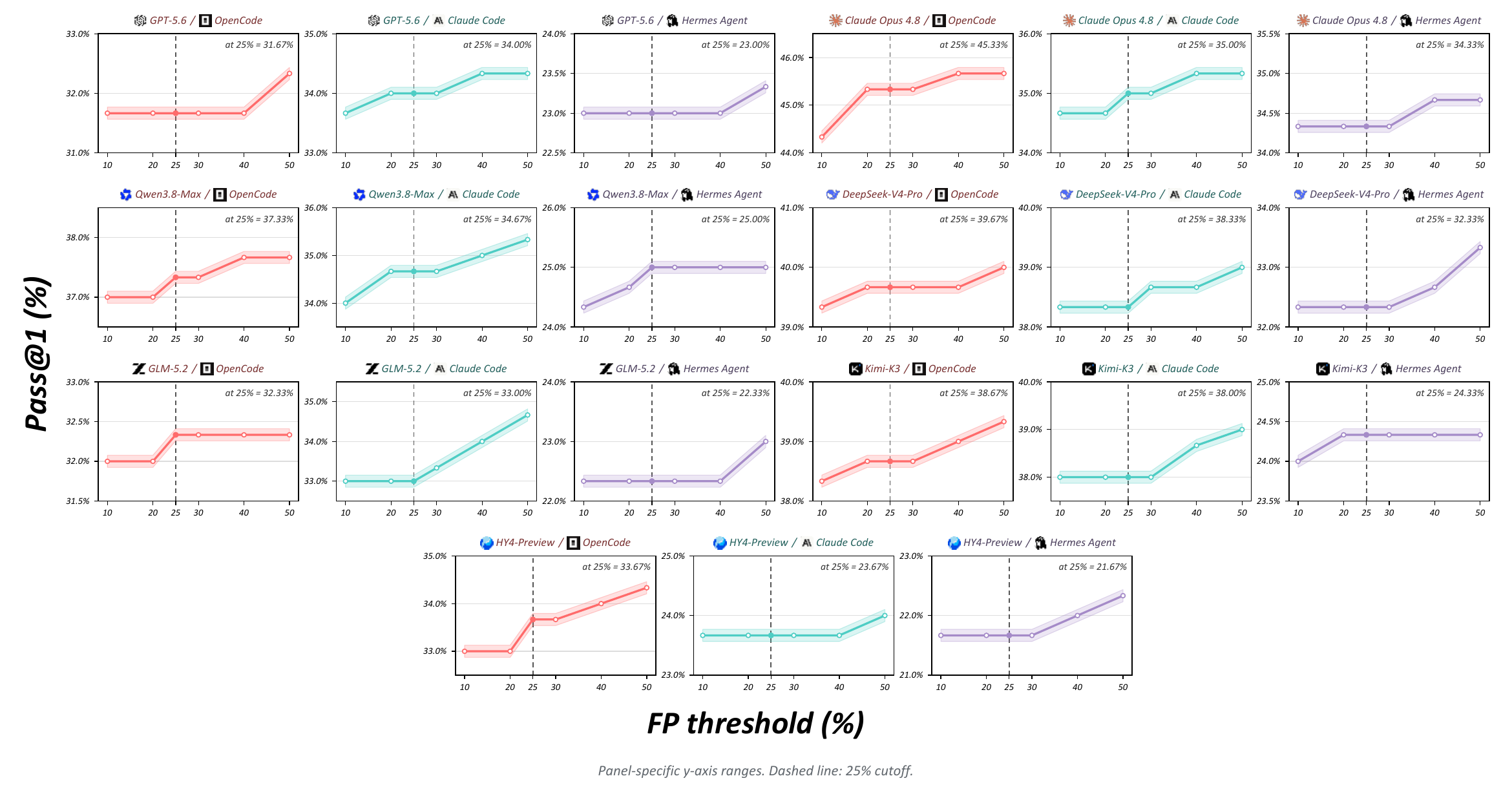}
\caption{Pass@1 versus FP threshold for 21 model--harness
pairs (300 fixed tasks each). Translucent bands highlight the curves.
Dashed lines mark the 25\% cutoff;
annotations show Pass@1 at this cutoff.}
\label{fig:fp-threshold-sensitivity}
\end{figure}

To quantify the influence of the FP threshold on checker
acceptance, we evaluate thresholds of 10\%, 20\%,
25\%, 30\%, 40\%, and 50\% across all 21 model--harness configurations
(Figure~\ref{fig:fp-threshold-sensitivity}). For checkers with adjudicated
alerts, the additional acceptance criterion requires
$100\,\mathrm{FP}/(\mathrm{TP}+\mathrm{FP})$ to be no greater than the
specified threshold.
Adjudication labels and all other eligibility criteria are held fixed,
with 300 tasks retained in the denominator for each configuration.

The 25\% threshold reproduces the acceptance rates for all 21
configurations in Table~\ref{tab:overall-performance}. The aggregate
acceptance rate increases from 32.05\% at 10\% to 32.30\% at 25\%
and 32.83\% at 50\%, yielding a net increase of 0.78 percentage points
over the evaluated range. Over the same range, the largest increase
for an individual configuration is 1.67 percentage points, observed
for GLM-5.2 with Claude Code. These results suggest that aggregate
acceptance is relatively stable across the evaluated FP thresholds
within the recorded cohort.

% Queue the temporal figure before the next page starts so it follows Figure 9.
\begin{figure}[t]
    \centering
    \setlength{\abovecaptionskip}{2pt}
    \includegraphics[width=\textwidth,trim=0 0 0 7bp,clip]{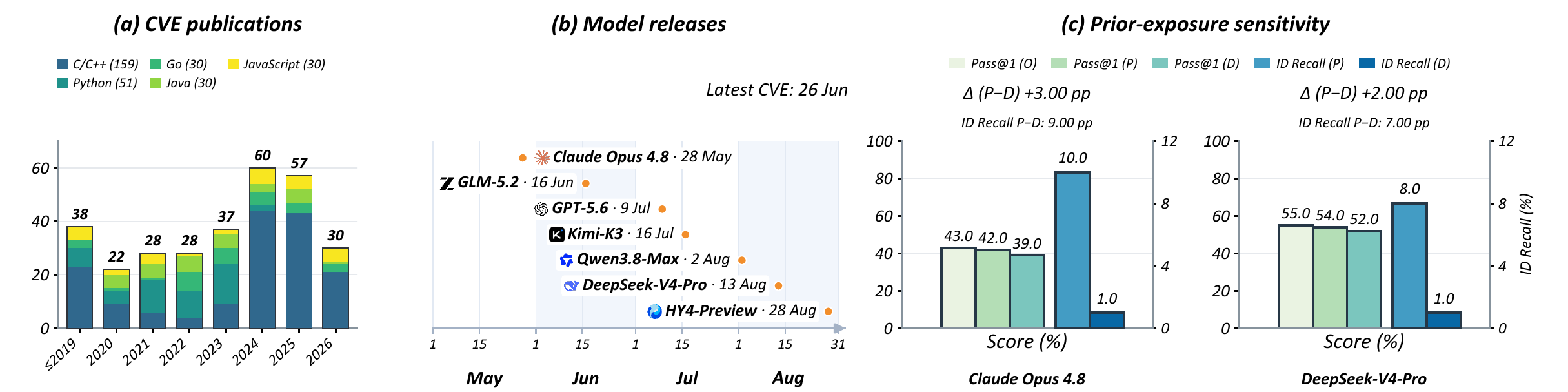}
    \caption{Temporal coverage and prior exposure:
    (a) CVE publication counts; (b) model release dates;
    (c) Pass@1 (O/P/D; left axes, 0--100\%) and exact CVE recall
    (P/D; right axes, 0--12\%).}
    \label{fig:temporal-contamination}
\end{figure}

\subsection{Analysis of Checker Synthesis Failures}
\label{sec:failure-analysis}
We analyze 4,360 selected failed runs across seven models and three harnesses
using tool traces, independent verification logs, and existing judge records.
The leading failure categories are budget exhaustion (25.32\%), inadequate
synthesis process (22.96\%), and inadequate semantic modeling (21.15\%).
These records expose gaps between analysis plans and implementations:
for example, a planned path-sensitive analysis becomes AST matching that
omits locking and control-flow conditions.
Excess false positives account for 15.96\% of failures, underscoring the need
to evaluate precision beyond the original vulnerable--fixed pair.
The remaining failures involve differential detection or localization
(8.17\%), patch-specific overfitting or anchoring (3.56\%), missing or invalid
artifacts (1.83\%), and compilation or registration (1.05\%).
Key challenges for checker synthesis include timely completion, faithful
implementation of analysis plans, and precise semantic modeling.

\subsection{Data Contamination and Prior Exposure}
\label{sec:data-contamination}

CVE publication postdates release for 3/300 tasks with Claude Opus 4.8,
one with GLM-5.2, and none with other models
(Figure~\ref{fig:temporal-contamination}(a,b)); code or patches may have appeared earlier.
On 100 language-stratified tasks, we compare original inputs (O), paraphrases
retaining identifiers (P), and versions of P with meaning-preserving identifier
substitutions (D)~\citep{fang2025lastingbench}.
Figure~\ref{fig:temporal-contamination}(c) shows P--D Pass@1 gaps of
$+3.00$ and $+2.00$ percentage points for Claude Opus 4.8 and DeepSeek-V4-Pro.
Exact CVE recall falls from 10.00\% to 1.00\% and from 8.00\% to 1.00\%,
respectively. The 7--9-point recall decline accompanies only a 2--3-point
Pass@1 drop, suggesting that recognizing familiar vulnerabilities yields
only limited gains under this intervention
(Appendix~\ref{app:contamination-sensitivity}).

\section{Conclusion}
\label{sec:conclusion}

We introduced \textsc{CheckerBench}, an executable benchmark for long-horizon
static-analysis checker synthesis with 300 tasks across five programming-language
ecosystems. \textsc{CheckerLab} independently rebuilds submitted checkers
and assesses diagnostic behavior on vulnerable and fixed revisions,
patch localization, and false positives. Across seven models and three agent
harnesses, the best configuration achieves 45.33\% Pass@1, leaving substantial
room for improvement within the evaluated repositories and analyzer
configurations. \textsc{CheckerBench} provides a reproducible testbed for agents
that turn vulnerability specifications into reliable, reusable checkers.

% Keep the remaining experiment figures ahead of the references.
\FloatBarrier
\restorefloatlayout
\bibliographystyle{abbrvnat}
\nocite{zeng2026recrmbench,su2026coalrag}
\bibliography{reference}

\clearpage
\appendix
% Appendix-only contents, following the compact directory style used in ICLR papers.
\begingroup
\hypersetup{hidelinks,linktoc=all}
\pdfbookmark[0]{Appendix Contents}{appendix-contents}
\section*{Appendix Contents}
\vspace{4pt}
\setcounter{tocdepth}{3}
\setlength{\parskip}{0pt}
\renewcommand{\baselinestretch}{1.08}\selectfont
\makeatletter
\renewcommand*{\l@section}[2]{%
  \addvspace{11pt}%
  {\bfseries\@dottedtocline{1}{0em}{1.8em}{#1}{#2}}}
\renewcommand*{\l@subsection}{\@dottedtocline{2}{1.8em}{2.7em}}
\renewcommand*{\l@subsubsection}{\@dottedtocline{3}{4.5em}{3.4em}}
\@starttoc{atoc}
\makeatother
\clearpage
\endgroup

% Record only appendix headings in the separate contents stream.
\makeatletter
\let\CheckerBenchOriginalAddContentsLine\addcontentsline
\renewcommand{\addcontentsline}[3]{%
  \CheckerBenchOriginalAddContentsLine{#1}{#2}{#3}%
  \def\CheckerBenchContentsExtension{#1}%
  \def\CheckerBenchContentsToc{toc}%
  \ifx\CheckerBenchContentsExtension\CheckerBenchContentsToc
    \ifcsname CheckerBenchContentsLevel@#2\endcsname
      \addtocontents{atoc}{%
        \protect\contentsline{#2}{#3}{\thepage}{\@currentHref}}%
    \fi
  \fi}
\expandafter\def\csname CheckerBenchContentsLevel@section\endcsname{}
\expandafter\def\csname CheckerBenchContentsLevel@subsection\endcsname{}
\expandafter\def\csname CheckerBenchContentsLevel@subsubsection\endcsname{}
\makeatother

% Use the main text's compact paragraph spacing in the appendix as well.
\setlength{\parskip}{4pt plus 1pt minus 1pt}
\setlength{\textfloatsep}{12pt plus 2pt minus 2pt}
\setlength{\floatsep}{10pt plus 2pt minus 2pt}
\setlength{\intextsep}{10pt plus 2pt minus 2pt}

% Let appendix long tables use the remaining page space and continue naturally.
% Flush earlier floats only when needed to preserve figure/table order.
\AddToHook{env/longtable/before}[checkerbench-appendix-order]{\FloatBarrier}
\setlength{\LTpre}{\smallskipamount}
\setlength{\LTpost}{\smallskipamount}
\setlength{\heavyrulewidth}{1.5pt}
\definecolor{appendixharness}{HTML}{EEEEEE}

\section{Extended Related Work}
\label{app:benchmark-feature-audit}

Table~\ref{tab:benchmark-comparison} applies the following operational
criteria. \emph{Repository Context} requires the evaluated agent or system to
receive a complete repository snapshot together with general-purpose
navigation capabilities. Isolated functions, patch-only inputs, pre-extracted
dependency views, and repositories used solely by an
external validation pipeline receive partial credit. \emph{Executable
Environment} requires a reproducible repository or analyzer state with
automated build and execution support. \emph{Multi-Harness Support} is assessed against the six coding-agent
harnesses integrated through \textsc{CheckerBench}'s adapters: \harnessname{claude-code}{Claude Code}~\citep{anthropic2026claudecode}, \harnessname{hermes-agent}{Hermes Agent}~\citep{nousresearch2026hermesagent},
\harnessname{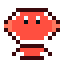}{OpenClaw}~\citep{openclaw2026}, \harnessname{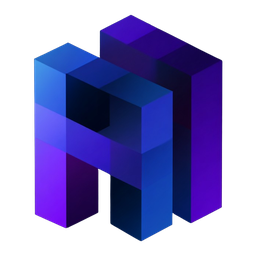}{PAI}~\citep{miessler2026pai},
\harnessname{open-code}{OpenCode}~\citep{anomaly2026opencode}, and \harnessname{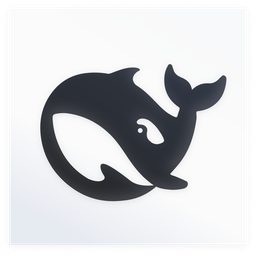}{DS Harness}~\citep{deepseek2026harness}. Full support requires documented,
reproducible integrations for all six harnesses under a shared task and scoring
contract; support for a subset receives partial credit. A generic patch-submission
format, compatibility that depends on an unreleased wrapper, or a
system-specific internal agent does not count as released harness support.

\emph{Long-Horizon Interaction} requires a repeated
observation--tool--edit--execution loop. Protocols that permit multi-step agent
execution or bounded retrieval without specifying the complete loop receive
partial credit. \emph{Checker Synthesis} requires the task output to be an
executable, reusable static-analysis rule or checker. \emph{Semantic
Verification} requires an independent rerun of the submitted checker on
vulnerable and fixed code together with patch-localized diagnostic evidence;
self-reported counts, generic repository tests, and vulnerability labels
receive partial credit. \emph{False-Positive Evaluation} requires residual
diagnostic control on the fixed revision, a production scan, or an equivalent
procedure that measures overly broad reporting. The audit reflects the task definitions, released
artifacts, and evaluation protocols documented by each work, not capabilities
that could be supplied by an external agent implementation.

\paragraph{Adjacent agent benchmarks.}
Recent benchmarks extend agentic software engineering beyond short issue
resolution tasks. NL2Repo-Bench starts with a natural-language specification
and an empty workspace, requiring agents to construct an installable,
multi-module Python library~\citep{ding2025nl2repo}. SWE-Bench Pro instead
draws complex issues from maintained repositories, where solutions often
require coordinated changes across multiple files~\citep{deng2025swebenchpro}.
LongCLI-Bench covers from-scratch development, feature addition, bug fixing,
and refactoring in a command-line environment, using requirement and
regression tests together with step-level scoring~\citep{feng2026longcli}.
SlopCodeBench follows agents as they repeatedly extend their own solutions
under evolving requirements, measuring checkpoint success alongside
structural erosion and code verbosity~\citep{orlanski2026slopcodebench}.

Other evaluations focus on the trajectory itself. Long-Horizon-Terminal-Bench
uses graded subtasks across software engineering and other terminal workflows
to measure partial progress during extended runs
~\citep{li2026longhorizonterminal}. UltraHorizon tests sustained exploration
in partially observable environments, probing planning, memory, and tool use
over long interactions~\citep{luo2025ultrahorizon}. These benchmarks expose
failure modes in sustained execution that also matter when an agent must
implement, compile, and refine a checker.

Binary-analysis benchmarks address a related security workflow.
BIN-BENCH evaluates long chains of tool-assisted binary analysis and examines
reasoning traces in addition to final outcomes~\citep{anonymous2026binbench}.
AgentRE-Bench asks agents to investigate unfamiliar binaries without source
code and scores their structured behavioral claims against fixed ground
truth~\citep{agentre2026bench}. CrackMeBench requires agents to recover a
binary's validation logic and submit an input or generator accepted by the
executable~\citep{david2026crackmebench}. Together, these works measure
software construction, intermediate progress, and binary reasoning.
\textsc{CheckerBench} evaluates a
distinct artifact: a reusable analyzer checker, independently rebuilt and
tested for vulnerable--fixed diagnostic contrast, patch localization, and
false positives.

\section{Dataset Construction Details}
\label{app:dataset-construction}

\subsection{Source Collection and Provenance}
\label{app:vulnerability-source}

We collect public CVE records from the National Vulnerability Database and
link them to fixing commits in open-source repositories written in C/C++,
Java, Python, JavaScript, and Go. A candidate must identify an exact fix,
a retrievable parent revision, and a patch that exposes a defect-bearing
program change. CVE identifiers provide instance-level provenance; CWE
labels support coverage analysis and stratified sampling. We discard
candidates with unresolved provenance, unavailable source, incompatible
redistribution terms, or no reusable static-analysis property. A stable
instance identifier binds each retained repository, revision pair, patch,
and downstream artifact.

\subsection{Build Recovery and Executable Scope}
\label{app:environment-reconstruction}

The following build-anchor and compilation-database details describe the CSA
construction adapter. CodeQL tasks retain their own pinned query-analysis
environment and reconstruction metadata. The construction catalog links each
vulnerability-fixing commit pair to its repository and to one build anchor per revision. Each anchor records a
compilation database, a build snapshot, and build-completeness metadata.
Construction first extracts the unified diff, commit message, and pre- and
post-fix functions. The vulnerable-side compilation database then filters
changed files to translation units for which the original include paths,
preprocessor definitions, language mode, and generated dependencies can be
recovered. Unmatched files remain in the provenance record as skipped inputs;
pairs with no analyzable changed translation unit do not enter synthesis.

For each retained pair, the builder creates isolated Git worktrees for the two
source revisions and overlays revision-compatible generated files from the
corresponding snapshots. Paths in compilation-database entries are rewritten
from the original repository and build roots to the worktree and snapshot
roots. Output directives, warnings-as-errors, disabled system includes, and
known incompatible GCC-specific options are removed or normalized before CSA
execution. Worktrees permit the two revisions to coexist and allow pairs to be
processed concurrently without changing the shared repository mirror. They are
removed after scanning, and stale worktrees from interrupted tasks are pruned
before subsequent construction runs.

After recovering both revisions, we extract the affected pre- and post-fix
functions and generate deterministic commands for compiling a checker and
scanning the pair. We reject an instance if either revision remains
unreconstructable, affected translation units cannot be analyzed after
bounded build recovery, or the pair does not support stable differential
analysis. The resulting source--build correspondence, build metadata, and
scan context remain available for auditing failures.

\subsection{Task Packaging and Shared References}
\label{app:task-packaging}

Each task is packaged in a randomly named workspace containing the patch,
affected functions, a minimal checker template, compile and scan scripts, and
read-only reference assets. The visible configuration contains repository,
revision, build, and scan-scope metadata but omits CVE and CWE identifiers.
Static templates and reference assets are hard-linked when possible and copied
otherwise; in either case they are made read-only. A separate provenance ledger
maps the random identifier back to the source pair. This design exposes all
inputs required for synthesis while preventing vulnerability labels from
becoming task shortcuts.

The shared library \(\mathcal K\) has two layers. Harness-specific skills
support patch interpretation, checker design, build recovery, functional
validation, and false-positive triage. A harness-independent, read-only
reference layer provides analyzer API documentation, official checker or
rule implementations (including CSA resources for C/C++
tasks~\citep{clang2026staticanalyzer,llvm2026csacheckers}), reusable
language-specific AST, data-flow, and state utilities, CWE mechanism
cards~\citep{mitre2026cwe}, and repository-specific conventions when available.
These assets provide implementation knowledge but do not encode the task's
defect-specific state-transition logic or diagnostic reporting predicate.
The agent must infer defect semantics, select analyzer extension points and
semantic or state representations, implement and register the checker, and
refine it using compiler and analyzer feedback. A completed run produces
checker source and plugin or query artifacts, a structured defect analysis,
and a scope statement. The declarations support analysis; only independent
execution establishes correctness.

\subsection{Human-Guided Pilot Rerolls}
\label{app:human-guided-rerolls}

Calibration fixes model and harness versions, each task's analyzer container
and language toolchain, workspace schema, enabled tools, shared skills, and
time, token, and interaction budgets for each attempt. Pilot runs cover the
inspect--implement--compile--scan--repair cycle. We retain observations,
edits, tool calls, and responses. One \emph{interaction step} is one agent
decision, independent of token count or low-level tool events.

For each candidate task, the construction protocol permits one initial
rollout and at most two additional rollouts after human review. Each attempt
uses an isolated workspace and the same task, analyzer configuration,
verification predicates, and acceptance thresholds. After a failed checker
submission, the reviewer identifies the failure from execution logs and
verification evidence, revises the prompt guidance, and records the reason
for the change before launching the next rollout. A prompt revision guides
the agent's next attempt; every resulting checker is independently checked.

The ledger links the attempt index (1--3), prompt version and changes, human
diagnosis, source checksum, complete trace, verification evidence, verdict,
and termination status. The first passing checker advances to quality
control and sampling. A task with no passing checker after attempt 3 is
excluded. The released checker--trace pairs therefore all pass the
construction-time checker-admission criteria. These pilot retries are
recorded separately from independent main-evaluation repeats and their
infrastructure-recovery attempts.

Human diagnosis can address compiler errors, missed or mislocalized
diagnostics, semantic mismatches, or false positives. Verifier internals
and task-specific reference checker source remain held out from the
calibration agent. Each submitted checker undergoes independent contrastive
and patch-relevance checks; accepted executable checkers additionally
undergo a subsystem scan on a prebuilt main-branch worktree and sampled
LLM warning triage (Appendix~\ref{app:quality-control}). All applicable
executable, semantic, scope, and false-positive gates must pass.

Final selection requires reproducible artifacts, complete traces, and
independent verification records. We retain passing trajectories with at
least 40 interaction steps and sample jointly over repository, CWE category,
observed horizon, and difficulty. This threshold selects sustained
interaction under the calibration configuration; it is not an intrinsic
lower bound on task difficulty. The construction ledger retains checker
source and traces for every attempt, independent verification and
false-positive evidence, human diagnoses and prompt revisions, attempt
counts, and either a passing checker--trace pair eligible for curation or
an exclusion record after the cap.

\section{Quality-Control and Contamination Audits}
\label{app:quality-control}

Automated admission checks cover revision ancestry, patch and artifact
integrity, the presence of build anchors, compilation-database coverage,
clean checker compilation, registration and plugin loading, deterministic
vulnerable--fixed scans, verifier isolation, and package completeness. The
CSA contrastive gate requires \(N_i^{-}>N_i^{+}\) and \(N_i^{+}<50\);
CodeQL retains its own frozen differential predicate. Passing the CSA
gate is insufficient by itself: code verification recomputes diagnostics and
requires at least one disappearing vulnerable-revision report whose source
location maps to a function modified by the patch. Hence empty scans,
unrelated diagnostic reductions, and agent-reported counts cannot satisfy the
hard gate.

For accepted executable checkers, a broader subsystem scan on a prebuilt
main-branch worktree supplies false-positive evidence. This worktree is neither
the vulnerable nor the fixed task revision. During construction-time
calibration, a triage model assesses sampled diagnostics. Failed
false-positive checks enter the same human-diagnosis and prompt-revision loop
as other admission failures, within the shared three-attempt cap
(Appendix~\ref{app:human-guided-rerolls}). Each new checker is rebuilt and
independently rechecked. The main evaluation
uses the frozen-candidate scan and single \modelname{claude}{Claude Opus 5} judge in
Section~\ref{sec:evaluation-framework}, without returning scores for repair. Manual audit
samples passing and failing submissions across projects, CWE families, and
horizon strata, and inspects both defect-mechanism fidelity and the
patch-localized true positive. Final removal counts, auditor agreement, and
contamination-overlap statistics will be reported with the frozen release.

\paragraph{Deduplication and split assignment.}
Manual review confirms that the patch supports the stated defect and that
disappearing reports lie in changed defect-bearing functions. We remove
exact duplicate revision pairs; separable units from the same CVE retain
a shared provenance link. Connected groups formed by repository identity,
CVE identity, shared revisions, and patch similarity are each assigned to
one split. Repository-disjoint and weakness-family-disjoint subsets are
additionally marked for generalization analysis.

\paragraph{Artifact isolation.}
Reference states, task-specific checker implementations, calibration
trajectories, verifier thresholds, and adjudication records remain outside
the agent-visible snapshot. The vulnerability patch is an intentional task
input. Potential training exposure to public vulnerability information is
distinct from evaluation-time leakage of checker implementations,
calibration traces, or verifier artifacts. The following audits examine
public dates and model release boundaries; artifact isolation addresses
the separate evaluation-time leakage risk.

\subsection{Temporal Coverage and Prior-Exposure Audit}
\label{app:temporal-audit}

\paragraph{Public evidence and date provenance.}
We archive CVE Program JSON records for the fixed task inventory, recording
the source URL, retrieval provenance, and SHA-256 checksum alongside each
task~\citep{cveprogram2026records}. The audit snapshot of September 15, 2026
covers all 300 tasks and 297 canonical CVEs. The original inventory identifier
CVE-2024-7773 is a rejected duplicate of CVE-2024-45436; the audit retains the
original identifier and uses the canonical record for dating. No task is
removed by this normalization. Publication dates range from June 14, 2010,
to June 26, 2026. Appendix~\ref{app:temporal-distribution} gives the complete
year distribution. The machine-readable ledger is
\path{figures/temporal/task_dates.csv}, accompanied by the archived records
and a public-evidence index.

The ledger separates CVE publication, the CNA's explicit public-disclosure
date, dated public events, and independently checked patch or checker
publication evidence. Discovery dates, CVE reservation dates, packaging dates,
and CVE identifier years do not establish public availability. Git author and
committer timestamps alone also do not establish when a patch became public.
For 48 tasks, the CVE record explicitly dates a public event before its own
publication. The earliest documented event establishes that information was
public by that date; it does not prove the absence of still earlier sources.
The full first-public patch and checker audit is unresolved, so the ledger
does not certify any task as previously unseen on the basis of publication
date alone.

\paragraph{Official model release dates.}
Table~\ref{tab:temporal-model-releases} uses public release dates documented in
provider announcements or release logs, checked on September 19--20, 2026,
with the Opus 4.8 source rechecked on October 4, 2026.
\modelname{openai}{GPT-5.6} Sol and \modelname{claude}{Claude Opus 4.8} were released on July 9 and May 28,
respectively~\citep{openai2026gpt56,anthropic2026opus48}.
The official records date \modelname{qwen}{Qwen3.8-Max} to August 2,
\modelname{deepseek}{DeepSeek-V4-Pro} (the evaluated 0813 version) to August 13, \modelname{glm}{GLM-5.2} to June 16, and \modelname{kimi}{Kimi-K3} to
July 16~\citep{qwen2026qwen38release,deepseek2026v4release,zai2026glm52release,kimi2026k3release}.
\modelname{hunyuan}{HY4-Preview} was released on August 28~\citep{tencent2026hy4preview}.
Dates follow the named model version. Public service or API availability
counts as release even if weights follow later; named public previews are
included and restricted partner previews excluded.
The ledger \path{figures/temporal/model_releases.json} records each source and
date convention. Counts distinguish CVE publication before, on, and after
the reported release day. These boundaries describe public availability,
not a model's training-data cutoff. The actual model version must be pinned
and matched to its release before temporal scoring.

\begin{table}[tp]
\centering\small
\caption{Official model release dates and task counts by CVE publication date. The CVE snapshot is September 15, 2026; model release sources were checked September 19--20, 2026, with Opus 4.8 rechecked October 4, 2026. ``On date'' counts publications on the release day. Later publication alone does not certify an unseen task.}
\label{tab:temporal-model-releases}
\setlength{\tabcolsep}{4pt}
\begin{tabular}{@{}llrrr@{}}
\toprule
Model & Release date & Before & On date & After \\
\midrule
\modelname{openai}{GPT-5.6} & 2026-07-09 & 300 & 0 & 0 \\
\modelname{claude}{Claude Opus 4.8} & 2026-05-28 & 295 & 2 & 3 \\
\modelname{qwen}{Qwen3.8-Max} & 2026-08-02 & 300 & 0 & 0 \\
\modelname{deepseek}{DeepSeek-V4-Pro} & 2026-08-13 & 300 & 0 & 0 \\
\modelname{glm}{GLM-5.2} & 2026-06-16 & 299 & 0 & 1 \\
\modelname{kimi}{Kimi-K3} & 2026-07-16 & 300 & 0 & 0 \\
\modelname{hunyuan}{HY4-Preview} & 2026-08-28 & 300 & 0 & 0 \\
\bottomrule
\end{tabular}
\end{table}

\paragraph{A patch-before-CVE example.}
CVE-2026-10722, a task in \texttt{cilium/ebpf}, was published on June 3, 2026.
Its CVE record references pull request \#2021, whose fix was already merged
on May 27, 2026~\citep{cilium2026btffix}. Thus the patch was public a week
before the CVE publication. A CVE-only filter can overstate the evidence for
novelty relative to a model release between those dates. We
retain both dates and require review of the earlier source; we do not infer
whether the model actually trained on it.

\subsection{Time-Window Evaluation Protocol}
\label{app:temporal-evaluation}

Following LiveCodeBench~\citep{jain2025livecodebench}, the temporal analysis
groups eligible tasks by audited public-availability windows and compares
Pass@1 for each harness--model pair using the same three independent
generation repeats as the main comparison. Eligibility requires a pinned model
version and its official release date together with an audit of public
vulnerability descriptions, patches, and task-specific checker sources.
Tasks with unresolved public-availability dates remain explicitly unclassified.
A common post-release comparison requires the same audited task subset after
all included models' releases. The current inventory has no CVE publications
after the latest of the seven release dates, so it supplies no common
post-release subset for all seven models. Model-specific comparisons report
their own eligible task counts and remain conditional on the earlier-source audit.

Windows are fixed from date and task metadata before inspecting outcomes.
We begin with half-year windows and merge adjacent windows until each
contains at least 20 distinct CVEs, without crossing a model's release
date. A remaining smaller subset is reported with its sample count and
interval, without fitting a temporal trend. Each point reports its distinct
task count and a 95\% confidence interval. For each window, we compute Pass@1
from the final-pass verdicts of the same eligible tasks separately in each
repeat and report the arithmetic mean of the three rates, weighting repeats
equally. Resampling keeps the three repeat outcomes for each task and all
tasks sharing a CVE together. We report language and static analysis
framework strata wherever both periods have
support, and disclose repository and CWE composition for every window;
unsupported strata are not treated as matched comparisons.
Construction-time horizon information serves only as a workload proxy.
A score decline across time is interpreted jointly with these composition
differences, rather than as proof of memorization.

The patch remains an intended synthesis input. Reference checker code,
calibration trajectories, and verifier artifacts remain held out from the
agent-visible package. The audit of generation traces separately records any
external retrieval of task-specific solutions or verification material;
time-window evidence alone cannot rule out such evaluation-time leakage.
The CVE publication counts in Figure~\ref{fig:temporal-contamination}(a)
describe coverage only and do not supply performance estimates.
The construction pilot trajectories do not supply independent temporal
performance estimates. Performance curves are populated only from completed
main-evaluation verdicts and audited task windows.
% Add the time-window Pass@1 figure here only after the audited subset,
% pinned model versions, and final main-evaluation verdicts are available.

\subsection{Paired Prior-Exposure Sensitivity Protocol}
\label{app:contamination-sensitivity}

\paragraph{Sampling.}
The sampling protocol was fixed before collecting supplementary model outcomes.
The initial candidate roster contains 100 distinct canonical CVEs: 53 C/C++,
17 Python, and 10 each in Go, Java, and JavaScript. Within each language, task
IDs are sorted and placed in a fixed pseudorandom order;
canonical-CVE duplicates are skipped across strata. The full candidate
order and original metadata hashes are retained. A task that fails
environment validation is replaced by the next
eligible task in its frozen language order, with the failure evidence and
replacement recorded before any generation or recognition probe. The final
roster is admitted only after all 100 tasks pass validation. Selection and
replacement never use model outcomes or probe recognition.

\paragraph{Conditions and execution.}
\modelname{claude}{Claude Opus 4.8} and \modelname{deepseek}{DeepSeek-V4-Pro} each run with \harnessname{open-code}{OpenCode} under original (O),
ordinary-paraphrase (P), and identity-cue-reduced (D) conditions. P rewrites
non-executable task prose while preserving requirements, source identifiers,
and the complete patch. D starts from P and consistently substitutes safely
modifiable internal identifiers with meaningful alternatives, and removes
residual provenance cues from visible metadata. Standard-library and required
external API names, types, constants, control/data flow, and the vulnerability
and fix mechanisms are preserved. Unsafe transformations involving reflection,
macros, or dynamic name lookup are rejected. This is a semantics-preserving
intervention, distinct from the answer-changing counterfactuals studied by
LastingBench~\citep{fang2025lastingbench}.

There is one independent generation per task--model--condition combination,
for 600 scheduled generations. This supplementary experiment does not reuse
historical Full trajectories. All three conditions share the same skill
library, executable feedback, 3,000-second active budget, model decoding
settings within each model, and frozen verification rules with \modelname{claude}{Claude Opus 5}
as judge. Condition order is randomized within task blocks. Workspaces and
sessions are independent, and probes never enter generation context. Model
API IDs, returned version metadata, access dates, harness version, and
environment fingerprints are archived. Mutable API aliases cannot establish
that a provider's underlying weights remained unchanged.

\paragraph{Transformation and admission checks.}
The before, after, and mainline source trees, patch, extracted context,
metadata, and tool-visible paths must agree with the same transformation.
For CodeQL, all three corresponding databases must be rebuilt from transformed
source; changing only prompts, source-archive text, or database display labels
does not qualify. Original Git history, source copies, private reverse maps,
reference checkers, and verifier material must be inaccessible to generation.
All conditions use preinstalled dependencies and local documentation, with
network access restricted to the model gateway.

Admission requires reverse-mapped syntax checks, human review of defect
and fix equivalence, successful builds or database extraction, and independent
reference-checker replay covering differential behavior, patch localization,
and mainline false positives. Validation logs and human review records bind to
the exact variant hashes. Missing evidence blocks admission. Infrastructure failures
follow the existing bounded recovery policy and are reported separately;
they are not imputed as model failures. Valid model failures and budget
exhaustion remain failures, without score-based retries.

\modelname{claude}{Claude Opus 4.8} drafts transformation proposals and \modelname{deepseek}{DeepSeek-V4-Pro} provides
preliminary critiques. These preparation calls are separate from the 600
synthesis runs and 400 recognition probes and request neither checker
solutions nor CVE identification. A human reviewer checks defect and fix
equivalence in the complete transformed environments; proposal review does not
admit an unbuilt environment. Final semantic approval must reference the exact
transformed environment hashes and retain the reviewer identity, review date,
supporting evidence, and decision.
Independent executable checks remain mandatory. Human review records are
retained with the experiment artifacts.

\paragraph{Closed-book recognition probes.}
Each task--model pair receives separate P and D probes, giving 400 scheduled
short calls. Each probe contains pre-fix code only: no fixing patch, CVE/CWE
label, source URL, or identifying document heading. A fixed extraction rule
selects at most 16,000 characters of code at complete line boundaries; P and D
use corresponding source spans, with equal clipping boundaries rather than
independently selected excerpts. Probe requests use temperature 0 and an
output limit of 256 tokens, with no tools or retrieval. The common instruction
is:
\begin{quote}\small
Identify the publicly documented CVE associated with the supplied pre-fix
code only if you recognize this specific instance. Do not infer an identifier
from the general vulnerability type. Return exactly one JSON object with the
key \texttt{cve\_id}: a single CVE identifier, or \texttt{"unknown"} if you do
not know. Do not use tools or external retrieval.
\end{quote}
ID Recall is the fraction of valid completed calls returning the canonical
CVE; documented duplicate identifiers are normalized through the frozen
ledger. Incorrect, unknown, refused, or malformed model answers do not receive
repair attempts; infrastructure-incomplete calls remain unobserved. A correct
identifier indicates instance familiarity, not memorization of a checker
solution. Failure to recall an identifier does not establish non-exposure.

\paragraph{Estimation and interpretation.}
The primary outcome is the existing final \texttt{full\_pass} verdict.
For each model, the primary effect is
\[
\Delta_{\mathrm{identity}}
=\frac{100}{N}\sum_{i=1}^{N}\bigl(Y_{i,P}-Y_{i,D}\bigr),\qquad N=100,
\]
in percentage points. O--P and O--D differences separate general paraphrase
sensitivity from the total intervention. The prespecified uncertainty analysis
uses 10,000 paired bootstrap draws, resampling canonical-CVE clusters
jointly across conditions, to obtain percentile 95\% intervals. A
supplementary analysis resamples repositories while retaining
within-repository task weights. The intervals
describe variation across sampled tasks/clusters under one realized
generation per condition; they do not separately estimate repeated-generation
variability. The two models are reported separately.

Probe P--D recall differences and descriptive P--D effects among tasks
recognized and unrecognized under P provide supporting evidence. A decline in
recognition together with a positive P--D effect supports sensitivity to
identifying cues; ordinary paraphrase effects and residual difficulty changes
remain alternative explanations. We prespecify a practical margin of
$\pm5$ percentage points. Only an entire 95\% interval inside that margin
supports a bounded effect under this intervention. A nonsignificant difference
alone does not support equivalence or absence of training contamination.
SWE-QA similarly interprets its older-versus-newer source-cohort gap as a
possible exposure association~\citep{peng2026sweqa}.

Build and Diff.\ SR retain the scheduled task denominator. Semantic scores
report their own valid-score counts. Incomplete conditions and contrasts
remain unavailable in the main comparison rather than silently changing
denominators. Figure~\ref{fig:temporal-contamination}(c) in the main text reports
aggregate Pass@1 and exact CVE recall for both models. The P--D effects are
derived from O/P/D Pass@1 of 43.00/42.00/39.00\% for \modelname{claude}{Claude Opus 4.8} and
55.00/54.00/52.00\% for \modelname{deepseek}{DeepSeek-V4-Pro}, yielding $+3.00$ and $+2.00$
percentage points, respectively. The corresponding P--D recall declines are
9.00 and 7.00 percentage points.
% Restore the diagnostic table after its condition-level measurements are available.
% \input{sections/generated/contamination-diagnostics}

\section{Dataset Composition}
\label{app:dataset-composition}

\subsection{Repository-Level Task Counts}
\label{app:repository-counts}

\textsc{CheckerBench} contains 300 tasks from 167 repositories: 159 C/C++ tasks use
CSA, while the 141 CodeQL tasks comprise Java (30), Python (51), JavaScript
(30), and Go (30). Table~\ref{tab:repository-task-counts} lists every
repository, its target language, static analysis framework, task count, and share.

Projects are counted at the repository level. Dataset prefixes are removed,
and repository identifiers are compared without case sensitivity. The
historical \texttt{python-imaging/Pillow} identifier is merged with
\texttt{python-pillow/Pillow}; \texttt{ImageMagick/ImageMagick} and
\texttt{ImageMagick/ImageMagick6} remain separate repositories. Target
language denotes the language analyzed by the task, which can differ from
the repository's dominant implementation language. The counts are taken from
the frozen task inventory and its repository-level aggregation.

Across the full corpus, 151 repositories contribute one task each, 7 contribute
two, 4 contribute three, and 4 contribute four; the Linux kernel repository
contributes 107 tasks (35.7\%).

% Generated by scripts/build_repository_table.py; do not edit rows by hand.
% Sources: figures/project_counts.csv and figures/task_inventory.csv.
\begingroup
\scriptsize
\setlength{\tabcolsep}{3pt}
\renewcommand{\arraystretch}{1.08}
\begin{longtable}{@{}>{\raggedright\arraybackslash}p{0.54\linewidth}
                        >{\raggedright\arraybackslash}p{0.13\linewidth}lrr@{}}
\caption{Complete repository-level task counts. Shares use all 300 tasks as
the denominator and are rounded to one decimal place.}
\label{tab:repository-task-counts}\\
\toprule
\textbf{Repository} & \textbf{Target language} & \textbf{\shortstack[l]{Static analysis\\framework}} & \textbf{Tasks} & \textbf{Share} \\
\midrule
\endfirsthead
\multicolumn{5}{@{}l}{\textit{Table \thetable\ continued}}\\
\toprule
\textbf{Repository} & \textbf{Target language} & \textbf{\shortstack[l]{Static analysis\\framework}} & \textbf{Tasks} & \textbf{Share} \\
\midrule
\endhead
\midrule
\multicolumn{5}{r@{}}{\textit{Continued on next page}}\\
\endfoot
\bottomrule
\endlastfoot
\nolinkurl{torvalds/linux} & C/C++ & CSA & 107 & 35.7\% \\
\nolinkurl{ffmpeg/ffmpeg} & C/C++ & CSA & 4 & 1.3\% \\
\nolinkurl{imagemagick/imagemagick} & C/C++ & CSA & 4 & 1.3\% \\
\nolinkurl{jerryscript-project/jerryscript} & C/C++ & CSA & 4 & 1.3\% \\
\nolinkurl{qemu/qemu} & C/C++ & CSA & 4 & 1.3\% \\
\nolinkurl{airlift/aircompressor} & Java & CodeQL & 3 & 1.0\% \\
\nolinkurl{freerdp/freerdp} & C/C++ & CSA & 3 & 1.0\% \\
\nolinkurl{gpac/gpac} & C/C++ & CSA & 3 & 1.0\% \\
\nolinkurl{libredwg/libredwg} & C/C++ & CSA & 3 & 1.0\% \\
\nolinkurl{ikus060/rdiffweb} & Python & CodeQL & 2 & 0.7\% \\
\nolinkurl{libarchive/libarchive} & C/C++ & CSA & 2 & 0.7\% \\
\nolinkurl{mlflow/mlflow} & Python & CodeQL & 2 & 0.7\% \\
\nolinkurl{openexr/openexr} & C/C++ & CSA & 2 & 0.7\% \\
\nolinkurl{python-pillow/pillow} & Python & CodeQL & 2 & 0.7\% \\
\nolinkurl{rweather/noise-java} & Java & CodeQL & 2 & 0.7\% \\
\nolinkurl{verdammelt/tnef} & C/C++ & CSA & 2 & 0.7\% \\
\nolinkurl{adaptivecomputing/torque} & C/C++ & CSA & 1 & 0.3\% \\
\nolinkurl{alex/rply} & Python & CodeQL & 1 & 0.3\% \\
\nolinkurl{andialbrecht/sqlparse} & Python & CodeQL & 1 & 0.3\% \\
\nolinkurl{ansible/ansible} & Python & CodeQL & 1 & 0.3\% \\
\nolinkurl{apache/incubator-answer} & JavaScript & CodeQL & 1 & 0.3\% \\
\nolinkurl{apache/netbeans-html4j} & Java & CodeQL & 1 & 0.3\% \\
\nolinkurl{apache/sling-org-apache-sling-jcr-base} & Java & CodeQL & 1 & 0.3\% \\
\nolinkurl{apache/thrift} & JavaScript & CodeQL & 1 & 0.3\% \\
\nolinkurl{appleboy/gorush} & Go & CodeQL & 1 & 0.3\% \\
\nolinkurl{argoproj/argo-events} & Go & CodeQL & 1 & 0.3\% \\
\nolinkurl{argoproj/argo-workflows} & Go & CodeQL & 1 & 0.3\% \\
\nolinkurl{auth0/nextjs-auth0} & JavaScript & CodeQL & 1 & 0.3\% \\
\nolinkurl{avo-hq/avo} & JavaScript & CodeQL & 1 & 0.3\% \\
\nolinkurl{aws/aws-sdk-go} & Go & CodeQL & 1 & 0.3\% \\
\nolinkurl{bettercap/bettercap} & Go & CodeQL & 1 & 0.3\% \\
\nolinkurl{bitly/oauth2_proxy} & Go & CodeQL & 1 & 0.3\% \\
\nolinkurl{bonigarcia/webdrivermanager} & Java & CodeQL & 1 & 0.3\% \\
\nolinkurl{bonitasoft/bonita-connector-webservice} & Java & CodeQL & 1 & 0.3\% \\
\nolinkurl{briancappello/flask-unchained} & Python & CodeQL & 1 & 0.3\% \\
\nolinkurl{brokercap/bifrost} & JavaScript & CodeQL & 1 & 0.3\% \\
\nolinkurl{bspkrs/mcpmappingviewer} & Java & CodeQL & 1 & 0.3\% \\
\nolinkurl{caronc/apprise} & Python & CodeQL & 1 & 0.3\% \\
\nolinkurl{ccxvii/mujs} & C/C++ & CSA & 1 & 0.3\% \\
\nolinkurl{cesnet/libyang} & C/C++ & CSA & 1 & 0.3\% \\
\nolinkurl{chaosblade-io/chaosblade} & Go & CodeQL & 1 & 0.3\% \\
\nolinkurl{churchcrm/crm} & JavaScript & CodeQL & 1 & 0.3\% \\
\nolinkurl{cilium/ebpf} & Go & CodeQL & 1 & 0.3\% \\
\nolinkurl{codehaus-plexus/plexus-utils} & Java & CodeQL & 1 & 0.3\% \\
\nolinkurl{containernetworking/plugins} & Go & CodeQL & 1 & 0.3\% \\
\nolinkurl{corazawaf/coraza} & Go & CodeQL & 1 & 0.3\% \\
\nolinkurl{cowtowncoder/java-merge-sort} & Java & CodeQL & 1 & 0.3\% \\
\nolinkurl{craftcms/cms} & JavaScript & CodeQL & 1 & 0.3\% \\
\nolinkurl{cryptpad/cryptpad} & JavaScript & CodeQL & 1 & 0.3\% \\
\nolinkurl{darrenofficial/dpaste} & Python & CodeQL & 1 & 0.3\% \\
\nolinkurl{datadog/guarddog} & Python & CodeQL & 1 & 0.3\% \\
\nolinkurl{decidim/decidim} & JavaScript & CodeQL & 1 & 0.3\% \\
\nolinkurl{decolua/9router} & JavaScript & CodeQL & 1 & 0.3\% \\
\nolinkurl{dedsecinside/torbot} & Python & CodeQL & 1 & 0.3\% \\
\nolinkurl{deeplook/svglib} & Python & CodeQL & 1 & 0.3\% \\
\nolinkurl{dhowden/tag} & Go & CodeQL & 1 & 0.3\% \\
\nolinkurl{dojo/dojo} & JavaScript & CodeQL & 1 & 0.3\% \\
\nolinkurl{duke-git/lancet} & Go & CodeQL & 1 & 0.3\% \\
\nolinkurl{dwisiswant0/apkleaks} & Python & CodeQL & 1 & 0.3\% \\
\nolinkurl{eladnava/mailgen} & JavaScript & CodeQL & 1 & 0.3\% \\
\nolinkurl{electerm/electerm} & JavaScript & CodeQL & 1 & 0.3\% \\
\nolinkurl{encode/starlette} & Python & CodeQL & 1 & 0.3\% \\
\nolinkurl{ethyca/fides} & Python & CodeQL & 1 & 0.3\% \\
\nolinkurl{fonttools/fonttools} & Python & CodeQL & 1 & 0.3\% \\
\nolinkurl{foxcpp/maddy} & Go & CodeQL & 1 & 0.3\% \\
\nolinkurl{gomarkdown/markdown} & Go & CodeQL & 1 & 0.3\% \\
\nolinkurl{googleapis/google-oauth-java-client} & Java & CodeQL & 1 & 0.3\% \\
\nolinkurl{goreleaser/goreleaser} & Go & CodeQL & 1 & 0.3\% \\
\nolinkurl{gphper/ginadmin} & Go & CodeQL & 1 & 0.3\% \\
\nolinkurl{gruntjs/grunt} & JavaScript & CodeQL & 1 & 0.3\% \\
\nolinkurl{hedgedoc/hedgedoc} & JavaScript & CodeQL & 1 & 0.3\% \\
\nolinkurl{horazont/xmpp-http-upload} & Python & CodeQL & 1 & 0.3\% \\
\nolinkurl{horovod/horovod} & Python & CodeQL & 1 & 0.3\% \\
\nolinkurl{huggingface/transformers} & Python & CodeQL & 1 & 0.3\% \\
\nolinkurl{hybridgroup/gobot} & Go & CodeQL & 1 & 0.3\% \\
\nolinkurl{hyperadev/dragonfly} & Java & CodeQL & 1 & 0.3\% \\
\nolinkurl{hyperledger/fabric} & Go & CodeQL & 1 & 0.3\% \\
\nolinkurl{ifmeorg/ifme} & JavaScript & CodeQL & 1 & 0.3\% \\
\nolinkurl{imagemagick/imagemagick6} & C/C++ & CSA & 1 & 0.3\% \\
\nolinkurl{jberet/jsr352} & Java & CodeQL & 1 & 0.3\% \\
\nolinkurl{jdhwpgmbca/pcapture} & Java & CodeQL & 1 & 0.3\% \\
\nolinkurl{jedisct1/pure-ftpd} & C/C++ & CSA & 1 & 0.3\% \\
\nolinkurl{jenkinsci/aws-global-configuration-plugin} & Java & CodeQL & 1 & 0.3\% \\
\nolinkurl{jenkinsci/azure-ad-plugin} & JavaScript & CodeQL & 1 & 0.3\% \\
\nolinkurl{jenkinsci/codedx-plugin} & Java & CodeQL & 1 & 0.3\% \\
\nolinkurl{jenkinsci/dependency-track-plugin} & Java & CodeQL & 1 & 0.3\% \\
\nolinkurl{jenkinsci/environment-manager-tools-plugin} & Java & CodeQL & 1 & 0.3\% \\
\nolinkurl{jenkinsci/xcode-plugin} & Java & CodeQL & 1 & 0.3\% \\
\nolinkurl{jupyter-server/jupyter_server} & Python & CodeQL & 1 & 0.3\% \\
\nolinkurl{keycloak/keycloak} & JavaScript & CodeQL & 1 & 0.3\% \\
\nolinkurl{keylime/keylime} & Python & CodeQL & 1 & 0.3\% \\
\nolinkurl{knative/func} & Go & CodeQL & 1 & 0.3\% \\
\nolinkurl{kyverno/kyverno} & Go & CodeQL & 1 & 0.3\% \\
\nolinkurl{langchain-ai/langchain} & Python & CodeQL & 1 & 0.3\% \\
\nolinkurl{libical/libical} & C/C++ & CSA & 1 & 0.3\% \\
\nolinkurl{libssh2/libssh2} & C/C++ & CSA & 1 & 0.3\% \\
\nolinkurl{libvnc/libvncserver} & C/C++ & CSA & 1 & 0.3\% \\
\nolinkurl{line/centraldogma} & JavaScript & CodeQL & 1 & 0.3\% \\
\nolinkurl{mansuf/mangadex-downloader} & Python & CodeQL & 1 & 0.3\% \\
\nolinkurl{matrix-org/matrix-react-sdk} & JavaScript & CodeQL & 1 & 0.3\% \\
\nolinkurl{matrix-org/sydent} & Python & CodeQL & 1 & 0.3\% \\
\nolinkurl{matrix-org/synapse} & Python & CodeQL & 1 & 0.3\% \\
\nolinkurl{mde/ejs} & JavaScript & CodeQL & 1 & 0.3\% \\
\nolinkurl{meetecho/janus-gateway} & JavaScript & CodeQL & 1 & 0.3\% \\
\nolinkurl{mholt/archiver} & Go & CodeQL & 1 & 0.3\% \\
\nolinkurl{michaelforney/samurai} & C/C++ & CSA & 1 & 0.3\% \\
\nolinkurl{misp/misp-maltego} & Python & CodeQL & 1 & 0.3\% \\
\nolinkurl{mity/md4c} & C/C++ & CSA & 1 & 0.3\% \\
\nolinkurl{mommyheather/advancedbackups} & Java & CodeQL & 1 & 0.3\% \\
\nolinkurl{moodle/moodle} & JavaScript & CodeQL & 1 & 0.3\% \\
\nolinkurl{mrvautin/expresscart} & JavaScript & CodeQL & 1 & 0.3\% \\
\nolinkurl{mwarning/kadnode} & C/C++ & CSA & 1 & 0.3\% \\
\nolinkurl{nautobot/nautobot-plugin-device-onboarding} & Python & CodeQL & 1 & 0.3\% \\
\nolinkurl{netplex/json-smart-v2} & Java & CodeQL & 1 & 0.3\% \\
\nolinkurl{netty/netty-incubator-codec-quic} & Java & CodeQL & 1 & 0.3\% \\
\nolinkurl{nltk/nltk} & Python & CodeQL & 1 & 0.3\% \\
\nolinkurl{nvbn/thefuck} & Python & CodeQL & 1 & 0.3\% \\
\nolinkurl{nvidia/nvflare} & Python & CodeQL & 1 & 0.3\% \\
\nolinkurl{nyuccl/psiturk} & Python & CodeQL & 1 & 0.3\% \\
\nolinkurl{oblac/jodd-http} & Java & CodeQL & 1 & 0.3\% \\
\nolinkurl{ofirdagan/cross-domain-local-storage} & JavaScript & CodeQL & 1 & 0.3\% \\
\nolinkurl{ollama/ollama} & Go & CodeQL & 1 & 0.3\% \\
\nolinkurl{openstack/ceilometer} & Python & CodeQL & 1 & 0.3\% \\
\nolinkurl{opensuse/libsolv} & C/C++ & CSA & 1 & 0.3\% \\
\nolinkurl{ossrs/srs} & Go & CodeQL & 1 & 0.3\% \\
\nolinkurl{outray-tunnel/outray} & JavaScript & CodeQL & 1 & 0.3\% \\
\nolinkurl{overleaf/overleaf} & JavaScript & CodeQL & 1 & 0.3\% \\
\nolinkurl{pallets/werkzeug} & Python & CodeQL & 1 & 0.3\% \\
\nolinkurl{panva/jose} & JavaScript & CodeQL & 1 & 0.3\% \\
\nolinkurl{paramiko/paramiko} & Python & CodeQL & 1 & 0.3\% \\
\nolinkurl{petl-developers/petl} & Python & CodeQL & 1 & 0.3\% \\
\nolinkurl{pgadmin-org/pgadmin4} & Python & CodeQL & 1 & 0.3\% \\
\nolinkurl{phpmyadmin/phpmyadmin} & JavaScript & CodeQL & 1 & 0.3\% \\
\nolinkurl{pomerium/pomerium} & Go & CodeQL & 1 & 0.3\% \\
\nolinkurl{prismjs/prism} & JavaScript & CodeQL & 1 & 0.3\% \\
\nolinkurl{prometheus/client_golang} & Go & CodeQL & 1 & 0.3\% \\
\nolinkurl{pytorch/serve} & Java & CodeQL & 1 & 0.3\% \\
\nolinkurl{pytorchlightning/pytorch-lightning} & Python & CodeQL & 1 & 0.3\% \\
\nolinkurl{rasahq/rasa} & Python & CodeQL & 1 & 0.3\% \\
\nolinkurl{redis/redis} & C/C++ & CSA & 1 & 0.3\% \\
\nolinkurl{revel/revel} & Go & CodeQL & 1 & 0.3\% \\
\nolinkurl{saitoha/libsixel} & C/C++ & CSA & 1 & 0.3\% \\
\nolinkurl{sajari/docconv} & Go & CodeQL & 1 & 0.3\% \\
\nolinkurl{sass/libsass} & C/C++ & CSA & 1 & 0.3\% \\
\nolinkurl{satori/go.uuid} & Go & CodeQL & 1 & 0.3\% \\
\nolinkurl{seccomp/libseccomp} & C/C++ & CSA & 1 & 0.3\% \\
\nolinkurl{sehmaschine/django-grappelli} & Python & CodeQL & 1 & 0.3\% \\
\nolinkurl{skvadrik/re2c} & C/C++ & CSA & 1 & 0.3\% \\
\nolinkurl{smallrye/smallrye-config} & Java & CodeQL & 1 & 0.3\% \\
\nolinkurl{snowflakedb/snowflake-connector-python} & Python & CodeQL & 1 & 0.3\% \\
\nolinkurl{spiral-project/ihatemoney} & Python & CodeQL & 1 & 0.3\% \\
\nolinkurl{stranger6667/pyanyapi} & Python & CodeQL & 1 & 0.3\% \\
\nolinkurl{strukturag/libde265} & C/C++ & CSA & 1 & 0.3\% \\
\nolinkurl{strukturag/libheif} & C/C++ & CSA & 1 & 0.3\% \\
\nolinkurl{tankywoo/simiki} & Python & CodeQL & 1 & 0.3\% \\
\nolinkurl{tdunning/pig-vector} & Java & CodeQL & 1 & 0.3\% \\
\nolinkurl{tenable/integration-jira-cloud} & Python & CodeQL & 1 & 0.3\% \\
\nolinkurl{txthinking/brook} & Go & CodeQL & 1 & 0.3\% \\
\nolinkurl{uclouvain/openjpeg} & C/C++ & CSA & 1 & 0.3\% \\
\nolinkurl{vadz/libtiff} & C/C++ & CSA & 1 & 0.3\% \\
\nolinkurl{vyperlang/vyper} & Python & CodeQL & 1 & 0.3\% \\
\nolinkurl{weaveworks/tf-controller} & Go & CodeQL & 1 & 0.3\% \\
\nolinkurl{wildfly/jboss-ejb-client} & Java & CodeQL & 1 & 0.3\% \\
\nolinkurl{xsuchy/templated-dictionary} & Python & CodeQL & 1 & 0.3\% \\
\nolinkurl{xuxueli/xxl-job} & Java & CodeQL & 1 & 0.3\% \\
\nolinkurl{zopefoundation/products.genericsetup} & Python & CodeQL & 1 & 0.3\% \\
\nolinkurl{zwczou/weixin-python} & Python & CodeQL & 1 & 0.3\% \\

\midrule
\textbf{Total: 167 repositories} & \textbf{5 ecosystems} & CSA / CodeQL & \textbf{300} & \textbf{100.0\%} \\
\end{longtable}
\endgroup

\subsection{Programming Languages and Static Analysis Frameworks}
\label{app:language-framework-counts}

Table~\ref{tab:language-framework-counts} gives the complete task distribution
by target programming language and static analysis framework. Each task
contributes once. Language denotes the code analyzed by the task, using the
same definition and frozen inventory as Table~\ref{tab:repository-task-counts}.

% Generated by scripts/build_distribution_tables.py.
\begin{table}[htp]
\centering
\scriptsize
\setlength{\tabcolsep}{4pt}
\renewcommand{\arraystretch}{1.08}
\caption{Complete task counts by target language and static analysis framework. Shares use all 300 tasks.}
\label{tab:language-framework-counts}
\begin{tabular}{@{}lrrrr@{}}
\toprule
\textbf{Target language} & \textbf{CSA} & \textbf{CodeQL} & \textbf{Tasks} & \textbf{Share} \\
\midrule
C/C++ & 159 & 0 & 159 & 53.0\% \\
Java & 0 & 30 & 30 & 10.0\% \\
Python & 0 & 51 & 51 & 17.0\% \\
JavaScript & 0 & 30 & 30 & 10.0\% \\
Go & 0 & 30 & 30 & 10.0\% \\
\midrule \textbf{Total} & \textbf{159} & \textbf{141} & \textbf{300} & \textbf{100.0\%} \\
\bottomrule
\end{tabular}
\end{table}

\subsection{Complete CWE Label Distribution}
\label{app:cwe-distribution}

Table~\ref{tab:cwe-counts} lists all 85 CWEs and the separate \emph{Other}
group. The source retains original task annotations, supplements three
previously unlabeled tasks with explicit CWE labels from saved public CVE
records, and assigns Other to the 14 tasks for which no explicit CWE label was
found. Labels are deduplicated within each task. Of the 300 tasks, 272 have
one CWE label, 14 have two, and 14 have none; 286 tasks therefore have an
explicit CWE label. The table's Other row contains the 14 tasks without an
explicit CWE label.

Each task contributes once to each of its labels. Table percentages use
all 300 tasks, so rows can overlap and their shares sum to 104.7\% including
Other. Figure~\ref{fig:benchmark-statistics}(c) instead partitions the 314
task--label pairs into pie sectors. Its five leading CWEs are shown with
their names. The pie's Other sector combines 213 pairs from the remaining
80 CWEs with the 14 unlabeled tasks, giving 227 pairs (72.3\%).
Ties are ordered by ascending CWE number.

The CWEs comprise 82 Weakness entries and three upstream Category entries
(CWE-189, CWE-264, and CWE-399), as recorded in the supplied CWE v4.20
catalog~\citep{mitre2026cwe}. Category annotations are retained without
remapping them to more specific weaknesses; CWE-264 is marked obsolete in
that catalog. CSA and CodeQL columns report task counts for the respective
static analysis frameworks.

% Generated by scripts/build_distribution_tables.py.
\begingroup
\scriptsize
\setlength{\tabcolsep}{4pt}
\renewcommand{\arraystretch}{1.08}
\begin{longtable}{@{}>{\raggedright\arraybackslash}p{0.13\linewidth}>{\raggedright\arraybackslash}p{0.46\linewidth}rrrr@{}}
\caption{Complete distribution of 85 CWEs and Other. Each task counts once per distinct label; shares use all 300 tasks. Multiple-label tasks appear in more than one row. $\dagger$ marks an upstream Category entry.}
\label{tab:cwe-counts}\\
\toprule
\textbf{CWE} & \textbf{Name} & \textbf{CSA} & \textbf{CodeQL} & \textbf{Tasks} & \textbf{Share} \\
\midrule
\endfirsthead
\multicolumn{6}{@{}l}{\textit{Table \thetable\ continued}}\\
\toprule
\textbf{CWE} & \textbf{Name} & \textbf{CSA} & \textbf{CodeQL} & \textbf{Tasks} & \textbf{Share} \\
\midrule
\endhead
\midrule
\multicolumn{6}{r@{}}{\textit{Continued on next page}}\\
\endfoot
\bottomrule
\endlastfoot
CWE-787 & Out-of-bounds Write & 25 & 1 & 26 & 8.7\% \\
CWE-125 & Out-of-bounds Read & 14 & 7 & 21 & 7.0\% \\
CWE-22 & Improper Limitation of a Pathname to a Restricted Directory ('Path Traversal') & 2 & 13 & 15 & 5.0\% \\
CWE-476 & NULL Pointer Dereference & 13 & 0 & 13 & 4.3\% \\
CWE-79 & Improper Neutralization of Input During Web Page Generation ('Cross-site Scripting') & 0 & 12 & 12 & 4.0\% \\
CWE-362 & Concurrent Execution using Shared Resource with Improper Synchronization ('Race Condition') & 9 & 3 & 12 & 4.0\% \\
CWE-416 & Use After Free & 12 & 0 & 12 & 4.0\% \\
CWE-400 & Uncontrolled Resource Consumption & 2 & 8 & 10 & 3.3\% \\
CWE-908 & Use of Uninitialized Resource & 9 & 0 & 9 & 3.0\% \\
CWE-667 & Improper Locking & 8 & 0 & 8 & 2.7\% \\
CWE-190 & Integer Overflow or Wraparound & 6 & 1 & 7 & 2.3\% \\
CWE-415 & Double Free & 7 & 0 & 7 & 2.3\% \\
CWE-20 & Improper Input Validation & 1 & 5 & 6 & 2.0\% \\
CWE-611 & Improper Restriction of XML External Entity Reference & 0 & 6 & 6 & 2.0\% \\
CWE-617 & Reachable Assertion & 6 & 0 & 6 & 2.0\% \\
CWE-119 & Improper Restriction of Operations within the Bounds of a Memory Buffer & 5 & 0 & 5 & 1.7\% \\
CWE-191 & Integer Underflow (Wrap or Wraparound) & 5 & 0 & 5 & 1.7\% \\
CWE-200 & Exposure of Sensitive Information to an Unauthorized Actor & 0 & 5 & 5 & 1.7\% \\
CWE-377 & Insecure Temporary File & 0 & 5 & 5 & 1.7\% \\
CWE-401 & Missing Release of Memory after Effective Lifetime & 5 & 0 & 5 & 1.7\% \\
CWE-601 & URL Redirection to Untrusted Site ('Open Redirect') & 0 & 5 & 5 & 1.7\% \\
CWE-770 & Allocation of Resources Without Limits or Throttling & 3 & 2 & 5 & 1.7\% \\
CWE-78 & Improper Neutralization of Special Elements used in an OS Command ('OS Command Injection') & 0 & 4 & 4 & 1.3\% \\
CWE-129 & Improper Validation of Array Index & 3 & 1 & 4 & 1.3\% \\
CWE-674 & Uncontrolled Recursion & 3 & 1 & 4 & 1.3\% \\
CWE-94 & Improper Control of Generation of Code ('Code Injection') & 0 & 3 & 3 & 1.0\% \\
CWE-189$^{\dagger}$ & Numeric Errors & 1 & 2 & 3 & 1.0\% \\
CWE-287 & Improper Authentication & 0 & 3 & 3 & 1.0\% \\
CWE-326 & Inadequate Encryption Strength & 0 & 3 & 3 & 1.0\% \\
CWE-1333 & Inefficient Regular Expression Complexity & 0 & 3 & 3 & 1.0\% \\
CWE-23 & Relative Path Traversal & 0 & 2 & 2 & 0.7\% \\
CWE-59 & Improper Link Resolution Before File Access ('Link Following') & 0 & 2 & 2 & 0.7\% \\
CWE-74 & Improper Neutralization of Special Elements in Output Used by a Downstream Component ('Injection') & 0 & 2 & 2 & 0.7\% \\
CWE-77 & Improper Neutralization of Special Elements used in a Command ('Command Injection') & 0 & 2 & 2 & 0.7\% \\
CWE-284 & Improper Access Control & 0 & 2 & 2 & 0.7\% \\
CWE-338 & Use of Cryptographically Weak Pseudo-Random Number Generator (PRNG) & 0 & 2 & 2 & 0.7\% \\
CWE-352 & Cross-Site Request Forgery (CSRF) & 0 & 2 & 2 & 0.7\% \\
CWE-502 & Deserialization of Untrusted Data & 0 & 2 & 2 & 0.7\% \\
CWE-532 & Insertion of Sensitive Information into Log File & 0 & 2 & 2 & 0.7\% \\
CWE-672 & Operation on a Resource after Expiration or Release & 2 & 0 & 2 & 0.7\% \\
CWE-772 & Missing Release of Resource after Effective Lifetime & 2 & 0 & 2 & 0.7\% \\
CWE-835 & Loop with Unreachable Exit Condition ('Infinite Loop') & 2 & 0 & 2 & 0.7\% \\
CWE-863 & Incorrect Authorization & 0 & 2 & 2 & 0.7\% \\
CWE-29 & Path Traversal: '\textbackslash{}..\textbackslash{}filename' & 0 & 1 & 1 & 0.3\% \\
CWE-88 & Improper Neutralization of Argument Delimiters in a Command ('Argument Injection') & 0 & 1 & 1 & 0.3\% \\
CWE-91 & XML Injection (aka Blind XPath Injection) & 0 & 1 & 1 & 0.3\% \\
CWE-120 & Buffer Copy without Checking Size of Input ('Classic Buffer Overflow') & 1 & 0 & 1 & 0.3\% \\
CWE-122 & Heap-based Buffer Overflow & 1 & 0 & 1 & 0.3\% \\
CWE-131 & Incorrect Calculation of Buffer Size & 1 & 0 & 1 & 0.3\% \\
CWE-193 & Off-by-one Error & 1 & 0 & 1 & 0.3\% \\
CWE-201 & Insertion of Sensitive Information Into Sent Data & 0 & 1 & 1 & 0.3\% \\
CWE-203 & Observable Discrepancy & 1 & 0 & 1 & 0.3\% \\
CWE-209 & Generation of Error Message Containing Sensitive Information & 0 & 1 & 1 & 0.3\% \\
CWE-250 & Execution with Unnecessary Privileges & 0 & 1 & 1 & 0.3\% \\
CWE-256 & Plaintext Storage of a Password & 0 & 1 & 1 & 0.3\% \\
CWE-264$^{\dagger}$ & Permissions, Privileges, and Access Controls & 0 & 1 & 1 & 0.3\% \\
CWE-266 & Incorrect Privilege Assignment & 0 & 1 & 1 & 0.3\% \\
CWE-295 & Improper Certificate Validation & 0 & 1 & 1 & 0.3\% \\
CWE-312 & Cleartext Storage of Sensitive Information & 0 & 1 & 1 & 0.3\% \\
CWE-327 & Use of a Broken or Risky Cryptographic Algorithm & 0 & 1 & 1 & 0.3\% \\
CWE-331 & Insufficient Entropy & 0 & 1 & 1 & 0.3\% \\
CWE-347 & Improper Verification of Cryptographic Signature & 0 & 1 & 1 & 0.3\% \\
CWE-364 & Signal Handler Race Condition & 1 & 0 & 1 & 0.3\% \\
CWE-366 & Race Condition within a Thread & 0 & 1 & 1 & 0.3\% \\
CWE-369 & Divide By Zero & 1 & 0 & 1 & 0.3\% \\
CWE-399$^{\dagger}$ & Resource Management Errors & 1 & 0 & 1 & 0.3\% \\
CWE-407 & Inefficient Algorithmic Complexity & 0 & 1 & 1 & 0.3\% \\
CWE-459 & Incomplete Cleanup & 1 & 0 & 1 & 0.3\% \\
CWE-460 & Improper Cleanup on Thrown Exception & 0 & 1 & 1 & 0.3\% \\
CWE-522 & Insufficiently Protected Credentials & 0 & 1 & 1 & 0.3\% \\
CWE-538 & Insertion of Sensitive Information into Externally-Accessible File or Directory & 0 & 1 & 1 & 0.3\% \\
CWE-552 & Files or Directories Accessible to External Parties & 0 & 1 & 1 & 0.3\% \\
CWE-613 & Insufficient Session Expiration & 0 & 1 & 1 & 0.3\% \\
CWE-639 & Authorization Bypass Through User-Controlled Key & 0 & 1 & 1 & 0.3\% \\
CWE-668 & Exposure of Resource to Wrong Sphere & 0 & 1 & 1 & 0.3\% \\
CWE-670 & Always-Incorrect Control Flow Implementation & 0 & 1 & 1 & 0.3\% \\
CWE-680 & Integer Overflow to Buffer Overflow & 1 & 0 & 1 & 0.3\% \\
CWE-692 & Incomplete Denylist to Cross-Site Scripting & 0 & 1 & 1 & 0.3\% \\
CWE-789 & Memory Allocation with Excessive Size Value & 0 & 1 & 1 & 0.3\% \\
CWE-826 & Premature Release of Resource During Expected Lifetime & 1 & 0 & 1 & 0.3\% \\
CWE-843 & Access of Resource Using Incompatible Type ('Type Confusion') & 1 & 0 & 1 & 0.3\% \\
CWE-862 & Missing Authorization & 0 & 1 & 1 & 0.3\% \\
CWE-918 & Server-Side Request Forgery (SSRF) & 0 & 1 & 1 & 0.3\% \\
CWE-1188 & Initialization of a Resource with an Insecure Default & 0 & 1 & 1 & 0.3\% \\
CWE-1336 & Improper Neutralization of Special Elements Used in a Template Engine & 0 & 1 & 1 & 0.3\% \\
Other & Other (no explicit CWE found) & 8 & 6 & 14 & 4.7\% \\
\midrule \multicolumn{2}{@{}l}{\textbf{Sum across label rows}} & \textbf{165} & \textbf{149} & \textbf{314} & \textbf{104.7\%} \\
\end{longtable}
\endgroup

\subsection{Calibration Horizon and Difficulty Distributions}
\label{app:calibration-distributions}

Table~\ref{tab:trajectory-distributions} expands the horizon and difficulty
statistics reported in Section~\ref{sec:benchmark-statistics}. The minimum horizon is a curation constraint: only accepted calibration
trajectories with at least 40 interaction steps are eligible for the final
300-instance sample. Difficulty scores are used by the dataset-selection
pipeline.
The median horizon is 60 steps and the maximum is 123. The associated
difficulty score has median 0.684 and range 0.414--0.846. These retained
construction-time fields use the original calibration metadata.

\begin{table}[htp]
    \centering
    \scriptsize
    \caption{Interaction-horizon and difficulty distributions of the
    calibration trajectories.}
    \label{tab:trajectory-distributions}
    \setlength{\tabcolsep}{8pt}
    \renewcommand{\arraystretch}{0.94}
    \begin{tabular}{@{}lrr@{}}
        \toprule[1.5pt]
        \textbf{Interval} & \textbf{Count} & \textbf{Share} \\
        \midrule
        \multicolumn{3}{@{}l}{\textit{Interaction steps}} \\
        $[40,60)$  & 144 & 48.0\% \\
        $[60,80)$  & 104 & 34.7\% \\
        $[80,\infty)$ & 52 & 17.3\% \\
        \midrule
        \multicolumn{3}{@{}l}{\textit{Difficulty score}} \\
        $[0.0,0.5)$ & 32  & 10.7\% \\
        $[0.5,0.6)$ & 74  & 24.7\% \\
        $[0.6,0.7)$ & 59  & 19.7\% \\
        $[0.7,0.8)$ & 130 & 43.3\% \\
        $[0.8,1.0)$ & 5   & 1.7\% \\
        \bottomrule[1.5pt]
    \end{tabular}
\end{table}

Table~\ref{tab:calibration-step-counts} reports every one-step bin underlying
the horizon intervals above, including zero-count bins. These counts use
construction-time interaction-step metadata; the message-level pilot
counts follow below.

% Generated by scripts/build_distribution_tables.py.
\begingroup
\scriptsize
\setlength{\tabcolsep}{4pt}
\renewcommand{\arraystretch}{0.90}
\begin{longtable}{@{}rrr@{\hspace{14pt}}rrr@{\hspace{14pt}}rrr@{}}
\caption{Complete one-step frequency counts in the retained construction-time interaction-step metadata. Read down each three-column block. All 84 bins, including zero-count bins, are shown; shares use 300 trajectories.}
\label{tab:calibration-step-counts}\\
\toprule
\textbf{Steps} & \textbf{Traces} & \textbf{Share} & \textbf{Steps} & \textbf{Traces} & \textbf{Share} & \textbf{Steps} & \textbf{Traces} & \textbf{Share} \\
\midrule
\endfirsthead
\multicolumn{9}{@{}l}{\textit{Table \thetable\ continued}}\\
\toprule
\textbf{Steps} & \textbf{Traces} & \textbf{Share} & \textbf{Steps} & \textbf{Traces} & \textbf{Share} & \textbf{Steps} & \textbf{Traces} & \textbf{Share} \\
\midrule
\endhead
\midrule
\multicolumn{9}{r@{}}{\textit{Continued on next page}}\\
\endfoot
\bottomrule
\endlastfoot
40 & 4 & 1.3\% & 68 & 2 & 0.7\% & 96 & 2 & 0.7\% \\
41 & 4 & 1.3\% & 69 & 7 & 2.3\% & 97 & 0 & 0.0\% \\
42 & 4 & 1.3\% & 70 & 8 & 2.7\% & 98 & 1 & 0.3\% \\
43 & 3 & 1.0\% & 71 & 5 & 1.7\% & 99 & 1 & 0.3\% \\
44 & 2 & 0.7\% & 72 & 5 & 1.7\% & 100 & 0 & 0.0\% \\
45 & 6 & 2.0\% & 73 & 6 & 2.0\% & 101 & 1 & 0.3\% \\
46 & 7 & 2.3\% & 74 & 5 & 1.7\% & 102 & 0 & 0.0\% \\
47 & 8 & 2.7\% & 75 & 5 & 1.7\% & 103 & 0 & 0.0\% \\
48 & 6 & 2.0\% & 76 & 5 & 1.7\% & 104 & 0 & 0.0\% \\
49 & 11 & 3.7\% & 77 & 6 & 2.0\% & 105 & 2 & 0.7\% \\
50 & 3 & 1.0\% & 78 & 1 & 0.3\% & 106 & 0 & 0.0\% \\
51 & 12 & 4.0\% & 79 & 4 & 1.3\% & 107 & 1 & 0.3\% \\
52 & 7 & 2.3\% & 80 & 1 & 0.3\% & 108 & 1 & 0.3\% \\
53 & 8 & 2.7\% & 81 & 1 & 0.3\% & 109 & 0 & 0.0\% \\
54 & 11 & 3.7\% & 82 & 9 & 3.0\% & 110 & 0 & 0.0\% \\
55 & 6 & 2.0\% & 83 & 2 & 0.7\% & 111 & 1 & 0.3\% \\
56 & 13 & 4.3\% & 84 & 3 & 1.0\% & 112 & 0 & 0.0\% \\
57 & 9 & 3.0\% & 85 & 1 & 0.3\% & 113 & 0 & 0.0\% \\
58 & 12 & 4.0\% & 86 & 3 & 1.0\% & 114 & 0 & 0.0\% \\
59 & 8 & 2.7\% & 87 & 3 & 1.0\% & 115 & 0 & 0.0\% \\
60 & 7 & 2.3\% & 88 & 3 & 1.0\% & 116 & 1 & 0.3\% \\
61 & 8 & 2.7\% & 89 & 4 & 1.3\% & 117 & 0 & 0.0\% \\
62 & 4 & 1.3\% & 90 & 4 & 1.3\% & 118 & 0 & 0.0\% \\
63 & 5 & 1.7\% & 91 & 0 & 0.0\% & 119 & 0 & 0.0\% \\
64 & 3 & 1.0\% & 92 & 1 & 0.3\% & 120 & 0 & 0.0\% \\
65 & 5 & 1.7\% & 93 & 1 & 0.3\% & 121 & 1 & 0.3\% \\
66 & 6 & 2.0\% & 94 & 2 & 0.7\% & 122 & 0 & 0.0\% \\
67 & 7 & 2.3\% & 95 & 1 & 0.3\% & 123 & 1 & 0.3\% \\

\end{longtable}
\endgroup

\subsection{Pilot Trajectory Statistics}
\label{app:pilot-trajectories}

\begin{figure}[htp]
    \centering
    % Crop the surrounding padding, including the space left by the footer.
    \includegraphics[width=\linewidth,trim=0 30bp 0 27bp,clip]{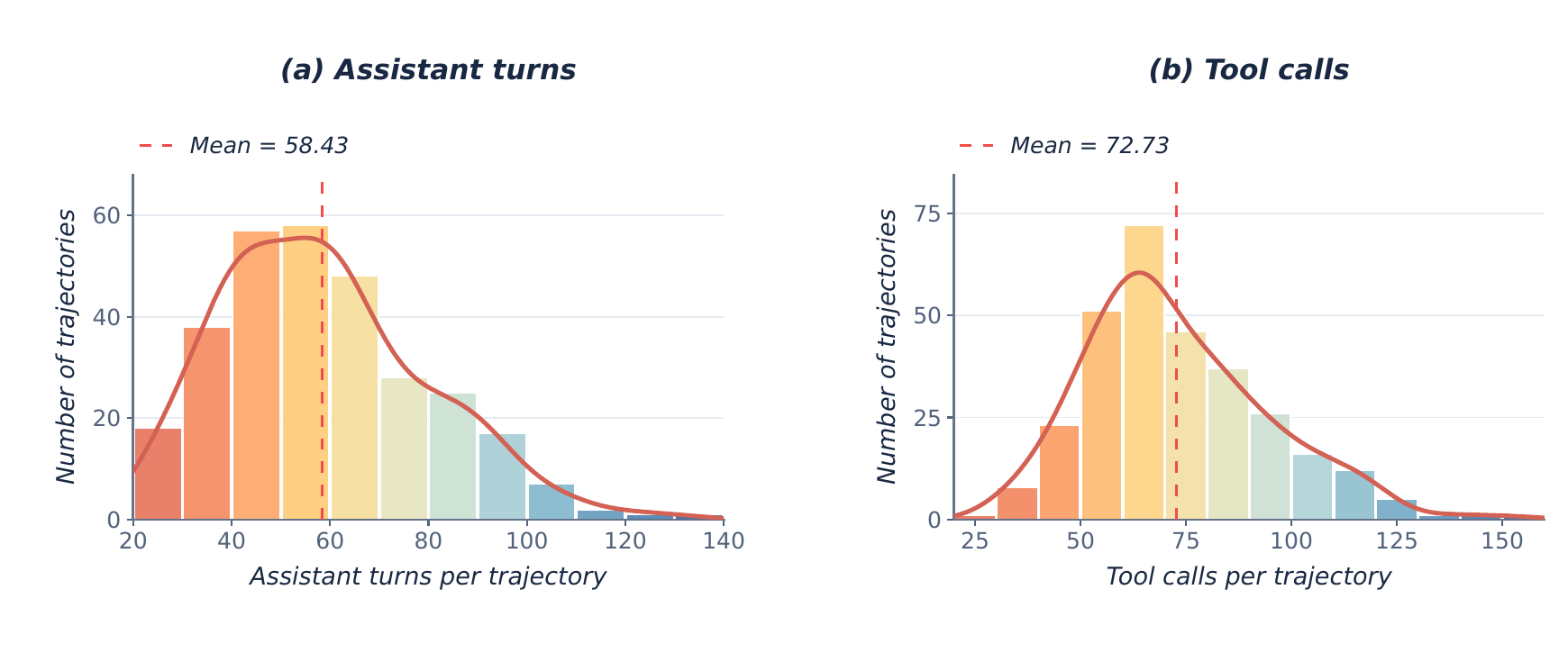}
    \caption{Distributions of (a) assistant turns and (b) tool calls across
    300 admitted pilot trajectories in \textsc{CheckerBench}. Histogram bins
    have width 10. Dashed red lines mark sample means; solid red curves show
    Gaussian kernel density estimates with Scott's bandwidth, scaled to
    histogram counts.}
    \label{fig:pilot-trajectory-distributions}
\end{figure}

Figure~\ref{fig:pilot-trajectory-distributions} summarizes the original pilot
traces for the 300 selected tasks (159 CSA and 141 CodeQL), with one admitted
trajectory per task. Selection follows the original checker-admission
records, including 13 trajectories whose checkers passed admission while
their agent runs ended at the spending limit. These statistics describe the
construction pilot; independent main-evaluation runs follow the final-pass
criteria in Section~\ref{sec:evaluation-framework}.

% Original supplied figure and audited data:
% output/trajectory_statistics_300_20260915/checkerbench300_pilot_turns_tools.pdf
% The publication PDF omits the repeated sample-size labels and footer.

An \emph{assistant turn} is a unique assistant message, deduplicated by
session and message identifier. A \emph{tool call} is a unique tool-use
invocation, deduplicated by session and invocation identifier; failed calls
and calls without a matching result are included. The tool-call total
includes explicit skill invocations. Assistant-turn counts are computed from
message events and use a different definition from the construction-time
interaction-step metadata in Table~\ref{tab:trajectory-distributions} and
the runner's native turn counter.

Table~\ref{tab:pilot-bin-counts} gives all histogram bins for both pilot
metrics. The two columns each account for the same 300 admitted pilot
trajectories under their respective event-counting definitions.

% Generated by scripts/build_distribution_tables.py.
\begin{table}[htp]
\centering
\scriptsize
\setlength{\tabcolsep}{4pt}
\renewcommand{\arraystretch}{1.08}
\caption{Complete bin counts for the pilot distributions in Figure~\ref{fig:pilot-trajectory-distributions}. Intervals are left-inclusive and right-exclusive. Each share uses 300 pilot trajectories.}
\label{tab:pilot-bin-counts}
\begin{tabular}{@{}lrrrr@{}}
\toprule
\textbf{Interval} & \textbf{Assistant turns} & \textbf{Share} & \textbf{Tool calls} & \textbf{Share} \\
\midrule
$[20,30)$ & 18 & 6.0\% & 1 & 0.3\% \\
$[30,40)$ & 38 & 12.7\% & 8 & 2.7\% \\
$[40,50)$ & 57 & 19.0\% & 23 & 7.7\% \\
$[50,60)$ & 58 & 19.3\% & 51 & 17.0\% \\
$[60,70)$ & 48 & 16.0\% & 72 & 24.0\% \\
$[70,80)$ & 28 & 9.3\% & 46 & 15.3\% \\
$[80,90)$ & 25 & 8.3\% & 37 & 12.3\% \\
$[90,100)$ & 17 & 5.7\% & 26 & 8.7\% \\
$[100,110)$ & 7 & 2.3\% & 16 & 5.3\% \\
$[110,120)$ & 2 & 0.7\% & 12 & 4.0\% \\
$[120,130)$ & 1 & 0.3\% & 5 & 1.7\% \\
$[130,140)$ & 1 & 0.3\% & 1 & 0.3\% \\
$[140,150)$ & 0 & 0.0\% & 1 & 0.3\% \\
$[150,160)$ & 0 & 0.0\% & 1 & 0.3\% \\
\midrule \textbf{Total} & \textbf{300} & \textbf{100.0\%} & \textbf{300} & \textbf{100.0\%} \\
\bottomrule
\end{tabular}
\end{table}

\FloatBarrier

\subsection{CVE Publication-Time Distribution}
\label{app:temporal-distribution}

Table~\ref{tab:temporal-publication-counts} gives every publication year,
including empty years. Counts use the 300 task records, so separable tasks
sharing a CVE contribute separately; uncertainty estimates for temporal
performance keep those tasks grouped. The 2026 bin is observed through the
September 15 metadata snapshot. The complete dating and boundary policies
are in Appendix~\ref{app:temporal-audit}.

\begin{table}[htp]
\centering\small
\caption{CVE publication years for the fixed 300 tasks. Dates use the canonical CVE record; the CVE identifier year is not used.}
\label{tab:temporal-publication-counts}
\begin{tabular}{lrrrr}
\toprule
Publication year & CSA & CodeQL & Tasks & Share \\
\midrule
2010 & 0 & 1 & 1 & 0.3\% \\
2011 & 1 & 0 & 1 & 0.3\% \\
2012 & 0 & 0 & 0 & 0.0\% \\
2013 & 1 & 1 & 2 & 0.7\% \\
2014 & 2 & 2 & 4 & 1.3\% \\
2015 & 0 & 0 & 0 & 0.0\% \\
2016 & 2 & 0 & 2 & 0.7\% \\
2017 & 7 & 3 & 10 & 3.3\% \\
2018 & 5 & 2 & 7 & 2.3\% \\
2019 & 5 & 6 & 11 & 3.7\% \\
2020 & 9 & 13 & 22 & 7.3\% \\
2021 & 6 & 22 & 28 & 9.3\% \\
2022 & 4 & 24 & 28 & 9.3\% \\
2023 & 9 & 28 & 37 & 12.3\% \\
2024 & 44 & 16 & 60 & 20.0\% \\
2025 & 43 & 14 & 57 & 19.0\% \\
2026 & 21 & 9 & 30 & 10.0\% \\
\midrule
Total & 159 & 141 & 300 & 100.0\% \\
\bottomrule
\end{tabular}
\end{table}

\FloatBarrier

\section{Evaluation and Inference Protocol}
\label{app:evaluation-protocol}

\subsection{System Configurations and Reproducibility}

The seven evaluated models are \modelname{openai}{GPT-5.6-Sol}~\citep{openai2026gpt56},
\modelname{claude}{Claude Opus 4.8}~\citep{anthropic2026opus48},
\modelname{qwen}{Qwen3.8-Max}~\citep{qwen2026qwen38release},
\modelname{deepseek}{DeepSeek-V4-Pro-0813}~\citep{deepseek2026v4}, \modelname{glm}{GLM-5.2}~\citep{zai2026glm52},
\modelname{kimi}{Kimi-K3}~\citep{kimi2026k3}, and \modelname{hunyuan}{HY4-Preview}~\citep{tencent2026hy4preview}.
We record harness and model versions, context and decoding settings, reasoning
effort, and enabled tools. All systems share pinned repositories, toolchains,
containers, resource limits, and the verifier. Released artifacts record
prompts, skills, tool schemas, commands, workspace diffs, diagnostics,
termination reasons, and submissions.

\subsection{Task Admission and Independent Repeats}

The evaluation manifest fixes 300 environments (159 CSA and 141 CodeQL),
three harnesses, seven models, and three independent generation repeats
per environment--harness--model combination. All three repeats cover the
same 300 tasks for each of the 21 harness--model pairs, yielding 6,300
logical runs per repeat and 18,900 in total. The reported results are the
arithmetic mean of the three repeat-level estimates, with equal weight
for each repeat. Recovery attempts are tracked separately.
A held-out reference checker is used to verify environment usability before
runs are admitted. Only environment--model combinations that pass preflight
are released, and blocked combinations remain visible in the coverage ledger.
The reference checker is never part of the synthesis inputs.

Each independent repeat uses a separate workspace and a 50-minute generation
budget. It receives the task instruction, patch, relevant code context, tools,
and enabled skills. Visible compilation and analysis feedback can support
multiple inspect--write--compile--repair cycles within this budget. The final
source is \texttt{checker.cpp} for CSA or CodeQL query source. Generation
completion status is recorded independently of downstream verification.

\subsection{Source and Evidence Freezing}

At termination, the ledger records the submitted source, SHA-256 digest,
complete trajectory, tool-call events, explicit skill invocations, elapsed
time, usage, and termination reason. Task, harness, model, container, tool,
and scorer versions are retained with independent-repeat and recovery-attempt
indices. Subsequent scans and judges consume this exact candidate. Compiled
artifacts can be regenerated from it, but verification recovery and a valid
low quality score cannot replace its source. Raw evidence is retained alongside
any normalized summaries and referenced by path and content hash.

\subsection{Backend Execution}

The reference execution image pins LLVM/Clang~18 and mounts the task workspace
at \texttt{/work}. Checker source is compiled with
\texttt{/opt/llvm/bin/clang++} in C++17 mode as a position-independent shared
object using \texttt{-fPIC}, \texttt{-shared}, \texttt{-fno-rtti},
\texttt{-fno-exceptions}, the pinned LLVM include directory, and
\texttt{-Wl,--allow-shlib-undefined}. The resulting object is a CSA plugin,
not a standalone executable. Each scan invokes
\texttt{/opt/llvm/bin/clang --analyze}, loads the plugin through the analyzer,
enables its registered checker name, and appends the translated
compilation-database flags and source file. Diagnostics are attributed to the
checker through its registered diagnostic label. The task-local validator and
the held-out verifier call the same generated scripts, but the verifier
rebuilds from source in a clean workspace.

For CodeQL, the verifier independently compiles and executes the submitted
query using the task's pinned CodeQL environment and revision-specific
analysis inputs. It enforces the frozen CodeQL rules for target and patch
relevance, vulnerable--fixed differential behavior, query scope, diagnostic
quality, and hardcoding. CSA-specific compiler commands, report signatures,
count limits, and plugin gates are not reused as CodeQL acceptance rules.
The backend and rule version accompany every reported gate result.

\subsection{Frozen Mainline Scan and Single-Judge Decisions}

After paired-revision verification, the same checker is scanned over a
broader mainline scope. The mainline code is distinct from the vulnerable and
fixed task revisions. At most ten warnings requiring judgment are sampled;
the selected warning IDs and supporting code context are retained.
Appendix~\ref{app:fp-sampling} specifies the selection rule, recorded warning
counts, and the distinction between the sampled gate and aggregate precision.
\modelname{claude}{Claude Opus 5}~\citep{anthropic2026opus5}
performs false-positive triage and, in separate stages, assesses process
quality and vulnerability semantics. All stages use this one judge model and
record the exact model configuration and rubric version. Judge decisions
cannot override a failed executable or risk gate.

Each stage commits its first valid decision, including valid low scores and
negative verdicts. Only unavailable or invalid evidence follows recovery;
quality-based resampling is prohibited. Mainline ``refinement'' is an
assessment of the frozen candidate, with no follow-up edits by the synthesis
model. Empty warning sets and missing or invalid judgments follow each
backend's frozen rules; an undefined sample statistic is not fabricated.

\subsection{Mainline Alert Selection and Sensitivity}
\label{app:fp-sampling}

\paragraph{Selection unit and ordering.}
The frozen implementation selects individual alert records from each
checker's saved mainline output. Selection is deterministic conditional on
that list; it is not uniform random sampling over
alerts, files, or alert types, and the selector does not rank by severity.
All records are judged when the list has at most ten entries. For a longer
list of length $N$, the zero-based selected indices are
\begin{equation}
\begin{aligned}
 I_{\mathrm{CSA}} &= \{k\lfloor N/10\rfloor : k=0,\ldots,9\},\\
 I_{\mathrm{CodeQL}} &=
 \{\operatorname{round}(k(N-1)/9) : k=0,\ldots,9\}.
\end{aligned}
\label{eq:fp-alert-selection}
\end{equation}
CSA retains checker-tagged warnings and deduplicates identical
path--line--column locations within each compiler invocation. Its collected
list follows scan-worker completion order, so a fresh scan need not reproduce
the same ordering. In particular, for 11--19 CSA warnings the rule takes the
first ten. CodeQL uses the normalized finding order and first persists at
most 100 findings; its ten-alert selector operates on that retained list.
Neither backend allocates a fixed quota per file or alert type. Replaying a
sample therefore requires the saved list and selected warning identities.

\paragraph{Warning volume before triage.}
Table~\ref{tab:fp-warning-volume} reports the archived audit
of 1,940 passed checkers from retained repeat~1, covering all 21
harness--model combinations. This historical cohort predates the subsequent
aggregate corrections to Table~\ref{tab:overall-performance}; it is not a
three-repeat total. Of its 1,335 recorded alerts, 978 were initially sampled
and 357 required supplementary judgment. Only 16 checkers had more than ten
recorded alerts; 1,566 had none. These counts cover the passed
cohort, not all generated checkers. CSA stops collecting after a compiler
invocation brings the collected total to at least 100 warnings, and one scan in this cohort
reached that cap. Consequently, recorded counts do not establish the number
of alerts an uncapped scan of the entire repository would produce.

% Recomputed from the frozen per-checker audit's source manifest, script,
% and warning-volume records.
\begin{table}[tp]
\centering\small
\caption{Recorded warning volume in the historical repeat-1 cohort of
passed checkers. Counts precede the ten-alert triage selection.}
\label{tab:fp-warning-volume}
\begin{tabular}{lrrrrr}
\toprule
Backend & Checkers & Zero alerts & $>10$ alerts & Recorded alerts & Sampled alerts \\
\midrule
CSA & 550 & 348 & 11 & 880 & 596 \\
CodeQL & 1,390 & 1,218 & 5 & 455 & 382 \\
\midrule
Total & 1,940 & 1,566 & 16 & 1,335 & 978 \\
\bottomrule
\end{tabular}
\end{table}

\paragraph{Alert-weighted and checker-weighted precision.}
The ten-alert sample supplies a per-checker acceptance gate.
The reported $\mathrm{FP}_{\mathrm{pass}}$ instead pools recorded alerts from
passed checkers, including supplementary judgments beyond that
sample. For a fixed system and repeat, let $\mathcal{C}_{+}$ contain those
checkers with $N_i=\mathrm{TP}_i+\mathrm{FP}_i>0$. The corresponding
alert-weighted and checker-weighted precision percentages are
\begin{equation}
 P_{\mathrm{micro}}=100\frac{\sum_{i\in\mathcal{C}_{+}}\mathrm{TP}_i}
                                {\sum_{i\in\mathcal{C}_{+}}N_i},
 \qquad
 P_{\mathrm{macro}}=\frac{100}{|\mathcal{C}_{+}|}
                    \sum_{i\in\mathcal{C}_{+}}\frac{\mathrm{TP}_i}{N_i}.
\label{eq:fp-precision-weighting}
\end{equation}
Thus $\mathrm{FP}_{\mathrm{pass}}=100-P_{\mathrm{micro}}$; larger alert sets
receive more weight in this metric. $P_{\mathrm{macro}}$ gives every
alert-bearing checker equal weight. Zero-alert checkers are reported
separately and enter neither precision denominator; a clean scan does not
establish perfect precision.

\paragraph{Observed sensitivity in the archived cohort.}
We recompute both estimators using the same post hoc \modelname{claude}{Opus~5} audit of all
1,335 saved alerts. It contains 1,063 TP and 272 FP labels.
We average the seven system-level estimates
equally within each harness. The precision ordering changes from \harnessname{open-code}{OpenCode},
\harnessname{claude-code}{Claude Code}, \harnessname{hermes-agent}{Hermes Agent} under alert weighting to \harnessname{claude-code}{Claude Code}, \harnessname{open-code}{OpenCode},
\harnessname{hermes-agent}{Hermes Agent} under checker weighting. Within \harnessname{claude-code}{Claude Code}, the precision
leader changes from \modelname{openai}{GPT-5.6} (93.33\% micro) to \modelname{hunyuan}{HY4-Preview} (96.11\% macro).
These are conditional precision rankings, distinct from Pass@1 rankings.

Adding the all-recorded-alert gate removes 81 of the 1,940 original passes.
The harness ordering by Pass@1 remains unchanged, but the leading
harness--model pair changes from \harnessname{claude-code}{Claude Code}/\modelname{kimi}{Kimi-K3} (122/300 originally)
to \harnessname{open-code}{OpenCode}/\modelname{deepseek}{DeepSeek-V4-Pro} (119/300 after the added gate).
The archived sampled gates and supplementary alert audit both used
\modelname{claude}{Claude Opus 5} judgments. The additional gate expands coverage from the
ten-alert sample to all recorded alerts while retaining the same judge.
Moreover, this is a passed-only audit of saved output: it does not
re-adjudicate originally failed checkers or recover alerts beyond scan caps.
A representative-subset comparison would need a common task subset spanning
repositories and backends, include checkers reaching the FP stage regardless
of their original verdict, and compare the sampled and complete outputs
under the same judge and rubric. That analysis is not established by this
audit; stability of the current full-evaluation rankings under exhaustive
adjudication remains unverified.

\subsection{Failure Accounting, Recovery, and Aggregation}

Final \texttt{passed} requires normal generation completion and every
backend-specific executable, mainline, process, semantic, mechanism, scope,
and anti-hardcoding gate. Valid generation failures, refusals, and candidate
code errors are failures. Infrastructure failures permit at most two
additional attempts under the recovery policy. When source already exists,
verification recovery reuses it and preserves prior valid stage decisions.
Each additional attempt is attached to its original logical run and does not
contribute an additional independent value to the reported estimate.

Unresolved infrastructure outcomes are recorded separately from valid model
failures, and evaluation coverage is reported with the scheduled manifest.
Incomplete repeat groups are reported as incomplete coverage; they are not
silently treated as completed failures or removed from the scheduled task set.
For each harness--model pair, we compute Pass@1 as the percentage of the
300 tasks whose checker passes the full evaluation in each independent
repeat. We report the arithmetic mean of these three rates, giving each
repeat equal weight (Section~\ref{sec:metrics}). Each independent run
contributes its own final verdict;
a successful repeat does not replace a failed repeat.
Final-success efficiency summaries average interaction
turns, tool calls, and explicit skill calls over passed trajectories in that
repeat, counting recorded invocation events including subagent activity.
The conditional means from the three repeats are averaged equally.
Each repeat's passed-trajectory count is retained.
If a retained repeat has no successful trajectory, its conditional efficiency
and the aggregate efficiency mean are \texttt{N/A}, not zero.
% Optional aggregate diagnostics: analyzer invocations, elapsed time, tokens, and cost.
Full failure and recovery records remain archived for audit and resource accounting.

\subsection{Statistical Reporting}

The reported main comparison, time-budget levels, and task subsets average
the same three independent repeats equally. The separate ablation study
uses one generation per task for each new variant and reuses the original
Full cohort (Appendix~\ref{app:ablation-details}). The subgroup
figure and tables report descriptive estimates without significance claims.
Ablation intervals resample paired task IDs jointly across configurations;
they reflect task composition and do not estimate uncertainty over repeated
model generations or independent CVE clusters.
Valid-score sample sizes and backend-specific breakdowns accompany diagnostic
means; passed-trajectory counts accompany efficiency summaries.

Build and Diff.\ SR use all 300 tasks as their denominator. Process and
Semantic average available valid scores on their respective scales.
For passed checkers, each recorded mainline alert is labeled TP
or FP. Within each repeat, $\mathrm{FP}_{\mathrm{pass}}$ pools these alerts as
$100\sum\mathrm{FP}/\sum(\mathrm{TP}+\mathrm{FP})$.
The reported $\mathrm{FP}_{\mathrm{pass}}$ and \#WN are equal-weighted means
of the three repeat-level proportions and warning counts, respectively.
Checkers with no alerts contribute no denominator. One retained mainline
scan reached its warning cap. Judge labels are not human ground truth.
The Average and Overall Average rows in
Table~\ref{tab:overall-performance} are unweighted means of the reported
metric values over the seven models within each harness and all 21
harness--model combinations, respectively.

\subsection{Human Validation of the LLM Judge}
\label{app:human-validation}

We validate \modelname{claude}{Claude Opus 5}'s false-positive triage, process-quality assessment,
and vulnerability-semantic assessment against independent human annotations
under the backend-specific rubrics in Appendix~\ref{app:verifier-rubric}.
The three annotators have backgrounds in computer science graduate study and
software development.

\paragraph{Sampling and judgment units.}
We sampled 100 of the 159 CSA tasks and 100 of the 141 CodeQL tasks.
The CodeQL allocation comprised 25 tasks each for Java, Python, JavaScript, and Go.
The task allocation followed these backend and language strata, with one
frozen checker submission per task and no score-boundary or outcome
oversampling. Each of the 200 submissions received one process and one
semantic judgment. For each backend, 80 of the 100 sampled tasks contributed
100 individual mainline warnings: 60 tasks contributed one warning each, and
20 contributed two each. The remaining 20 tasks contributed none. This yielded
200 warning judgments, 400 submission-stage judgments, and 600 judge--human pairs.
No item was excluded or weighted. Model, harness, repeat, and pass/fail status
were retained for audit, without allocation quotas.

\paragraph{Human labels and reference.}
All three annotators labeled every item independently, giving 1,800 initial
human labels. Their view contained the warning and source context for
false-positive triage; checker source and trajectory evidence for process
quality; and the fixing patch plus vulnerable and fixed code for semantic
quality. Judge labels and rationales, model and harness identities, and final
verdicts were hidden during initial labeling. A majority of the three binary
labels defined the human reference. For process and semantics, the binary
label records whether the score and applicable discrete gates meet the
backend's rubric: process at least 80/100, CSA semantics at least 7/10, and
CodeQL semantics at least 8/10. The table covers these stage decisions;
numeric-score error and individual rubric-criterion agreement are outside
its scope.

\paragraph{Label distribution and agreement.}
Table~\ref{tab:human-validation} gives separate judge and majority-human
positive counts. Positive means a false positive at the warning stage and an
accepted decision at the process and semantic stages. Each row contains 100
valid pairs. Judge--human agreement uses unweighted Cohen's $\kappa$;
human--human agreement uses three-rater Fleiss' $\kappa$ on the original
independent binary labels. Exact matches range from 89 to 95 per row.
The $2\times2$ cells in table order are
$(21,3,2,74)$, $(23,2,5,70)$, $(67,3,5,25)$, $(76,3,2,19)$,
$(59,7,4,30)$, and $(65,5,3,27)$, with cells ordered as
judge-positive/human-positive, judge-positive/human-negative,
judge-negative/human-positive, and judge-negative/human-negative.
The minority vote is positive on $(3,5,4,3,7,5)$ split items in table order;
the remaining split items have a negative minority vote.

\begin{table}[htbp]
\centering
\small
\caption{Human validation of the three categorical judge stages, with 100
items per row. J+/H+ gives judge and majority-human positive counts; exact
agreement is a count. Split gives items with a nonunanimous initial human
vote.}
\label{tab:human-validation}
\setlength{\tabcolsep}{4pt}
\renewcommand{\arraystretch}{1.08}
\begin{tabular}{@{}llccccc@{}}
\toprule
\textbf{Backend} & \textbf{Stage} & \textbf{J+/H+} & \textbf{Exact} &
\textbf{J--H $\kappa$} & \textbf{H--H $\kappa$} & \textbf{Split} \\
\midrule
CSA & FP triage & 24/23 & 95 & 0.861 & 0.867 & 7 \\
CodeQL & FP triage & 25/28 & 93 & 0.821 & 0.835 & 10 \\
CSA & Process & 70/72 & 92 & 0.806 & 0.852 & 9 \\
CodeQL & Process & 79/78 & 95 & 0.852 & 0.883 & 6 \\
CSA & Semantic & 66/63 & 89 & 0.760 & 0.800 & 14 \\
CodeQL & Semantic & 70/68 & 92 & 0.813 & 0.832 & 11 \\
\midrule
\textbf{Pooled} & \textbf{All stages} & \textbf{334/332} & \textbf{556} &
\textbf{0.852} & \textbf{0.872} & \textbf{57} \\
\bottomrule
\end{tabular}
\end{table}

\paragraph{Adjudication.}
The annotation set has 57 nonunanimous human items. Majority voting resolved
all 57; no item was tied or removed from the denominator. The majority label
changed only the human audit reference and left benchmark verdicts intact.
Fleiss' $\kappa$ uses the 1,800 initial labels before majority resolution.

\paragraph{CSA versus CodeQL and interpretation of $\kappa=0.85$.}
Across the six rows, the 556 matching labels among 600 pairs give exact
agreement of $0.9267$. The judge-positive total is 334 and the human-positive
total is 332, so chance agreement is $0.5060$ and pooled Cohen's
$\kappa=(0.9267-0.5060)/(1-0.5060)=0.852$ to three decimal places.
These calculations give the pooled $\kappa=0.85$ reported in the main text
after rounding to two decimal places. Reproducing the CSA and CodeQL results
requires the frozen task IDs, checker and warning IDs, judge labels, all three
independent human labels, and the adjudicated references. Task-clustered
uncertainty intervals likewise require those records.

\section{Post-Synthesis Verifier Rubric}
\label{app:verifier-rubric}

CSA and CodeQL retain separate frozen executable and scoring policies.
Their common final requirements are normal generation completion, independent
executable evidence, mainline false-positive control, process and semantic
quality, and backend-specific mechanism, scope, and anti-hardcoding checks.
Figure~\ref{fig:evaluation-architecture} gives the numerical thresholds. The
CSA implementation details below describe the \texttt{post-synthesis-v4}
rubric; CodeQL retains its own backend requirements.

\subsection{Independent Replay and Diagnostic Predicates}

Before collecting trajectories, we instantiate and freeze each task's
verifier \(\mathcal V_i\) from its recovered revision pair and pinned
environment \(\mathcal E_i\). Given checker source \(q_i\), the verifier
ignores agent-produced binaries, rebuilds or compiles the candidate in a
clean workspace, and scans the task-specific scope in \(c_i^-\) and
\(c_i^+\). CSA uses a loadable checker plugin; CodeQL uses submitted query
source in its pinned analysis environment. Version-pinned scripts preserve
each backend's context, while isolation prevents artifact leakage.

Replay produces diagnostic multisets \(W_i^-(q_i)\) and \(W_i^+(q_i)\),
with \(N_i^\pm(q_i)=|W_i^\pm(q_i)|\). Executable validity combines three
binary indicators for build success, diagnostic contrast, and patch relevance:
\begin{equation}
\begin{aligned}
 \operatorname{Valid}_i(q_i)&=B_i(q_i)\,C_i(q_i)\,L_i(q_i),\\
 B_i(q_i)&=\mathbf 1[\operatorname{Build}_{\alpha_i}(q_i,\mathcal E_i)],\\
 C_i(q_i)&=\mathcal C_{\alpha_i,i}(W_i^-,W_i^+),\\
 L_i(q_i)&=\mathcal L_{\alpha_i,i}(W_i^-,W_i^+,\Delta_i).
\end{aligned}
\label{eq:verifier-validity}
\end{equation}
Here, \(\mathcal C_{\alpha_i,i}\) is the frozen differential-detection
predicate and \(\mathcal L_{\alpha_i,i}\) is its target/patch-relevance
predicate. CSA contrast requires \(N_i^->N_i^+\) and
\(N_i^+<T_{\mathrm{res}}\), with \(T_{\mathrm{res}}=50\) fixed before
collection. Reports are matched using the line-insensitive signature
\[
 \kappa_i(w)=(\operatorname{checker}(w),\operatorname{file}(w),
              \operatorname{func}(w),\operatorname{msg}(w)).
\]
At least one disappearing report must lie in a patch-modified function.
CodeQL applies its own frozen query, differential, and relevance rules;
CSA's count and matching predicates are not substituted for them. Replay
retains normalized locations, diff-hunk distances, residual diagnostics,
and in- and out-of-patch reports as evidence.

Let \(O_i\) denote the completed objective gate, including required
artifacts, independent replay, and the backend's scope and diagnostic-quality
checks. CSA uses the operational G6 gate below; CodeQL uses its own frozen
objective gate. In both cases,
\(O_i\Rightarrow\operatorname{Valid}_i(q_i)=1\). Executable validity,
process quality, semantic fidelity, and mainline false-positive control
are non-compensating requirements.

\subsection{CSA Deterministic Objective Gate}

This gate invokes no LLM. The source and rebuilt plugin are
\texttt{checker.cpp} and \texttt{checker.so}. The full frozen CSA contrast and
localization predicates must hold in addition to the recorded gate evidence.
\begin{description}
\item[G0.] The task work directory and interaction trace exist.
\item[G1.] The checker source exists and contains at least 50 bytes.
\item[G2.] Independent compilation produces the loadable analyzer artifact.
\item[G3.] \texttt{verify\_result.json} exists and the vulnerable revision
produces at least one diagnostic, $n_{\mathrm{buggy}}>0$.
\item[G4.] Vulnerable and fixed diagnostics satisfy the frozen CSA contrast
rule; equal positive counts are rejected.
\item[G5.] Patch relevance is at least 0.5 and diagnostic sparsity is at least 0.3.
\item[G6.] All G0--G5 requirements pass.
\end{description}
Only G6 makes a run eligible for a nonzero CSA process score. Four additional
objective diagnostics---Differential Detection (15 points), Diagnostic
Sharpness (15), False-Positive and Scope Control (15), and basic CSA
Implementation Quality (4)---are reported for diagnosis. These 49 points are
excluded from $S_p$ to avoid counting executability again as process quality.

\subsection{CSA Process and Semantic Quality}

The process stage examines the patch, vulnerable and fixed functions,
\texttt{state.json}, scope statement, frozen source, synthesis summary, and
validation evidence. Its raw score $Q_i$ comprises Patch Semantic Understanding
(12 points), Detection Plan Quality (12), Plan-to-Code Consistency (12), and
Validation Quality (6), for $Q_{\max}=42$. Repeated reads, redundant commands,
excessive edits, and non-progressing build loops incur a waste penalty
$p_i^{\mathrm{waste}}\in[0,5]$. The process floor is 80/100.

The semantic stage first infers the root cause, trigger and safe conditions,
defect class, propagation requirements, and minimum analyzer mechanism from
the patch and vulnerable functions. It then compares the frozen checker with
those requirements. The ordered mechanism ladder is
\[
\textit{syntax-only}\rightarrow\textit{call-match}\rightarrow
\textit{path-sensitive}\rightarrow\textit{dataflow}\rightarrow
\textit{cross-procedure}.
\]
Mechanism adequacy is classified as \texttt{adequate}, \texttt{partial},
\texttt{inadequate}, or \texttt{overkill}; semantic correctness as
\texttt{yes}, \texttt{partial}, or \texttt{no}. For ten-point correctness and
mechanism-fit subscores $s_{c,i},s_{m,i}$, the process and semantic scores are
\begin{equation}
 S_{p,i}=\mathbf{1}[O_i]\,
 \frac{100\max(0,Q_i-p_i^{\mathrm{waste}})}{Q_{\max}},
 \qquad S_{s,i}=0.6s_{c,i}+0.4s_{m,i}.
 \label{eq:layered-scores}
\end{equation}
The CSA semantic floor is 7/10.
CSA additionally requires \texttt{mechanism\_adapted=adequate},
\texttt{semantic\_correct} in \{\texttt{yes},\texttt{partial}\}, valid parsing,
and no semantic-anchor risk. A patch-adjacent AST match may therefore fail
semantic assessment despite satisfying a diagnostic contrast.

\subsection{CodeQL Quality Gates}
CodeQL follows the layered admission structure described above while retaining backend-specific executable, differential, relevance, scope, and hardcoding predicates.
Its query is independently rebuilt and checked on the vulnerable and fixed revisions before scoring, and its refinement scan uses the shared mainline false-positive gate, with a default sampled FP cutoff of 25\% in the main comparison.
Queries that fail or incompletely satisfy these prerequisites are not scored on process or semantic quality.

The CodeQL process-quality score is a 100-point score with a process floor of 80.
Its dimensions are vulnerability modeling (18 points), observable-evidence planning (14), mechanism selection (18), plan-to-code consistency (18), validation and recovery (14), efficiency and finalization (10), and generalization discipline (8).
The semantic-quality score is a 10-point score in half-point increments with a floor of 8; it assesses the vulnerable sink, trigger and dataflow, safety barrier, CodeQL mechanism, and report-scope generalization.
In addition to the numeric floor, automatic admission requires correct vulnerability and safety-barrier modeling, sound dataflow or control-flow support, acceptable scope generalization, an adequate mechanism, and no patch hardcoding or project-specific scope.

These CodeQL-specific conditions refine the common final requirements without changing the non-compensating relation between executable validity, false-positive control, process quality, and semantic quality.

\subsection{Mainline False Positives and Final Verdict}

The frozen candidate's mainline scan and at most ten judged warnings provide
false-positive evidence. In the main comparison, the sampled FP rate must be
at most 25\% for both CSA and CodeQL. For a nonempty, validly judged sample,
FP rate is the fraction classified as false positives.
Sample composition, warning-level decisions, and the
backend's complete refinement outcome are stored. A backend-recognized clean
scan follows its frozen empty-sample policy.

Let $F_i$ denote normal generation completion; $J_i$ the required valid
judge decisions; $R_i$ the mainline false-positive gate, including the
backend's full refinement rules; $M_i$ adequate mechanism fit;
$C_i^{\mathrm{sem}}$ the required semantic-correctness classification;
and $H_i,A_i$ hardcoding and semantic-anchor risks. With
$t_{\mathrm{CSA}}=7$ and $t_{\mathrm{CodeQL}}=8$, the final verdict is
\begin{equation}
\begin{aligned}
 \textsc{Pass}_i={}&F_i\land O_i\land J_i\land R_i
 \land[S_{p,i}\geq80]\land[S_{s,i}\geq t_{\alpha_i}]\\
 &\land M_i\land C_i^{\mathrm{sem}}\land\neg(H_i\lor A_i).
\end{aligned}
\label{eq:verdict-gate}
\end{equation}
Section~\ref{sec:fp-threshold-sensitivity} varies only the cutoff within $R_i$,
replacing its default 25\% value while retaining the saved warning samples,
judge decisions, and all remaining acceptance requirements.
The main evaluation uses \modelname{claude}{Claude Opus 5}~\citep{anthropic2026opus5} for staged
false-positive, process, and semantic judgments and retains each stage's
first valid decision. A valid low score is a failure, not a reason to ask
again.

A valid failed gate or low score yields a failed quality outcome; no new
checker is requested. An unavailable judge or missing verification evidence
cannot grant a pass and follows the bounded infrastructure-recovery rules,
without discarding valid earlier decisions. Only a normally completed
generation whose candidate satisfies every required gate receives final
\texttt{passed}. Objective-only scoring cannot issue that final outcome.

\section{Representative CSA Checker Patterns}
\label{app:csa-patterns}

This appendix summarizes a selected set of implementation patterns from
official LLVM~18 Clang Static Analyzer (CSA) checkers~\citep{llvm2026csacheckers}.
The catalog applies only to the CSA-backed C/C++ subset of
\textsc{CheckerBench}; tasks in other language ecosystems use the
corresponding analyzer-native rule APIs. We retain patterns that expose
distinct synthesis challenges rather than reproducing complete checker
implementations.

\begin{table}[H]
    \centering
    \small
    \caption{Selected CSA checker patterns used as implementation references
    for C/C++ tasks, from stateless local checks to path-sensitive lifecycle
    analyses with persistent program state.}
    \label{tab:csa-pattern-catalog}
    \setlength{\tabcolsep}{4pt}
    \renewcommand{\arraystretch}{1.12}
    \begin{tabularx}{\linewidth}{@{}
        >{\RaggedRight\arraybackslash}p{0.28\linewidth}
        >{\RaggedRight\arraybackslash}p{0.31\linewidth}
        >{\RaggedRight\arraybackslash}X@{}}
        \toprule
        \textbf{Pattern / official checker} &
        \textbf{Callbacks / state} &
        \textbf{Synthesis challenge} \\
        \midrule
        Constraint violation\par\texttt{DivZeroChecker} &
        \texttt{PreStmt}\par\emph{State:} none &
        Split feasible states and report only a definitely zero denominator. \\
        \addlinespace[3pt]
        Undefined value\par\texttt{UndefResultChecker} &
        \texttt{PostStmt}\par\emph{State:} none &
        Identify the undefined operand and retain useful diagnostic provenance. \\
        \addlinespace[3pt]
        Resource lifecycle\par\texttt{SimpleStreamChecker} &
        \texttt{PostCall}, \texttt{PreCall}, \texttt{DeadSymbols},
        \texttt{PointerEscape}\par\emph{State:} symbol-to-state map &
        Track opened and closed resources, double close, escape, and leaks. \\
        \addlinespace[3pt]
        Nullability propagation\par\texttt{NullabilityChecker} &
        Binding, calls, returns, locations, assumptions, and events\par
        \emph{State:} two maps and an invariant flag &
        Propagate annotations and constraints across multiple semantic events. \\
        \addlinespace[3pt]
        Memory management\par\texttt{MallocChecker} &
        Calls, locations, dead symbols, escapes, and assumptions\par
        \emph{State:} region-state map plus auxiliary maps and sets &
        Model allocation families, deallocation compatibility, leaks,
        double free, and use after free. \\
        \bottomrule
    \end{tabularx}
\end{table}

\paragraph{Stateless constraint checks.}
\texttt{DivZeroChecker} and \texttt{UndefResultChecker} illustrate the
smallest useful checker structures. They register a single statement callback,
query the symbolic value already maintained by CSA, and emit a report without
introducing persistent state. The former uses a dual feasibility assumption to
separate zero and nonzero denominators; the latter inspects an undefined result
after expression evaluation and attributes it to the relevant operand. These
patterns test whether an agent can select the correct callback phase and
translate a local defect predicate into analyzer constraints.

\paragraph{Finite-state resource tracking.}
\texttt{SimpleStreamChecker} represents the minimal lifecycle pattern. A
program-state map associates each resource symbol with an opened or closed
state. Post-call handling records acquisition, pre-call handling validates and
updates release, dead-symbol handling reports leaks, and pointer-escape
handling stops tracking resources whose ownership can no longer be established.
This pattern requires the synthesized checker to coordinate several callbacks
while preserving a compact state machine.

\paragraph{Multi-callback semantic propagation.}
\texttt{NullabilityChecker} and \texttt{MallocChecker} represent the complex
end of the catalog. They combine call, statement, location, assumption,
liveness, and escape events with multiple state traits. Their purpose here is
not to provide templates for direct copying, but to expose the design decisions
required for repository-level synthesis: choosing semantic events, defining
stable state keys, suppressing infeasible or duplicate reports, and attaching
path evidence to diagnostics.

\paragraph{Common reporting contract.}
Across the selected patterns, a checker obtains symbolic values from the
current analysis state, creates a sink or non-fatal error node as appropriate,
constructs a path-sensitive bug report, marks relevant symbols and source
ranges, optionally installs a visitor for path notes, and finally emits the
report. \textsc{CheckerBench} evaluates the resulting behavior through independent
rebuilding and vulnerable--fixed diagnostic contrast rather than by matching
these reference implementations.

\section{Additional Experimental Results}
\label{app:additional-results}

\subsection[KNighter Comparison Details]{\harnessname{knighter}{KNighter} Comparison Details}
\label{app:knighter-comparison}

\paragraph{Tasks and comparison.}
The comparison in Section~\ref{sec:knighter-comparison} uses the 61
Linux-kernel tasks from \harnessname{knighter}{KNighter}'s original benchmark~\citep{yang2025knighter}.
Each task takes a bug-fixing commit as input and requires a CSA checker that
distinguishes buggy code from its patched version. We evaluate our OpenCode-based
method, \harnessname{knighter}{KNighter}, and our method without skills, all using
\modelname{deepseek}{DeepSeek-V4-Pro}.
Table~\ref{tab:knighter-comparison} reports valid-checker counts and rates.

\paragraph{Time budget.}
All three configurations use the same 50-minute (3,000-second) active
generation budget per task. The budget covers model responses, tool execution,
and checker-repair iterations across the complete workflow, including
\harnessname{knighter}{KNighter}'s internal agent coordination. Recorded admission queues and
maintenance pauses, as well as subsequent independent verification, are
excluded under the timing convention in Appendix~\ref{app:time-budget-sensitivity}.
The submitted checker is frozen when generation terminates.

\paragraph{Accessible documentation.}
Our method and \harnessname{knighter}{KNighter} receive the same fixing patch, source-code context,
repository revisions, and pinned CSA environment. Both can consult the same
read-only Linux/CSA documentation, analyzer API references, and official
checker examples described in Appendix~\ref{app:task-packaging}.
These shared resources exclude held-out reference checkers and task-specific
checker solutions.

\paragraph{Skill library.}
Both systems have access to the same reusable skill and strategy content for
patch interpretation, checker design, build recovery, and validation.
The content is exposed through each system's native interface.
The \emph{w/o skills} variant of our method removes these skill packages and strategy
references while retaining the same task inputs, factual API documentation,
tools, and execution budget.

\paragraph{Task prompts.}
The task-facing instructions standardize the synthesis objective, supplied
patch and code context, available resources, required checker artifact, and
validation requirements. System-specific role prompts and internal
orchestration retain each harness's workflow, while the task information and
reusable knowledge made available to our method and \harnessname{knighter}{KNighter} are held constant.

\paragraph{Validation criteria.}
All three configurations use the same independent verifier, scan scopes,
and acceptance predicates. The verifier rebuilds each frozen checker from
source in the pinned CSA environment and applies the same build-success,
vulnerable--fixed diagnostic-contrast, and patch-relevance checks
(Equation~\ref{eq:verifier-validity}). Agent-reported success does not determine
the verdict. Table~\ref{tab:knighter-comparison} reports the number of valid
checkers and the corresponding percentage over the same 61 tasks.

% Source: output/ablation-final-20260920/results/*.csv and manifest.json.
\subsection{Ablation Outcomes and Coverage}
\label{app:ablation-details}

Each configuration covers the same 300 tasks: 159 CSA and 141 CodeQL tasks.
The three new variants have 900 frozen candidates in total; Full reuses
the original 300 candidates. We independently evaluated the 21 historical
artifacts whose assessment had stopped at the generation deadline. All 21
fail an artifact requirement, including three with no submitted source;
Full therefore has 146/300 artifact passes and 119/300 full passes
(39.67\%). No candidate was regenerated or edited during this completion.

\paragraph{Configurations and generation.}
Each new configuration uses one generation per task. \emph{w/o Skills}
jointly removes skill packages and strategy references while retaining
task inputs, repository tools, and factual API documentation.
\emph{w/o Execution Feedback} retains skills but returns neutral receipts
for build and analyzer calls; actual outputs remain outside the agent
workspace. Both use \harnessname{open-code}{OpenCode} with a 3,000-second active budget.
\emph{w/o Harness} makes one model call without tools or skills, using the
full patch and deterministically extracted code context. The new variants
share the model route, sampling settings, and 16,384-token output limit
per response; one-shot generation has no feedback or repair call.

\paragraph{Endpoints and comparability.}
Artifact pass applies the compilation, differential, semantic,
anti-hardcoding, and false-positive gates without requiring Process or
normal agent termination. Frozen artifacts are evaluated even when
generation reaches the time limit. The new runs retain their frozen
\modelname{claude}{Claude Opus 5} judgments for Process, Semantic, and task gates; a separate
\modelname{claude}{Opus~5} audit of all saved alerts supplies the reported
$\mathrm{FP}_{\mathrm{pass}}$ without changing the original verdicts.
Historical Full differs
from the new runs in execution date, client context limit, and workspace/tool
isolation, so the comparisons do not isolate causal effects of individual
components. The one-shot result also reflects its strict output protocol:
237/300 requests yield invalid or incomplete structured submissions, and
another 53 fail compilation.

\begin{table}[tp]
\centering\small
\caption{Artifact pass and differences from Full. All rates use the full
task denominator; CSA and CodeQL use 159 and 141 tasks. Intervals use
10,000 paired task bootstrap samples.}
\label{tab:ablation-artifact}
\setlength{\tabcolsep}{3pt}
\begin{tabular*}{\textwidth}{@{\extracolsep{\fill}}lrrrr@{}}
\toprule
Configuration & All (\%) & $\Delta$ pp [95\% CI] & CSA (\%) & CodeQL (\%) \\
\midrule
Full & 48.67 & Reference & 44.03 & 53.90 \\
w/o Skills & 18.67 & $-30.00\;[-36.33,-23.67]$ & 15.72 & 21.99 \\
w/o Execution Feedback & 20.33 & $-28.33\;[-34.67,-22.00]$ & 17.61 & 23.40 \\
w/o Harness & 1.33 & $-47.33\;[-53.00,-41.33]$ & 0.00 & 2.84 \\
\bottomrule
\end{tabular*}
\end{table}

Using the displayed rates in Table~\ref{tab:component-ablation}, the Pass@1
differences from Full are $-31.34$, $-26.34$, and $-39.34$ percentage points,
respectively. The artifact-pass intervals in Table~\ref{tab:ablation-artifact}
describe variation over paired tasks for one generation per task; they do
not cover model resampling, shared-CVE dependence, or differences between
historical and new execution conditions.

Process/Semantic valid sample counts are 239/239 for Full, 119/122 for
w/o Skills, 100/100 for w/o Execution Feedback, and 10/10 for w/o Harness.
These conditional means are not averages over all 300 tasks. Three
unavailable Process judgments are excluded instead of using heuristic
fallback scores. Missing evidence is never treated as a pass. One failed
candidate has seven false positives among ten sampled alerts, exceeding its
25\% threshold.

\modelname{claude}{Opus~5} independently adjudicates every saved alert of the 25, 40, and 1
passed new checkers (16, 64, and 0 alerts, respectively). All 80 labels
are resolved after source-evidence completion; the Full FP figure reuses
the 78-alert \modelname{claude}{Opus~5} audit from the main experiment. These post hoc labels
define $\mathrm{FP}_{\mathrm{pass}}$, not the original sampled FP gates.
The false-positive counts are 2/78 (2.56\%) for Full, 1/16 for w/o Skills, and 16/64
for w/o Execution Feedback. One-shot has no saved alerts, so its N/A entry
does not denote a zero false-positive proportion.
No-alert checkers add nothing to the alert denominator. The metric is a
false-discovery proportion over saved alerts, not a repository-wide
false-positive rate.

Because the ablation variants have substantially lower pass rates and therefore fewer successful trajectories, Table~\ref{tab:component-ablation} averages turns and tool calls over all 300 tasks, including failed trajectories. This differs from Table~\ref{tab:overall-performance}, which reports interaction metrics over successful trajectories only. Table~\ref{tab:ablation-common-cost}
additionally restricts each comparison to tasks whose artifacts pass in
both Full and the corresponding variant.

\begin{table}[tp]
\centering\small
\caption{Interaction cost on tasks passed by both artifacts. Each pair
uses its own common task subset; these means should not be compared across
rows as if the subsets were identical.}
\label{tab:ablation-common-cost}
\setlength{\tabcolsep}{3pt}
\begin{tabular*}{\textwidth}{@{\extracolsep{\fill}}lrrrrr@{}}
\toprule
Variant & Tasks & Full turns & Variant turns & Full tools & Variant tools \\
\midrule
w/o Skills & 42 & 49.29 & 64.02 & 63.21 & 89.24 \\
w/o Execution Feedback & 43 & 45.21 & 63.19 & 60.63 & 84.09 \\
w/o Harness & 3 & 22.00 & 1.00 & 39.33 & 0.00 \\
\bottomrule
\end{tabular*}
\end{table}

\subsection{Time-Budget Sensitivity}
\label{app:time-budget-sensitivity}

Tables~\ref{tab:budget-sensitivity}--\ref{tab:budget-sensitivity-hermes}
provide the complete time-budget results for the analysis in
Section~\ref{sec:budget-sensitivity}. The tables cover all seven models and
three harnesses at 10, 20, 30, 40, and 50 minutes of active generation time,
using the completed-within-budget policy below.

\paragraph{Timing and eligibility.}
We retrospectively apply the cutoffs to existing trajectories, retaining each
task's current evaluated attempt. Active generation time includes model
responses and tool execution, but excludes recorded API and compute admission
queues, maintenance pauses, and subsequent external verification and judging.
Only final artifacts completed within a cutoff contribute their existing
evaluation outcomes; unfinished runs count as unsuccessful for Pass@1, Build,
and Diff.\ SR, with all 300 tasks retained in each denominator.
Intermediate artifacts are not evaluated, and agents are not rerun with shorter
deadlines. Realized execution time and trajectory length can differ from the
budget because agents may terminate early, batch tool operations, or spend time
waiting for responses and recovering from build failures.

\paragraph{Conditional metrics and model differences.}
Process and Semantic average available valid scores from eligible completed
runs; Avg.\ Turns and Avg.\ Tool Calls average eligible passed runs.
$\mathrm{FP}_{\mathrm{pass}}$ pools recorded alerts from those passed runs,
using the same adjudicated labels as the main table. These statistics
describe changing subsets of completed runs, not within-trajectory quality
improvements. Averaged across harnesses, the 40-to-50-minute Pass@1 gains are
largest for \modelname{qwen}{Qwen3.8-Max} (3.67 percentage points) and \modelname{hunyuan}{HY4-Preview} (3.33), compared
with 0.44 for \modelname{openai}{GPT-5.6} and 3.00 for \modelname{claude}{Claude Opus 4.8}. These differences describe
when the retained trajectories finish successfully.

% The 50-minute FP values include author-verified corrections in
% figures/performance/table3_fp_corrections.csv.
\begingroup
\scriptsize
\setlength{\tabcolsep}{0.5pt}
\renewcommand{\arraystretch}{1.12}
\begin{longtable}{@{}lr*{8}{r}@{}}
\caption{Time-budget sensitivity on \textsc{CheckerBench}: \harnessname{open-code}{OpenCode}.}\label{tab:budget-sensitivity}\\
\toprule[1.5pt]
\textbf{Model} & \shortstack{\textbf{Budget}\\\textbf{(min)}}
& \textbf{Pass@1 (\%)} $\uparrow$ & \textbf{Build (\%)} $\uparrow$
& \textbf{Diff. SR (\%)} $\uparrow$
& $\mathbf{FP}_{\mathrm{pass}}$ \textbf{(\%)} $\downarrow$
& \textbf{Process} $\uparrow$ & \textbf{Semantic} $\uparrow$
& \shortstack{\textbf{Avg.}\\\textbf{Turns} $\downarrow$}
& \shortstack{\textbf{Avg.}\\\textbf{Tool Calls} $\downarrow$} \\
\midrule
\endfirsthead
\multicolumn{10}{@{}l}{\textit{Table \thetable\ continued}}\\
\toprule[1.5pt]
\textbf{Model} & \shortstack{\textbf{Budget}\\\textbf{(min)}}
& \textbf{Pass@1 (\%)} $\uparrow$ & \textbf{Build (\%)} $\uparrow$
& \textbf{Diff. SR (\%)} $\uparrow$
& $\mathbf{FP}_{\mathrm{pass}}$ \textbf{(\%)} $\downarrow$
& \textbf{Process} $\uparrow$ & \textbf{Semantic} $\uparrow$
& \shortstack{\textbf{Avg.}\\\textbf{Turns} $\downarrow$}
& \shortstack{\textbf{Avg.}\\\textbf{Tool Calls} $\downarrow$} \\
\midrule
\endhead
\midrule
\multicolumn{10}{r@{}}{\textit{Continued on next page}}\\
\endfoot
\bottomrule[1.5pt]
\multicolumn{10}{@{}p{\textwidth}@{}}{\scriptsize
Budgets threshold existing trajectories by active generation time, excluding
recorded resource-admission queues and maintenance pauses. External verification
and judging are outside the generation clock. Only final artifacts completed by
the cutoff contribute; unfinished runs contribute no success to Pass@1, Build,
or Diff.\ SR, whose denominator remains 300. Process and Semantic average
available valid scores among eligible completed runs. Turns and Tool Calls
average only eligible passed runs. $\mathrm{FP}_{\mathrm{pass}}$ is the pooled
alert-level FP fraction for those passed runs, using the same adjudicated
labels as the main table. A dash denotes no eligible score
or a zero alert denominator. Per-cell sample sizes accompany the CSV.
These are retrospective completion curves, not separately rerun deadline experiments.
}\\
\endlastfoot
\modelname{openai}{GPT-5.6} & 10 & 12.00 & 25.33 & 25.00 & 0.00 & 84.57 & 8.26 & 23.50 & 54.53 \\*
 & 20 & 23.33 & 64.67 & 58.67 & 0.00 & 82.20 & 7.83 & 26.89 & 57.60 \\*
 & 30 & 28.33 & 85.00 & 75.33 & 0.00 & 80.96 & 7.75 & 28.40 & 58.85 \\*
 & 40 & 31.33 & 95.33 & 83.00 & 0.00 & 80.71 & 7.68 & 29.22 & 60.06 \\*
 & \textbf{50} & 31.67 & 98.33 & 85.33 & 0.00 & 80.22 & 7.68 & 29.19 & 60.13 \\
\addlinespace[3pt]
\modelname{claude}{Claude Opus 4.8} & 10 & 8.33 & 9.67 & 9.67 & 0.00 & 88.28 & 9.04 & 26.20 & 39.36 \\*
 & 20 & 24.00 & 31.67 & 31.33 & 0.00 & 89.42 & 8.84 & 33.60 & 50.25 \\*
 & 30 & 36.00 & 53.67 & 53.00 & 5.36 & 88.11 & 8.51 & 39.41 & 57.24 \\*
 & 40 & 42.33 & 70.33 & 69.67 & 14.04 & 87.07 & 8.35 & 42.78 & 62.07 \\*
 & \textbf{50} & 45.33 & 81.67 & 81.00 & 13.79 & 86.42 & 8.32 & 44.78 & 64.35 \\
\addlinespace[3pt]
\modelname{qwen}{Qwen3.8-Max} & 10 & 6.33 & 7.67 & 7.67 & \textemdash & 92.15 & 9.22 & 24.11 & 31.68 \\*
 & 20 & 22.33 & 33.33 & 33.00 & 0.00 & 91.72 & 8.69 & 31.15 & 41.39 \\*
 & 30 & 29.33 & 50.33 & 50.00 & 0.00 & 91.16 & 8.50 & 33.18 & 44.00 \\*
 & 40 & 35.00 & 63.00 & 62.67 & 6.00 & 90.58 & 8.49 & 34.84 & 46.10 \\*
 & \textbf{50} & 37.33 & 69.33 & 69.00 & 5.77 & 90.15 & 8.46 & 35.96 & 47.54 \\
\addlinespace[3pt]
\modelname{deepseek}{DeepSeek-V4-Pro} & 10 & 13.33 & 22.67 & 21.67 & 0.00 & 89.98 & 8.40 & 34.25 & 47.88 \\*
 & 20 & 26.00 & 55.67 & 54.33 & 5.26 & 88.05 & 8.18 & 39.64 & 52.94 \\*
 & 30 & 34.33 & 75.33 & 73.67 & 5.00 & 86.34 & 8.19 & 44.62 & 58.42 \\*
 & 40 & 38.67 & 87.67 & 85.67 & 2.82 & 85.32 & 8.16 & 49.52 & 63.78 \\*
 & \textbf{50} & 39.67 & 92.67 & 90.67 & 2.56 & 84.89 & 8.12 & 51.09 & 65.28 \\
\addlinespace[3pt]
\modelname{glm}{GLM-5.2} & 10 & 3.67 & 4.33 & 4.33 & 0.00 & 91.61 & 9.58 & 24.82 & 43.73 \\*
 & 20 & 18.00 & 34.33 & 33.00 & 14.63 & 88.84 & 8.20 & 33.98 & 53.26 \\*
 & 30 & 27.33 & 59.67 & 58.33 & 10.34 & 87.41 & 8.15 & 38.62 & 58.17 \\*
 & 40 & 31.33 & 73.00 & 71.00 & 10.34 & 86.90 & 8.12 & 40.23 & 59.55 \\*
 & \textbf{50} & 32.33 & 77.33 & 75.33 & 10.34 & 86.73 & 8.11 & 41.21 & 61.02 \\
\addlinespace[3pt]
\modelname{kimi}{Kimi-K3} & 10 & 8.33 & 9.67 & 9.67 & 0.00 & 92.15 & 9.34 & 22.32 & 33.28 \\*
 & 20 & 22.33 & 40.33 & 40.00 & 0.00 & 90.99 & 8.53 & 26.10 & 36.84 \\*
 & 30 & 30.67 & 62.00 & 60.33 & 2.56 & 89.67 & 8.36 & 28.93 & 39.91 \\*
 & 40 & 36.67 & 76.00 & 74.33 & 1.75 & 88.90 & 8.34 & 31.32 & 42.35 \\*
 & \textbf{50} & 38.67 & 83.33 & 81.67 & 1.59 & 88.48 & 8.30 & 32.42 & 43.53 \\
\addlinespace[3pt]
\modelname{hunyuan}{HY4-Preview} & 10 & 5.33 & 7.67 & 7.00 & 0.00 & 90.07 & 8.76 & 25.06 & 32.44 \\*
 & 20 & 19.00 & 34.33 & 33.67 & 18.18 & 89.70 & 8.40 & 33.04 & 41.02 \\*
 & 30 & 28.67 & 55.67 & 55.00 & 14.29 & 88.20 & 8.32 & 38.64 & 47.22 \\*
 & 40 & 32.00 & 65.33 & 64.67 & 13.98 & 87.54 & 8.31 & 40.32 & 49.15 \\*
 & \textbf{50} & 33.67 & 72.67 & 71.33 & 12.38 & 86.90 & 8.32 & 41.28 & 50.15 \\
\end{longtable}
\endgroup

\begingroup
\scriptsize
\setlength{\tabcolsep}{0.5pt}
\renewcommand{\arraystretch}{1.12}
\begin{longtable}{@{}lr*{8}{r}@{}}
\caption{Time-budget sensitivity on \textsc{CheckerBench}: \harnessname{claude-code}{Claude Code}.}\label{tab:budget-sensitivity-claude}\\
\toprule[1.5pt]
\textbf{Model} & \shortstack{\textbf{Budget}\\\textbf{(min)}}
& \textbf{Pass@1 (\%)} $\uparrow$ & \textbf{Build (\%)} $\uparrow$
& \textbf{Diff. SR (\%)} $\uparrow$
& $\mathbf{FP}_{\mathrm{pass}}$ \textbf{(\%)} $\downarrow$
& \textbf{Process} $\uparrow$ & \textbf{Semantic} $\uparrow$
& \shortstack{\textbf{Avg.}\\\textbf{Turns} $\downarrow$}
& \shortstack{\textbf{Avg.}\\\textbf{Tool Calls} $\downarrow$} \\
\midrule
\endfirsthead
\multicolumn{10}{@{}l}{\textit{Table \thetable\ continued}}\\
\toprule[1.5pt]
\textbf{Model} & \shortstack{\textbf{Budget}\\\textbf{(min)}}
& \textbf{Pass@1 (\%)} $\uparrow$ & \textbf{Build (\%)} $\uparrow$
& \textbf{Diff. SR (\%)} $\uparrow$
& $\mathbf{FP}_{\mathrm{pass}}$ \textbf{(\%)} $\downarrow$
& \textbf{Process} $\uparrow$ & \textbf{Semantic} $\uparrow$
& \shortstack{\textbf{Avg.}\\\textbf{Turns} $\downarrow$}
& \shortstack{\textbf{Avg.}\\\textbf{Tool Calls} $\downarrow$} \\
\midrule
\endhead
\bottomrule[1.5pt]
\multicolumn{10}{r@{}}{\textit{Continued on next page}}\\
\endfoot
\bottomrule[1.5pt]
\multicolumn{10}{@{}p{\textwidth}@{}}{\scriptsize
Budgets threshold existing trajectories by active generation time, excluding
recorded resource-admission queues and maintenance pauses. External verification
and judging are outside the generation clock. Only final artifacts completed by
the cutoff contribute; unfinished runs contribute no success to Pass@1, Build,
or Diff.\ SR, whose denominator remains 300. Process and Semantic average
available valid scores among eligible completed runs. Turns and Tool Calls
average only eligible passed runs. $\mathrm{FP}_{\mathrm{pass}}$ is the pooled
alert-level FP fraction for those passed runs, using the same adjudicated
labels as the main table. A dash denotes no eligible score
or a zero alert denominator. Per-cell sample sizes accompany the CSV.
These are retrospective completion curves, not separately rerun deadline experiments.
}\\
\endlastfoot
\modelname{openai}{GPT-5.6} & 10 & 10.00 & 19.33 & 19.00 & 0.00 & 88.09 & 8.24 & 28.37 & 51.77 \\*
 & 20 & 23.67 & 55.67 & 53.33 & 0.00 & 82.77 & 7.73 & 31.34 & 54.06 \\*
 & 30 & 29.67 & 76.33 & 69.00 & 2.63 & 81.89 & 7.70 & 34.12 & 57.94 \\*
 & 40 & 33.00 & 92.00 & 81.33 & 2.44 & 81.38 & 7.67 & 35.18 & 59.87 \\*
 & \textbf{50} & 34.00 & 93.67 & 83.00 & 2.44 & 81.41 & 7.69 & 35.71 & 60.58 \\
\addlinespace[3pt]
\modelname{claude}{Claude Opus 4.8} & 10 & 9.00 & 10.00 & 9.00 & 0.00 & 87.40 & 9.28 & 17.52 & 30.11 \\*
 & 20 & 18.33 & 25.33 & 24.00 & 0.00 & 88.30 & 8.85 & 23.36 & 35.98 \\*
 & 30 & 28.67 & 45.33 & 44.00 & 2.94 & 87.77 & 8.47 & 28.49 & 41.53 \\*
 & 40 & 32.67 & 55.33 & 54.00 & 2.50 & 87.26 & 8.30 & 30.23 & 43.16 \\*
 & \textbf{50} & 35.00 & 63.67 & 62.33 & 2.33 & 86.88 & 8.21 & 32.60 & 45.63 \\
\addlinespace[3pt]
\modelname{qwen}{Qwen3.8-Max} & 10 & 7.00 & 8.67 & 8.67 & 0.00 & 92.41 & 9.12 & 24.00 & 29.33 \\*
 & 20 & 23.00 & 39.33 & 39.00 & 4.76 & 90.96 & 8.50 & 27.90 & 34.01 \\*
 & 30 & 27.33 & 51.33 & 51.00 & 5.88 & 90.33 & 8.41 & 30.12 & 36.60 \\*
 & 40 & 31.33 & 61.00 & 60.67 & 4.88 & 89.86 & 8.35 & 32.17 & 39.36 \\*
 & \textbf{50} & 34.67 & 67.33 & 67.00 & 4.55 & 89.58 & 8.38 & 33.91 & 41.57 \\
\addlinespace[3pt]
\modelname{deepseek}{DeepSeek-V4-Pro} & 10 & 14.00 & 22.67 & 22.33 & 0.00 & 89.43 & 8.56 & 35.07 & 48.45 \\*
 & 20 & 23.67 & 49.33 & 49.00 & 0.00 & 87.26 & 8.29 & 39.44 & 53.23 \\*
 & 30 & 31.00 & 73.67 & 72.67 & 0.00 & 85.86 & 8.26 & 42.24 & 56.84 \\*
 & 40 & 36.67 & 86.00 & 84.33 & 1.79 & 85.61 & 8.29 & 45.71 & 60.81 \\*
 & \textbf{50} & 38.33 & 91.33 & 89.67 & 1.72 & 85.25 & 8.27 & 47.17 & 62.50 \\
\addlinespace[3pt]
\modelname{glm}{GLM-5.2} & 10 & 1.00 & 1.33 & 1.33 & \textemdash & 95.75 & 9.38 & 21.67 & 40.33 \\*
 & 20 & 12.67 & 23.00 & 22.33 & 0.00 & 93.37 & 8.89 & 30.92 & 41.89 \\*
 & 30 & 23.00 & 47.00 & 45.33 & 0.00 & 91.55 & 8.64 & 34.01 & 45.87 \\*
 & 40 & 28.00 & 62.67 & 61.00 & 0.00 & 90.55 & 8.48 & 36.69 & 47.92 \\*
 & \textbf{50} & 33.00 & 76.33 & 74.67 & 0.00 & 89.67 & 8.39 & 38.75 & 49.93 \\
\addlinespace[3pt]
\modelname{kimi}{Kimi-K3} & 10 & 11.00 & 13.33 & 13.33 & 0.00 & 92.51 & 9.33 & 24.88 & 31.21 \\*
 & 20 & 22.67 & 47.67 & 47.00 & 0.00 & 90.11 & 8.29 & 27.32 & 33.54 \\*
 & 30 & 30.33 & 66.33 & 65.33 & 3.28 & 88.49 & 8.27 & 30.00 & 36.31 \\*
 & 40 & 35.67 & 78.33 & 77.00 & 2.86 & 88.04 & 8.29 & 31.46 & 37.83 \\*
 & \textbf{50} & 38.00 & 83.67 & 82.33 & 2.74 & 87.79 & 8.34 & 31.81 & 38.61 \\
\pagebreak
\modelname{hunyuan}{HY4-Preview} & 10 & 2.00 & 2.33 & 2.33 & 0.00 & 91.99 & 9.14 & 24.00 & 32.33 \\*
 & 20 & 8.33 & 13.33 & 13.00 & 0.00 & 93.03 & 8.66 & 34.04 & 43.00 \\*
 & 30 & 16.00 & 29.67 & 29.00 & 0.00 & 91.86 & 8.63 & 43.00 & 53.85 \\*
 & 40 & 19.67 & 41.00 & 39.33 & 0.00 & 91.42 & 8.44 & 47.02 & 58.29 \\*
 & \textbf{50} & 23.67 & 49.33 & 47.33 & 0.00 & 90.83 & 8.35 & 50.66 & 62.23 \\
\end{longtable}
\endgroup

\begingroup
\scriptsize
\setlength{\tabcolsep}{0.5pt}
\renewcommand{\arraystretch}{0.9}
\begin{longtable}{@{}lr*{8}{r}@{}}
\caption{Time-budget sensitivity on \textsc{CheckerBench}: \harnessname{hermes-agent}{Hermes Agent}.}\label{tab:budget-sensitivity-hermes}\\
\toprule[1.5pt]
\textbf{Model} & \shortstack{\textbf{Budget}\\\textbf{(min)}}
& \textbf{Pass@1 (\%)} $\uparrow$ & \textbf{Build (\%)} $\uparrow$
& \textbf{Diff. SR (\%)} $\uparrow$
& $\mathbf{FP}_{\mathrm{pass}}$ \textbf{(\%)} $\downarrow$
& \textbf{Process} $\uparrow$ & \textbf{Semantic} $\uparrow$
& \shortstack{\textbf{Avg.}\\\textbf{Turns} $\downarrow$}
& \shortstack{\textbf{Avg.}\\\textbf{Tool Calls} $\downarrow$} \\
\midrule
\endfirsthead
\multicolumn{10}{@{}l}{\textit{Table \thetable\ continued}}\\
\toprule[1.5pt]
\textbf{Model} & \shortstack{\textbf{Budget}\\\textbf{(min)}}
& \textbf{Pass@1 (\%)} $\uparrow$ & \textbf{Build (\%)} $\uparrow$
& \textbf{Diff. SR (\%)} $\uparrow$
& $\mathbf{FP}_{\mathrm{pass}}$ \textbf{(\%)} $\downarrow$
& \textbf{Process} $\uparrow$ & \textbf{Semantic} $\uparrow$
& \shortstack{\textbf{Avg.}\\\textbf{Turns} $\downarrow$}
& \shortstack{\textbf{Avg.}\\\textbf{Tool Calls} $\downarrow$} \\
\midrule
\endhead
\midrule
\multicolumn{10}{r@{}}{\textit{Continued on next page}}\\
\endfoot
\bottomrule[1.5pt]
\multicolumn{10}{@{}p{\textwidth}@{}}{\scriptsize
Budgets threshold existing trajectories by active generation time, excluding
recorded resource-admission queues and maintenance pauses. External verification
and judging are outside the generation clock. Only final artifacts completed by
the cutoff contribute; unfinished runs contribute no success to Pass@1, Build,
or Diff.\ SR, whose denominator remains 300. Process and Semantic average
available valid scores among eligible completed runs. Turns and Tool Calls
average only eligible passed runs. $\mathrm{FP}_{\mathrm{pass}}$ is the pooled
alert-level FP fraction for those passed runs, using the same adjudicated
labels as the main table. A dash denotes no eligible score
or a zero alert denominator. Per-cell sample sizes accompany the CSV.
These are retrospective completion curves, not separately rerun deadline experiments.
}\\
\endlastfoot
\rowcolor{appendixharness}
\multicolumn{10}{c}{\harnessname{hermes-agent}{\textbf{\textit{Hermes Agent}}}} \\
\modelname{openai}{GPT-5.6} & 10 & 9.67 & 27.67 & 20.00 & \textemdash & 83.85 & 7.97 & 24.00 & 44.21 \\*
 & 20 & 19.33 & 71.67 & 48.67 & 0.00 & 81.95 & 7.67 & 25.55 & 46.62 \\*
 & 30 & 22.33 & 89.67 & 59.33 & 0.00 & 80.49 & 7.57 & 25.90 & 47.19 \\*
 & 40 & 23.00 & 94.67 & 62.33 & 0.00 & 80.22 & 7.57 & 25.94 & 46.91 \\*
 & \textbf{50} & 23.00 & 100.00 & 64.00 & 0.00 & 79.85 & 7.50 & 25.94 & 46.91 \\
\addlinespace[2pt]
\modelname{claude}{Claude Opus 4.8} & 10 & 6.00 & 8.00 & 8.00 & 0.00 & 88.23 & 9.23 & 21.00 & 30.50 \\*
 & 20 & 16.00 & 26.33 & 26.00 & 0.00 & 88.30 & 8.59 & 28.31 & 41.40 \\*
 & 30 & 26.33 & 47.00 & 46.67 & 0.00 & 88.15 & 8.48 & 38.24 & 53.27 \\*
 & 40 & 30.67 & 62.67 & 62.33 & 0.00 & 87.23 & 8.41 & 42.08 & 58.26 \\*
 & \textbf{50} & 34.33 & 80.33 & 78.67 & 5.45 & 85.58 & 8.32 & 44.84 & 61.89 \\
\addlinespace[2pt]
\modelname{qwen}{Qwen3.8-Max} & 10 & 3.00 & 3.67 & 3.67 & \textemdash & 93.73 & 9.05 & 20.56 & 28.67 \\*
 & 20 & 10.00 & 15.00 & 14.67 & 0.00 & 93.15 & 9.07 & 29.37 & 38.43 \\*
 & 30 & 15.67 & 25.33 & 25.00 & 0.00 & 92.03 & 8.76 & 34.09 & 44.11 \\*
 & 40 & 19.67 & 34.33 & 34.00 & 5.88 & 91.35 & 8.62 & 36.93 & 47.54 \\*
 & \textbf{50} & 25.00 & 44.00 & 42.33 & 11.11 & 90.92 & 8.53 & 40.55 & 51.95 \\
\addlinespace[2pt]
\modelname{deepseek}{DeepSeek-V4-Pro} & 10 & 9.33 & 15.00 & 15.00 & 0.00 & 90.90 & 8.78 & 29.21 & 50.00 \\*
 & 20 & 22.00 & 49.33 & 48.00 & 0.00 & 88.11 & 8.34 & 34.56 & 56.45 \\*
 & 30 & 27.67 & 70.67 & 69.00 & 0.00 & 86.73 & 8.09 & 37.27 & 59.41 \\*
 & 40 & 30.33 & 85.00 & 82.33 & 0.00 & 85.93 & 8.07 & 38.24 & 60.56 \\*
 & \textbf{50} & 32.33 & 94.00 & 90.67 & 0.00 & 85.29 & 8.03 & 39.82 & 62.37 \\
\addlinespace[2pt]
\modelname{glm}{GLM-5.2} & 10 & 3.67 & 6.00 & 6.00 & \textemdash & 90.69 & 8.59 & 25.18 & 43.36 \\*
 & 20 & 12.33 & 28.00 & 27.33 & 0.00 & 87.49 & 8.27 & 35.41 & 55.84 \\*
 & 30 & 19.00 & 53.00 & 51.67 & 0.00 & 85.54 & 7.99 & 39.93 & 60.49 \\*
 & 40 & 21.67 & 70.00 & 68.67 & 0.00 & 83.98 & 7.90 & 41.37 & 62.00 \\*
 & \textbf{50} & 22.33 & 79.67 & 78.00 & 0.00 & 82.87 & 7.81 & 42.85 & 63.22 \\
\addlinespace[2pt]
\modelname{kimi}{Kimi-K3} & 10 & 7.67 & 14.67 & 14.33 & 0.00 & 90.14 & 8.38 & 21.96 & 29.43 \\*
 & 20 & 16.33 & 41.00 & 39.33 & 3.64 & 87.42 & 8.30 & 25.90 & 34.55 \\*
 & 30 & 22.67 & 67.67 & 63.00 & 4.76 & 84.78 & 8.18 & 31.13 & 40.41 \\*
 & 40 & 24.00 & 78.67 & 72.67 & 4.76 & 83.07 & 8.06 & 31.65 & 41.04 \\*
 & \textbf{50} & 24.33 & 87.00 & 79.67 & 4.76 & 82.62 & 8.08 & 32.73 & 42.12 \\
\addlinespace[2pt]
\modelname{hunyuan}{HY4-Preview} & 10 & 1.33 & 1.33 & 1.33 & \textemdash & 94.00 & 9.00 & 24.50 & 35.25 \\*
 & 20 & 7.67 & 12.00 & 12.00 & \textemdash & 92.19 & 8.91 & 31.39 & 42.83 \\*
 & 30 & 12.67 & 25.33 & 25.00 & \textemdash & 89.98 & 8.69 & 37.32 & 48.32 \\*
 & 40 & 17.33 & 38.33 & 37.00 & 0.00 & 90.34 & 8.53 & 44.44 & 56.00 \\*
 & \textbf{50} & 21.67 & 57.67 & 52.67 & 0.00 & 88.42 & 8.44 & 49.69 & 62.06 \\
\end{longtable}
\endgroup

% Generated by scripts/build_performance_analysis.py; do not edit numeric cells.
\subsection{Language, Repository, and CWE Performance}
\label{app:model-performance-analysis}
These breakdowns average the same three independent repeats as
Table~\ref{tab:overall-performance} and Figure~\ref{fig:model-performance-analysis}.
Each model--harness pair has the same 300 tasks in each repeat.
The main heatmaps average the three harness-specific estimates equally;
the tables below retain each harness.
The abbreviated model names in the CWE tables follow the full names
in Table~\ref{tab:overall-performance}.
All rates are percentages and are descriptive; small task subsets do not
establish reliable model rankings. The downloadable CSVs preserve full precision.
\paragraph{Languages and analyzers.}
C/C++ uses CSA (159 tasks); Python (51), Go (30), Java (30), and
JavaScript (30) use CodeQL. Language and analyzer effects cannot be separated.

\begingroup
\scriptsize
\setlength{\tabcolsep}{3pt}
\setlength{\LTcapwidth}{\textwidth}
\renewcommand{\arraystretch}{0.9}
\begin{longtable}{@{}p{.24\textwidth}*{5}{>{\centering\arraybackslash}p{.13\textwidth}}@{}}
\caption{Language-specific Pass@1 (\%) by harness. All tasks in each language remain in the denominator.}\label{tab:language-performance}\\
\toprule[1.5pt]
\textbf{Model} & \textbf{C/C++} & \textbf{Python} & \textbf{Go} & \textbf{Java} & \textbf{JavaScript} \\
\midrule
\endfirsthead
\multicolumn{6}{c}{\tablename\ \thetable{} (continued)}\\
\toprule[1.5pt]
\textbf{Model} & \textbf{C/C++} & \textbf{Python} & \textbf{Go} & \textbf{Java} & \textbf{JavaScript} \\
\midrule
\endhead
\midrule
\multicolumn{6}{r}{Continued on next page}\\
\endfoot
\bottomrule[1.5pt]
\endlastfoot
\rowcolor{appendixharness}
\multicolumn{6}{c}{\harnessname{open-code}{\textbf{\textit{OpenCode}}}}\\
\modelname{openai}{GPT-5.6} & 12.58 & 62.75 & 46.67 & 46.67 & 50.00 \\
\modelname{claude}{Claude Opus 4.8} & 26.42 & 72.55 & 60.00 & 70.00 & 60.00 \\
\modelname{qwen}{Qwen3.8-Max} & 15.72 & 72.55 & 46.67 & 53.33 & 66.67 \\
\modelname{deepseek}{DeepSeek-V4-Pro} & 28.30 & 58.82 & 53.33 & 40.00 & 53.33 \\
\modelname{glm}{GLM-5.2} & 16.35 & 60.78 & 53.33 & 46.67 & 33.33 \\
\modelname{kimi}{Kimi-K3} & 27.04 & 68.63 & 36.67 & 50.00 & 40.00 \\
\modelname{hunyuan}{HY4-Preview} & 13.84 & 54.90 & 60.00 & 53.33 & 56.67 \\
\midrule
\rowcolor{appendixharness}
\multicolumn{6}{c}{\harnessname{claude-code}{\textbf{\textit{Claude Code}}}}\\
\modelname{openai}{GPT-5.6} & 22.01 & 62.75 & 30.00 & 46.67 & 40.00 \\
\modelname{claude}{Claude Opus 4.8} & 21.38 & 62.75 & 53.33 & 40.00 & 36.67 \\
\modelname{qwen}{Qwen3.8-Max} & 14.47 & 54.90 & 60.00 & 66.67 & 50.00 \\
\modelname{deepseek}{DeepSeek-V4-Pro} & 28.30 & 56.86 & 46.67 & 43.33 & 46.67 \\
\modelname{glm}{GLM-5.2} & 20.13 & 64.71 & 43.33 & 43.33 & 26.67 \\
\modelname{kimi}{Kimi-K3} & 23.90 & 60.78 & 46.67 & 56.67 & 46.67 \\
\modelname{hunyuan}{HY4-Preview} & 5.66 & 52.94 & 36.67 & 43.33 & 36.67 \\
\midrule
\rowcolor{appendixharness}
\multicolumn{6}{c}{\harnessname{hermes-agent}{\textbf{\textit{Hermes Agent}}}}\\
\modelname{openai}{GPT-5.6} & 10.69 & 41.18 & 36.67 & 30.00 & 36.67 \\
\modelname{claude}{Claude Opus 4.8} & 22.01 & 60.78 & 40.00 & 40.00 & 43.33 \\
\modelname{qwen}{Qwen3.8-Max} & 3.77 & 60.78 & 40.00 & 43.33 & 43.33 \\
\modelname{deepseek}{DeepSeek-V4-Pro} & 16.98 & 54.90 & 60.00 & 43.33 & 36.67 \\
\modelname{glm}{GLM-5.2} & 5.66 & 50.98 & 43.33 & 33.33 & 30.00 \\
\modelname{kimi}{Kimi-K3} & 11.32 & 43.14 & 30.00 & 36.67 & 43.33 \\
\modelname{hunyuan}{HY4-Preview} & 2.52 & 45.10 & 36.67 & 46.67 & 43.33 \\
\end{longtable}
\endgroup

% Let the longtable use the remaining page space and break naturally.
\begingroup
\scriptsize
\setlength{\tabcolsep}{3pt}
\setlength{\LTcapwidth}{\textwidth}
\renewcommand{\arraystretch}{0.9}
\begin{longtable}{@{}p{.26\textwidth}*{3}{>{\centering\arraybackslash}p{.225\textwidth}}@{}}
\caption{Repository sensitivity by harness (\%). Task equal uses 300 tasks; Repo equal averages 167 repository rates; No Linux uses 193 tasks in 166 repositories.}\label{tab:repository-performance}\\
\toprule[1.5pt]
\textbf{Model} & \textbf{Task equal} & \textbf{Repo equal} & \textbf{No Linux} \\
\midrule
\endfirsthead
\multicolumn{4}{c}{\tablename\ \thetable{} (continued)}\\
\toprule[1.5pt]
\textbf{Model} & \textbf{Task equal} & \textbf{Repo equal} & \textbf{No Linux} \\
\midrule
\endhead
\midrule
\multicolumn{4}{r}{Continued on next page}\\
\endfoot
\bottomrule[1.5pt]
\endlastfoot
\rowcolor{appendixharness}
\multicolumn{4}{c}{\harnessname{open-code}{\textbf{\textit{OpenCode}}}}\\
\modelname{openai}{GPT-5.6} & 31.67 & 46.23 & 42.49 \\
\modelname{claude}{Claude Opus 4.8} & 45.33 & 57.32 & 53.37 \\
\modelname{qwen}{Qwen3.8-Max} & 37.33 & 53.73 & 49.74 \\
\modelname{deepseek}{DeepSeek-V4-Pro} & 39.67 & 49.17 & 46.11 \\
\modelname{glm}{GLM-5.2} & 32.33 & 43.63 & 39.38 \\
\modelname{kimi}{Kimi-K3} & 38.67 & 47.92 & 44.56 \\
\modelname{hunyuan}{HY4-Preview} & 33.67 & 48.60 & 43.52 \\
\midrule
\rowcolor{appendixharness}
\multicolumn{4}{c}{\harnessname{claude-code}{\textbf{\textit{Claude Code}}}}\\
\modelname{openai}{GPT-5.6} & 34.00 & 42.60 & 39.90 \\
\modelname{claude}{Claude Opus 4.8} & 35.00 & 44.51 & 40.41 \\
\modelname{qwen}{Qwen3.8-Max} & 34.67 & 49.44 & 45.60 \\
\modelname{deepseek}{DeepSeek-V4-Pro} & 38.33 & 46.68 & 43.52 \\
\modelname{glm}{GLM-5.2} & 33.00 & 42.50 & 38.86 \\
\modelname{kimi}{Kimi-K3} & 38.00 & 49.01 & 44.56 \\
\modelname{hunyuan}{HY4-Preview} & 23.67 & 38.06 & 33.68 \\
\midrule
\rowcolor{appendixharness}
\multicolumn{4}{c}{\harnessname{hermes-agent}{\textbf{\textit{Hermes Agent}}}}\\
\modelname{openai}{GPT-5.6} & 23.00 & 32.88 & 31.09 \\
\modelname{claude}{Claude Opus 4.8} & 34.33 & 43.70 & 40.41 \\
\modelname{qwen}{Qwen3.8-Max} & 25.00 & 40.54 & 36.79 \\
\modelname{deepseek}{DeepSeek-V4-Pro} & 32.33 & 42.64 & 38.34 \\
\modelname{glm}{GLM-5.2} & 22.33 & 35.51 & 31.61 \\
\modelname{kimi}{Kimi-K3} & 24.33 & 34.25 & 31.09 \\
\modelname{hunyuan}{HY4-Preview} & 21.67 & 35.99 & 32.12 \\
\end{longtable}
\endgroup

\paragraph{Interaction turns.}
Figure~\ref{fig:model-performance-analysis}(c) compares all 21 combinations.
Pass@1 uses all 300 tasks; Avg.\ Turns uses passed trajectories, including
subagents. Different task sets do not establish a causal effect of more turns.
Passed-trajectory counts appear in the companion summary.

\paragraph{Complete CWE breakdown.}
The following tables include all 85 explicit CWE labels and the 14 tasks
without an explicit label (Other). A task contributes once to each distinct
CWE label it carries; 14 tasks have two labels. Consequently the row counts
sum to 314, not 300. Other does not pool known low-frequency CWEs.
The main figure selects the eight explicit labels with at least ten tasks
using counts alone, before examining performance. This display threshold
is not a claim of statistical precision. CWE groups also differ in language,
analyzer, and repository composition.
\begingroup
\scriptsize
\setlength{\tabcolsep}{3.5pt}
\renewcommand{\arraystretch}{1.08}
\begin{longtable}{@{}lrrrrrrrr@{}}
\caption{Complete CWE-specific Pass@1 (\%) for \harnessname{open-code}{OpenCode}. $n$ counts distinct tasks, not harnesses or repeats.}\label{tab:cwe-performance-opencode}\\
\toprule
CWE & $n$ & \modelname{openai}{GPT} & \modelname{claude}{Claude} & \modelname{qwen}{Qwen} & \modelname{deepseek}{DeepSeek} & \modelname{glm}{GLM} & \modelname{kimi}{Kimi} & \modelname{hunyuan}{HY4} \\
\midrule
\endfirsthead
\multicolumn{9}{c}{\tablename\ \thetable{} (continued)}\\
\toprule
CWE & $n$ & \modelname{openai}{GPT} & \modelname{claude}{Claude} & \modelname{qwen}{Qwen} & \modelname{deepseek}{DeepSeek} & \modelname{glm}{GLM} & \modelname{kimi}{Kimi} & \modelname{hunyuan}{HY4} \\
\midrule
\endhead
\midrule
\multicolumn{9}{r}{Continued on next page}\\
\endfoot
\bottomrule
\endlastfoot
CWE-787 & 26 & 7.69 & 26.92 & 19.23 & 30.77 & 15.38 & 26.92 & 3.85 \\
CWE-125 & 21 & 23.81 & 28.57 & 23.81 & 23.81 & 19.05 & 33.33 & 42.86 \\
CWE-22 & 15 & 60.00 & 73.33 & 53.33 & 60.00 & 53.33 & 53.33 & 26.67 \\
CWE-476 & 13 & 7.69 & 30.77 & 15.38 & 30.77 & 15.38 & 23.08 & 7.69 \\
CWE-79 & 12 & 33.33 & 66.67 & 58.33 & 41.67 & 41.67 & 33.33 & 66.67 \\
CWE-362 & 12 & 8.33 & 25.00 & 8.33 & 8.33 & 8.33 & 25.00 & 16.67 \\
CWE-416 & 12 & 25.00 & 25.00 & 8.33 & 50.00 & 25.00 & 25.00 & 25.00 \\
CWE-400 & 10 & 40.00 & 50.00 & 40.00 & 50.00 & 50.00 & 70.00 & 60.00 \\
CWE-908 & 9 & 11.11 & 33.33 & 0.00 & 33.33 & 11.11 & 33.33 & 22.22 \\
CWE-667 & 8 & 12.50 & 12.50 & 12.50 & 0.00 & 37.50 & 12.50 & 0.00 \\
CWE-190 & 7 & 14.29 & 14.29 & 0.00 & 28.57 & 0.00 & 0.00 & 14.29 \\
CWE-415 & 7 & 28.57 & 14.29 & 28.57 & 28.57 & 28.57 & 28.57 & 0.00 \\
CWE-20 & 6 & 33.33 & 83.33 & 33.33 & 33.33 & 16.67 & 50.00 & 66.67 \\
CWE-611 & 6 & 83.33 & 83.33 & 66.67 & 33.33 & 50.00 & 66.67 & 50.00 \\
CWE-617 & 6 & 0.00 & 16.67 & 16.67 & 33.33 & 16.67 & 16.67 & 0.00 \\
CWE-119 & 5 & 0.00 & 0.00 & 0.00 & 0.00 & 0.00 & 20.00 & 0.00 \\
CWE-191 & 5 & 0.00 & 40.00 & 0.00 & 20.00 & 0.00 & 60.00 & 20.00 \\
CWE-200 & 5 & 40.00 & 20.00 & 60.00 & 60.00 & 40.00 & 60.00 & 60.00 \\
CWE-377 & 5 & 80.00 & 100.00 & 60.00 & 100.00 & 100.00 & 100.00 & 60.00 \\
CWE-401 & 5 & 20.00 & 80.00 & 20.00 & 60.00 & 60.00 & 40.00 & 60.00 \\
CWE-601 & 5 & 40.00 & 40.00 & 40.00 & 20.00 & 20.00 & 60.00 & 60.00 \\
CWE-770 & 5 & 40.00 & 20.00 & 40.00 & 0.00 & 20.00 & 20.00 & 40.00 \\
CWE-78 & 4 & 25.00 & 25.00 & 75.00 & 50.00 & 25.00 & 50.00 & 25.00 \\
CWE-129 & 4 & 0.00 & 25.00 & 25.00 & 25.00 & 25.00 & 0.00 & 25.00 \\
CWE-674 & 4 & 0.00 & 25.00 & 50.00 & 25.00 & 0.00 & 0.00 & 25.00 \\
CWE-94 & 3 & 100.00 & 100.00 & 66.67 & 100.00 & 66.67 & 100.00 & 100.00 \\
CWE-189 & 3 & 66.67 & 100.00 & 33.33 & 100.00 & 66.67 & 0.00 & 33.33 \\
CWE-287 & 3 & 33.33 & 66.67 & 66.67 & 66.67 & 0.00 & 66.67 & 66.67 \\
CWE-326 & 3 & 100.00 & 100.00 & 66.67 & 100.00 & 33.33 & 66.67 & 66.67 \\
CWE-1333 & 3 & 66.67 & 100.00 & 66.67 & 66.67 & 66.67 & 66.67 & 33.33 \\
CWE-23 & 2 & 50.00 & 100.00 & 100.00 & 50.00 & 50.00 & 50.00 & 100.00 \\
CWE-59 & 2 & 50.00 & 100.00 & 100.00 & 100.00 & 100.00 & 50.00 & 50.00 \\
CWE-74 & 2 & 50.00 & 50.00 & 100.00 & 50.00 & 50.00 & 100.00 & 50.00 \\
CWE-77 & 2 & 50.00 & 50.00 & 100.00 & 100.00 & 100.00 & 50.00 & 100.00 \\
CWE-284 & 2 & 50.00 & 50.00 & 50.00 & 50.00 & 50.00 & 50.00 & 0.00 \\
CWE-338 & 2 & 50.00 & 100.00 & 50.00 & 50.00 & 100.00 & 50.00 & 100.00 \\
CWE-352 & 2 & 100.00 & 100.00 & 100.00 & 50.00 & 0.00 & 100.00 & 50.00 \\
CWE-502 & 2 & 100.00 & 100.00 & 100.00 & 100.00 & 50.00 & 100.00 & 100.00 \\
CWE-532 & 2 & 50.00 & 100.00 & 100.00 & 0.00 & 50.00 & 50.00 & 50.00 \\
CWE-672 & 2 & 50.00 & 50.00 & 50.00 & 100.00 & 50.00 & 100.00 & 50.00 \\
CWE-772 & 2 & 0.00 & 50.00 & 100.00 & 50.00 & 0.00 & 0.00 & 0.00 \\
CWE-835 & 2 & 0.00 & 50.00 & 0.00 & 0.00 & 0.00 & 50.00 & 0.00 \\
CWE-863 & 2 & 100.00 & 100.00 & 100.00 & 50.00 & 100.00 & 50.00 & 50.00 \\
CWE-29 & 1 & 100.00 & 100.00 & 100.00 & 100.00 & 0.00 & 0.00 & 0.00 \\
CWE-88 & 1 & 0.00 & 0.00 & 100.00 & 100.00 & 100.00 & 100.00 & 0.00 \\
CWE-91 & 1 & 100.00 & 100.00 & 100.00 & 100.00 & 100.00 & 100.00 & 100.00 \\
CWE-120 & 1 & 0.00 & 0.00 & 0.00 & 0.00 & 0.00 & 0.00 & 0.00 \\
CWE-122 & 1 & 0.00 & 0.00 & 0.00 & 0.00 & 0.00 & 0.00 & 0.00 \\
CWE-131 & 1 & 100.00 & 0.00 & 0.00 & 0.00 & 0.00 & 0.00 & 0.00 \\
CWE-193 & 1 & 0.00 & 100.00 & 100.00 & 0.00 & 0.00 & 0.00 & 0.00 \\
CWE-201 & 1 & 0.00 & 0.00 & 0.00 & 0.00 & 100.00 & 0.00 & 100.00 \\
CWE-203 & 1 & 0.00 & 100.00 & 100.00 & 0.00 & 0.00 & 0.00 & 0.00 \\
CWE-209 & 1 & 100.00 & 100.00 & 100.00 & 100.00 & 100.00 & 100.00 & 100.00 \\
CWE-250 & 1 & 0.00 & 0.00 & 100.00 & 0.00 & 100.00 & 0.00 & 0.00 \\
CWE-256 & 1 & 0.00 & 100.00 & 100.00 & 0.00 & 100.00 & 100.00 & 0.00 \\
CWE-264 & 1 & 100.00 & 0.00 & 0.00 & 100.00 & 0.00 & 100.00 & 100.00 \\
CWE-266 & 1 & 0.00 & 0.00 & 0.00 & 0.00 & 0.00 & 0.00 & 0.00 \\
CWE-295 & 1 & 100.00 & 100.00 & 100.00 & 100.00 & 100.00 & 100.00 & 100.00 \\
CWE-312 & 1 & 0.00 & 100.00 & 100.00 & 0.00 & 100.00 & 100.00 & 0.00 \\
CWE-327 & 1 & 100.00 & 100.00 & 0.00 & 100.00 & 100.00 & 100.00 & 100.00 \\
CWE-331 & 1 & 0.00 & 0.00 & 0.00 & 0.00 & 0.00 & 0.00 & 0.00 \\
CWE-347 & 1 & 100.00 & 0.00 & 100.00 & 100.00 & 100.00 & 0.00 & 100.00 \\
CWE-364 & 1 & 0.00 & 0.00 & 0.00 & 0.00 & 0.00 & 0.00 & 0.00 \\
CWE-366 & 1 & 100.00 & 100.00 & 0.00 & 100.00 & 100.00 & 0.00 & 0.00 \\
CWE-369 & 1 & 0.00 & 0.00 & 0.00 & 0.00 & 0.00 & 0.00 & 0.00 \\
CWE-399 & 1 & 0.00 & 0.00 & 0.00 & 0.00 & 0.00 & 0.00 & 0.00 \\
CWE-407 & 1 & 0.00 & 0.00 & 100.00 & 100.00 & 100.00 & 0.00 & 0.00 \\
CWE-459 & 1 & 100.00 & 100.00 & 0.00 & 0.00 & 0.00 & 100.00 & 0.00 \\
CWE-460 & 1 & 0.00 & 0.00 & 100.00 & 0.00 & 0.00 & 100.00 & 0.00 \\
CWE-522 & 1 & 100.00 & 100.00 & 100.00 & 0.00 & 100.00 & 0.00 & 100.00 \\
CWE-538 & 1 & 0.00 & 0.00 & 100.00 & 0.00 & 100.00 & 0.00 & 100.00 \\
CWE-552 & 1 & 0.00 & 0.00 & 0.00 & 0.00 & 0.00 & 0.00 & 0.00 \\
CWE-613 & 1 & 0.00 & 100.00 & 100.00 & 0.00 & 0.00 & 0.00 & 100.00 \\
CWE-639 & 1 & 100.00 & 0.00 & 100.00 & 0.00 & 0.00 & 0.00 & 0.00 \\
CWE-668 & 1 & 100.00 & 100.00 & 100.00 & 100.00 & 100.00 & 100.00 & 100.00 \\
CWE-670 & 1 & 100.00 & 0.00 & 100.00 & 0.00 & 100.00 & 0.00 & 0.00 \\
CWE-680 & 1 & 0.00 & 0.00 & 0.00 & 0.00 & 0.00 & 0.00 & 0.00 \\
CWE-692 & 1 & 0.00 & 0.00 & 0.00 & 100.00 & 0.00 & 0.00 & 0.00 \\
CWE-789 & 1 & 0.00 & 100.00 & 100.00 & 100.00 & 100.00 & 0.00 & 100.00 \\
CWE-826 & 1 & 100.00 & 0.00 & 0.00 & 100.00 & 0.00 & 100.00 & 100.00 \\
CWE-843 & 1 & 0.00 & 100.00 & 0.00 & 100.00 & 100.00 & 100.00 & 0.00 \\
CWE-862 & 1 & 100.00 & 100.00 & 100.00 & 0.00 & 100.00 & 100.00 & 0.00 \\
CWE-918 & 1 & 0.00 & 100.00 & 100.00 & 100.00 & 0.00 & 0.00 & 100.00 \\
CWE-1188 & 1 & 100.00 & 100.00 & 0.00 & 100.00 & 0.00 & 0.00 & 100.00 \\
CWE-1336 & 1 & 100.00 & 100.00 & 100.00 & 100.00 & 100.00 & 100.00 & 100.00 \\
Other & 14 & 35.71 & 57.14 & 35.71 & 50.00 & 42.86 & 57.14 & 50.00 \\
\end{longtable}
\endgroup

\begingroup
\scriptsize
\setlength{\tabcolsep}{3.5pt}
\renewcommand{\arraystretch}{1.08}
\begin{longtable}{@{}lrrrrrrrr@{}}
\caption{Complete CWE-specific Pass@1 (\%) for \harnessname{claude-code}{Claude Code}. $n$ counts distinct tasks, not harnesses or repeats.}\label{tab:cwe-performance-claude-code}\\
\toprule
CWE & $n$ & \modelname{openai}{GPT} & \modelname{claude}{Claude} & \modelname{qwen}{Qwen} & \modelname{deepseek}{DeepSeek} & \modelname{glm}{GLM} & \modelname{kimi}{Kimi} & \modelname{hunyuan}{HY4} \\
\midrule
\endfirsthead
\multicolumn{9}{c}{\tablename\ \thetable{} (continued)}\\
\toprule
CWE & $n$ & \modelname{openai}{GPT} & \modelname{claude}{Claude} & \modelname{qwen}{Qwen} & \modelname{deepseek}{DeepSeek} & \modelname{glm}{GLM} & \modelname{kimi}{Kimi} & \modelname{hunyuan}{HY4} \\
\midrule
\endhead
\midrule
\multicolumn{9}{r}{Continued on next page}\\
\endfoot
\bottomrule
\endlastfoot
CWE-787 & 26 & 26.92 & 26.92 & 26.92 & 26.92 & 26.92 & 23.08 & 3.85 \\
CWE-125 & 21 & 14.29 & 9.52 & 28.57 & 28.57 & 28.57 & 38.10 & 4.76 \\
CWE-22 & 15 & 46.67 & 60.00 & 60.00 & 40.00 & 46.67 & 60.00 & 40.00 \\
CWE-476 & 13 & 30.77 & 23.08 & 7.69 & 38.46 & 15.38 & 15.38 & 0.00 \\
CWE-79 & 12 & 25.00 & 25.00 & 50.00 & 41.67 & 33.33 & 25.00 & 33.33 \\
CWE-362 & 12 & 25.00 & 16.67 & 25.00 & 25.00 & 16.67 & 16.67 & 8.33 \\
CWE-416 & 12 & 33.33 & 25.00 & 25.00 & 25.00 & 25.00 & 25.00 & 0.00 \\
CWE-400 & 10 & 40.00 & 40.00 & 40.00 & 40.00 & 40.00 & 40.00 & 50.00 \\
CWE-908 & 9 & 22.22 & 33.33 & 11.11 & 33.33 & 33.33 & 33.33 & 0.00 \\
CWE-667 & 8 & 25.00 & 25.00 & 0.00 & 37.50 & 25.00 & 25.00 & 12.50 \\
CWE-190 & 7 & 14.29 & 0.00 & 14.29 & 14.29 & 14.29 & 0.00 & 0.00 \\
CWE-415 & 7 & 14.29 & 14.29 & 28.57 & 14.29 & 14.29 & 14.29 & 14.29 \\
CWE-20 & 6 & 50.00 & 16.67 & 66.67 & 16.67 & 16.67 & 33.33 & 33.33 \\
CWE-611 & 6 & 66.67 & 100.00 & 83.33 & 50.00 & 66.67 & 83.33 & 83.33 \\
CWE-617 & 6 & 33.33 & 0.00 & 33.33 & 16.67 & 16.67 & 16.67 & 0.00 \\
CWE-119 & 5 & 20.00 & 0.00 & 0.00 & 0.00 & 0.00 & 0.00 & 0.00 \\
CWE-191 & 5 & 40.00 & 20.00 & 0.00 & 40.00 & 0.00 & 40.00 & 0.00 \\
CWE-200 & 5 & 40.00 & 60.00 & 40.00 & 0.00 & 40.00 & 40.00 & 40.00 \\
CWE-377 & 5 & 100.00 & 100.00 & 60.00 & 100.00 & 60.00 & 100.00 & 80.00 \\
CWE-401 & 5 & 20.00 & 80.00 & 20.00 & 60.00 & 40.00 & 40.00 & 20.00 \\
CWE-601 & 5 & 40.00 & 40.00 & 40.00 & 60.00 & 40.00 & 40.00 & 60.00 \\
CWE-770 & 5 & 0.00 & 40.00 & 20.00 & 20.00 & 0.00 & 0.00 & 20.00 \\
CWE-78 & 4 & 25.00 & 50.00 & 50.00 & 75.00 & 50.00 & 75.00 & 25.00 \\
CWE-129 & 4 & 25.00 & 0.00 & 25.00 & 50.00 & 50.00 & 50.00 & 25.00 \\
CWE-674 & 4 & 0.00 & 0.00 & 25.00 & 25.00 & 0.00 & 75.00 & 25.00 \\
CWE-94 & 3 & 100.00 & 66.67 & 100.00 & 100.00 & 66.67 & 66.67 & 66.67 \\
CWE-189 & 3 & 33.33 & 66.67 & 66.67 & 0.00 & 66.67 & 0.00 & 33.33 \\
CWE-287 & 3 & 33.33 & 0.00 & 33.33 & 33.33 & 33.33 & 33.33 & 66.67 \\
CWE-326 & 3 & 66.67 & 66.67 & 66.67 & 100.00 & 66.67 & 66.67 & 0.00 \\
CWE-1333 & 3 & 33.33 & 33.33 & 33.33 & 33.33 & 66.67 & 33.33 & 66.67 \\
CWE-23 & 2 & 50.00 & 100.00 & 50.00 & 50.00 & 0.00 & 50.00 & 0.00 \\
CWE-59 & 2 & 100.00 & 100.00 & 50.00 & 100.00 & 50.00 & 100.00 & 0.00 \\
CWE-74 & 2 & 50.00 & 50.00 & 100.00 & 50.00 & 50.00 & 50.00 & 50.00 \\
CWE-77 & 2 & 100.00 & 100.00 & 100.00 & 50.00 & 50.00 & 100.00 & 100.00 \\
CWE-284 & 2 & 50.00 & 50.00 & 50.00 & 50.00 & 0.00 & 50.00 & 0.00 \\
CWE-338 & 2 & 50.00 & 0.00 & 0.00 & 50.00 & 0.00 & 50.00 & 50.00 \\
CWE-352 & 2 & 50.00 & 100.00 & 0.00 & 100.00 & 50.00 & 100.00 & 0.00 \\
CWE-502 & 2 & 100.00 & 100.00 & 50.00 & 100.00 & 100.00 & 50.00 & 100.00 \\
CWE-532 & 2 & 50.00 & 50.00 & 50.00 & 50.00 & 50.00 & 50.00 & 50.00 \\
CWE-672 & 2 & 50.00 & 50.00 & 50.00 & 0.00 & 50.00 & 0.00 & 0.00 \\
CWE-772 & 2 & 50.00 & 50.00 & 50.00 & 0.00 & 0.00 & 50.00 & 0.00 \\
CWE-835 & 2 & 0.00 & 0.00 & 0.00 & 0.00 & 0.00 & 0.00 & 0.00 \\
CWE-863 & 2 & 50.00 & 50.00 & 100.00 & 50.00 & 100.00 & 50.00 & 100.00 \\
CWE-29 & 1 & 0.00 & 100.00 & 0.00 & 100.00 & 100.00 & 100.00 & 100.00 \\
CWE-88 & 1 & 0.00 & 0.00 & 0.00 & 100.00 & 0.00 & 100.00 & 0.00 \\
CWE-91 & 1 & 100.00 & 100.00 & 100.00 & 100.00 & 100.00 & 100.00 & 100.00 \\
CWE-120 & 1 & 0.00 & 0.00 & 0.00 & 0.00 & 0.00 & 0.00 & 0.00 \\
CWE-122 & 1 & 0.00 & 0.00 & 0.00 & 0.00 & 100.00 & 0.00 & 0.00 \\
CWE-131 & 1 & 0.00 & 0.00 & 0.00 & 0.00 & 0.00 & 0.00 & 0.00 \\
CWE-193 & 1 & 0.00 & 0.00 & 100.00 & 100.00 & 100.00 & 0.00 & 0.00 \\
CWE-201 & 1 & 0.00 & 100.00 & 0.00 & 0.00 & 0.00 & 0.00 & 0.00 \\
CWE-203 & 1 & 0.00 & 0.00 & 0.00 & 0.00 & 0.00 & 0.00 & 100.00 \\
CWE-209 & 1 & 100.00 & 0.00 & 100.00 & 0.00 & 100.00 & 100.00 & 100.00 \\
CWE-250 & 1 & 0.00 & 0.00 & 100.00 & 100.00 & 100.00 & 100.00 & 0.00 \\
CWE-256 & 1 & 0.00 & 0.00 & 100.00 & 0.00 & 0.00 & 0.00 & 0.00 \\
CWE-264 & 1 & 100.00 & 100.00 & 0.00 & 100.00 & 100.00 & 100.00 & 100.00 \\
CWE-266 & 1 & 0.00 & 0.00 & 0.00 & 0.00 & 0.00 & 0.00 & 0.00 \\
CWE-295 & 1 & 100.00 & 100.00 & 100.00 & 100.00 & 100.00 & 100.00 & 100.00 \\
CWE-312 & 1 & 0.00 & 0.00 & 100.00 & 0.00 & 0.00 & 0.00 & 0.00 \\
CWE-327 & 1 & 100.00 & 100.00 & 0.00 & 0.00 & 100.00 & 100.00 & 0.00 \\
CWE-331 & 1 & 0.00 & 0.00 & 100.00 & 0.00 & 0.00 & 0.00 & 0.00 \\
CWE-347 & 1 & 0.00 & 0.00 & 0.00 & 0.00 & 100.00 & 100.00 & 100.00 \\
CWE-364 & 1 & 0.00 & 0.00 & 0.00 & 0.00 & 0.00 & 0.00 & 0.00 \\
CWE-366 & 1 & 100.00 & 100.00 & 0.00 & 0.00 & 100.00 & 0.00 & 100.00 \\
CWE-369 & 1 & 0.00 & 0.00 & 0.00 & 0.00 & 0.00 & 0.00 & 0.00 \\
CWE-399 & 1 & 0.00 & 0.00 & 0.00 & 0.00 & 0.00 & 0.00 & 0.00 \\
CWE-407 & 1 & 100.00 & 0.00 & 0.00 & 0.00 & 0.00 & 0.00 & 0.00 \\
CWE-459 & 1 & 0.00 & 100.00 & 0.00 & 0.00 & 100.00 & 100.00 & 0.00 \\
CWE-460 & 1 & 0.00 & 0.00 & 0.00 & 0.00 & 0.00 & 0.00 & 0.00 \\
CWE-522 & 1 & 0.00 & 100.00 & 100.00 & 100.00 & 0.00 & 100.00 & 0.00 \\
CWE-538 & 1 & 0.00 & 100.00 & 100.00 & 0.00 & 100.00 & 0.00 & 100.00 \\
CWE-552 & 1 & 0.00 & 0.00 & 0.00 & 0.00 & 100.00 & 0.00 & 0.00 \\
CWE-613 & 1 & 0.00 & 0.00 & 100.00 & 100.00 & 0.00 & 100.00 & 100.00 \\
CWE-639 & 1 & 0.00 & 0.00 & 0.00 & 100.00 & 0.00 & 100.00 & 0.00 \\
CWE-668 & 1 & 100.00 & 100.00 & 100.00 & 100.00 & 100.00 & 100.00 & 100.00 \\
CWE-670 & 1 & 100.00 & 100.00 & 0.00 & 100.00 & 100.00 & 0.00 & 0.00 \\
CWE-680 & 1 & 0.00 & 0.00 & 0.00 & 0.00 & 0.00 & 0.00 & 0.00 \\
CWE-692 & 1 & 100.00 & 0.00 & 100.00 & 0.00 & 0.00 & 0.00 & 0.00 \\
CWE-789 & 1 & 100.00 & 0.00 & 100.00 & 100.00 & 100.00 & 100.00 & 100.00 \\
CWE-826 & 1 & 0.00 & 100.00 & 0.00 & 0.00 & 0.00 & 100.00 & 0.00 \\
CWE-843 & 1 & 0.00 & 0.00 & 0.00 & 100.00 & 0.00 & 0.00 & 0.00 \\
CWE-862 & 1 & 100.00 & 100.00 & 100.00 & 100.00 & 0.00 & 100.00 & 0.00 \\
CWE-918 & 1 & 0.00 & 0.00 & 100.00 & 100.00 & 0.00 & 100.00 & 100.00 \\
CWE-1188 & 1 & 100.00 & 100.00 & 100.00 & 100.00 & 0.00 & 0.00 & 0.00 \\
CWE-1336 & 1 & 100.00 & 100.00 & 100.00 & 100.00 & 100.00 & 100.00 & 100.00 \\
Other & 14 & 35.71 & 35.71 & 35.71 & 64.29 & 35.71 & 50.00 & 21.43 \\
\end{longtable}
\endgroup

\begingroup
\scriptsize
\setlength{\tabcolsep}{3.5pt}
\renewcommand{\arraystretch}{1.08}
\begin{longtable}{@{}lrrrrrrrr@{}}
\caption{Complete CWE-specific Pass@1 (\%) for \harnessname{hermes-agent}{Hermes Agent}. $n$ counts distinct tasks, not harnesses or repeats.}\label{tab:cwe-performance-hermes-agent}\\
\toprule
CWE & $n$ & \modelname{openai}{GPT} & \modelname{claude}{Claude} & \modelname{qwen}{Qwen} & \modelname{deepseek}{DeepSeek} & \modelname{glm}{GLM} & \modelname{kimi}{Kimi} & \modelname{hunyuan}{HY4} \\
\midrule
\endfirsthead
\multicolumn{9}{c}{\tablename\ \thetable{} (continued)}\\
\toprule
CWE & $n$ & \modelname{openai}{GPT} & \modelname{claude}{Claude} & \modelname{qwen}{Qwen} & \modelname{deepseek}{DeepSeek} & \modelname{glm}{GLM} & \modelname{kimi}{Kimi} & \modelname{hunyuan}{HY4} \\
\midrule
\endhead
\midrule
\multicolumn{9}{r}{Continued on next page}\\
\endfoot
\bottomrule
\endlastfoot
CWE-787 & 26 & 19.23 & 19.23 & 19.23 & 26.92 & 7.69 & 11.54 & 0.00 \\
CWE-125 & 21 & 19.05 & 9.52 & 14.29 & 14.29 & 4.76 & 9.52 & 14.29 \\
CWE-22 & 15 & 33.33 & 40.00 & 40.00 & 46.67 & 40.00 & 33.33 & 26.67 \\
CWE-476 & 13 & 15.38 & 30.77 & 0.00 & 7.69 & 7.69 & 30.77 & 7.69 \\
CWE-79 & 12 & 33.33 & 33.33 & 75.00 & 33.33 & 33.33 & 25.00 & 58.33 \\
CWE-362 & 12 & 8.33 & 8.33 & 16.67 & 8.33 & 16.67 & 0.00 & 0.00 \\
CWE-416 & 12 & 16.67 & 33.33 & 0.00 & 25.00 & 25.00 & 16.67 & 0.00 \\
CWE-400 & 10 & 40.00 & 30.00 & 30.00 & 20.00 & 30.00 & 30.00 & 30.00 \\
CWE-908 & 9 & 0.00 & 33.33 & 0.00 & 22.22 & 11.11 & 11.11 & 11.11 \\
CWE-667 & 8 & 0.00 & 25.00 & 0.00 & 12.50 & 0.00 & 0.00 & 0.00 \\
CWE-190 & 7 & 14.29 & 0.00 & 0.00 & 14.29 & 0.00 & 0.00 & 14.29 \\
CWE-415 & 7 & 14.29 & 28.57 & 0.00 & 28.57 & 0.00 & 14.29 & 0.00 \\
CWE-20 & 6 & 16.67 & 33.33 & 16.67 & 16.67 & 33.33 & 16.67 & 33.33 \\
CWE-611 & 6 & 50.00 & 83.33 & 33.33 & 66.67 & 50.00 & 50.00 & 50.00 \\
CWE-617 & 6 & 16.67 & 16.67 & 0.00 & 0.00 & 0.00 & 16.67 & 16.67 \\
CWE-119 & 5 & 0.00 & 0.00 & 0.00 & 0.00 & 0.00 & 0.00 & 0.00 \\
CWE-191 & 5 & 20.00 & 60.00 & 0.00 & 40.00 & 0.00 & 20.00 & 0.00 \\
CWE-200 & 5 & 20.00 & 60.00 & 60.00 & 60.00 & 20.00 & 20.00 & 20.00 \\
CWE-377 & 5 & 80.00 & 60.00 & 100.00 & 80.00 & 80.00 & 100.00 & 80.00 \\
CWE-401 & 5 & 0.00 & 60.00 & 0.00 & 80.00 & 20.00 & 20.00 & 20.00 \\
CWE-601 & 5 & 20.00 & 80.00 & 20.00 & 20.00 & 20.00 & 0.00 & 40.00 \\
CWE-770 & 5 & 0.00 & 0.00 & 20.00 & 20.00 & 0.00 & 20.00 & 0.00 \\
CWE-78 & 4 & 25.00 & 50.00 & 75.00 & 50.00 & 75.00 & 25.00 & 75.00 \\
CWE-129 & 4 & 0.00 & 25.00 & 0.00 & 25.00 & 0.00 & 0.00 & 0.00 \\
CWE-674 & 4 & 0.00 & 25.00 & 25.00 & 25.00 & 50.00 & 0.00 & 25.00 \\
CWE-94 & 3 & 100.00 & 66.67 & 33.33 & 100.00 & 66.67 & 33.33 & 66.67 \\
CWE-189 & 3 & 33.33 & 33.33 & 0.00 & 66.67 & 33.33 & 0.00 & 33.33 \\
CWE-287 & 3 & 33.33 & 33.33 & 33.33 & 66.67 & 33.33 & 33.33 & 66.67 \\
CWE-326 & 3 & 66.67 & 100.00 & 66.67 & 66.67 & 66.67 & 100.00 & 33.33 \\
CWE-1333 & 3 & 33.33 & 66.67 & 66.67 & 66.67 & 66.67 & 66.67 & 33.33 \\
CWE-23 & 2 & 50.00 & 50.00 & 50.00 & 100.00 & 50.00 & 0.00 & 0.00 \\
CWE-59 & 2 & 50.00 & 100.00 & 100.00 & 100.00 & 100.00 & 100.00 & 50.00 \\
CWE-74 & 2 & 50.00 & 50.00 & 100.00 & 50.00 & 50.00 & 50.00 & 100.00 \\
CWE-77 & 2 & 100.00 & 100.00 & 50.00 & 100.00 & 50.00 & 50.00 & 100.00 \\
CWE-284 & 2 & 50.00 & 0.00 & 0.00 & 50.00 & 50.00 & 50.00 & 0.00 \\
CWE-338 & 2 & 0.00 & 50.00 & 50.00 & 50.00 & 50.00 & 50.00 & 50.00 \\
CWE-352 & 2 & 50.00 & 100.00 & 0.00 & 0.00 & 50.00 & 100.00 & 0.00 \\
CWE-502 & 2 & 100.00 & 100.00 & 100.00 & 100.00 & 100.00 & 100.00 & 50.00 \\
CWE-532 & 2 & 50.00 & 50.00 & 50.00 & 50.00 & 100.00 & 50.00 & 100.00 \\
CWE-672 & 2 & 0.00 & 50.00 & 0.00 & 50.00 & 50.00 & 0.00 & 0.00 \\
CWE-772 & 2 & 0.00 & 0.00 & 50.00 & 0.00 & 0.00 & 50.00 & 0.00 \\
CWE-835 & 2 & 0.00 & 0.00 & 0.00 & 0.00 & 0.00 & 0.00 & 0.00 \\
CWE-863 & 2 & 50.00 & 50.00 & 100.00 & 50.00 & 50.00 & 50.00 & 50.00 \\
CWE-29 & 1 & 0.00 & 0.00 & 0.00 & 0.00 & 0.00 & 0.00 & 0.00 \\
CWE-88 & 1 & 0.00 & 0.00 & 100.00 & 0.00 & 100.00 & 0.00 & 100.00 \\
CWE-91 & 1 & 100.00 & 100.00 & 100.00 & 100.00 & 100.00 & 0.00 & 0.00 \\
CWE-120 & 1 & 0.00 & 0.00 & 0.00 & 0.00 & 0.00 & 0.00 & 0.00 \\
CWE-122 & 1 & 0.00 & 0.00 & 0.00 & 0.00 & 0.00 & 0.00 & 0.00 \\
CWE-131 & 1 & 100.00 & 100.00 & 0.00 & 100.00 & 0.00 & 0.00 & 0.00 \\
CWE-193 & 1 & 0.00 & 100.00 & 0.00 & 0.00 & 0.00 & 0.00 & 0.00 \\
CWE-201 & 1 & 0.00 & 0.00 & 0.00 & 0.00 & 100.00 & 0.00 & 0.00 \\
CWE-203 & 1 & 0.00 & 0.00 & 0.00 & 0.00 & 0.00 & 0.00 & 0.00 \\
CWE-209 & 1 & 0.00 & 0.00 & 100.00 & 100.00 & 0.00 & 100.00 & 100.00 \\
CWE-250 & 1 & 0.00 & 0.00 & 100.00 & 100.00 & 0.00 & 0.00 & 0.00 \\
CWE-256 & 1 & 0.00 & 0.00 & 0.00 & 0.00 & 100.00 & 0.00 & 0.00 \\
CWE-264 & 1 & 0.00 & 100.00 & 0.00 & 100.00 & 0.00 & 100.00 & 0.00 \\
CWE-266 & 1 & 0.00 & 0.00 & 0.00 & 0.00 & 0.00 & 100.00 & 0.00 \\
CWE-295 & 1 & 100.00 & 100.00 & 100.00 & 100.00 & 100.00 & 100.00 & 100.00 \\
CWE-312 & 1 & 0.00 & 0.00 & 0.00 & 0.00 & 100.00 & 0.00 & 0.00 \\
CWE-327 & 1 & 100.00 & 0.00 & 0.00 & 100.00 & 0.00 & 0.00 & 100.00 \\
CWE-331 & 1 & 0.00 & 0.00 & 100.00 & 100.00 & 0.00 & 0.00 & 0.00 \\
CWE-347 & 1 & 0.00 & 100.00 & 0.00 & 0.00 & 0.00 & 0.00 & 0.00 \\
CWE-364 & 1 & 0.00 & 0.00 & 0.00 & 0.00 & 0.00 & 0.00 & 0.00 \\
CWE-366 & 1 & 100.00 & 0.00 & 100.00 & 0.00 & 0.00 & 100.00 & 0.00 \\
CWE-369 & 1 & 0.00 & 0.00 & 0.00 & 0.00 & 0.00 & 0.00 & 0.00 \\
CWE-399 & 1 & 0.00 & 0.00 & 0.00 & 0.00 & 0.00 & 0.00 & 0.00 \\
CWE-407 & 1 & 0.00 & 0.00 & 0.00 & 100.00 & 0.00 & 100.00 & 100.00 \\
CWE-459 & 1 & 0.00 & 0.00 & 0.00 & 100.00 & 0.00 & 0.00 & 0.00 \\
CWE-460 & 1 & 0.00 & 0.00 & 0.00 & 0.00 & 0.00 & 0.00 & 0.00 \\
CWE-522 & 1 & 100.00 & 100.00 & 100.00 & 0.00 & 0.00 & 100.00 & 0.00 \\
CWE-538 & 1 & 0.00 & 0.00 & 100.00 & 0.00 & 0.00 & 100.00 & 100.00 \\
CWE-552 & 1 & 0.00 & 0.00 & 0.00 & 0.00 & 0.00 & 0.00 & 0.00 \\
CWE-613 & 1 & 0.00 & 0.00 & 0.00 & 100.00 & 100.00 & 0.00 & 0.00 \\
CWE-639 & 1 & 0.00 & 100.00 & 0.00 & 0.00 & 0.00 & 100.00 & 100.00 \\
CWE-668 & 1 & 100.00 & 100.00 & 100.00 & 100.00 & 100.00 & 100.00 & 100.00 \\
CWE-670 & 1 & 0.00 & 100.00 & 0.00 & 0.00 & 0.00 & 0.00 & 100.00 \\
CWE-680 & 1 & 0.00 & 0.00 & 0.00 & 0.00 & 0.00 & 0.00 & 0.00 \\
CWE-692 & 1 & 0.00 & 0.00 & 0.00 & 0.00 & 0.00 & 0.00 & 100.00 \\
CWE-789 & 1 & 100.00 & 0.00 & 100.00 & 100.00 & 0.00 & 100.00 & 0.00 \\
CWE-826 & 1 & 0.00 & 0.00 & 0.00 & 0.00 & 0.00 & 0.00 & 0.00 \\
CWE-843 & 1 & 0.00 & 0.00 & 0.00 & 100.00 & 0.00 & 0.00 & 0.00 \\
CWE-862 & 1 & 0.00 & 0.00 & 0.00 & 0.00 & 0.00 & 0.00 & 100.00 \\
CWE-918 & 1 & 100.00 & 0.00 & 0.00 & 0.00 & 0.00 & 100.00 & 0.00 \\
CWE-1188 & 1 & 100.00 & 100.00 & 0.00 & 0.00 & 0.00 & 100.00 & 0.00 \\
CWE-1336 & 1 & 100.00 & 100.00 & 100.00 & 100.00 & 100.00 & 100.00 & 100.00 \\
Other & 14 & 14.29 & 57.14 & 28.57 & 35.71 & 21.43 & 14.29 & 14.29 \\
\end{longtable}
\endgroup

\end{document}